\documentclass[a4paper,12pt,twoside,openright]{book}

\usepackage[T1]{fontenc}
\usepackage[utf8x]{inputenc}
\usepackage[english]{babel}
\usepackage{lmodern}

\usepackage{graphicx}
\usepackage{fancyhdr}
\usepackage{indentfirst}
\usepackage{setspace}
\usepackage{amsfonts}
\usepackage{amsmath}
\usepackage{amssymb}
\usepackage{cite}
\usepackage{booktabs}
\usepackage{tabularx}
\usepackage{caption}
\usepackage{subcaption}
\usepackage{float}
\usepackage{color}
\usepackage{epstopdf}
\usepackage{anysize}
\usepackage{float}
\usepackage{nccmath}
\usepackage{diagbox}
\usepackage{multirow}
\usepackage{rotating}
\usepackage{mathrsfs}
\usepackage{slashed}
\usepackage{capt-of}
\usepackage{lastpage}
\usepackage{appendix}
\usepackage{epigraph}
\usepackage{enumerate}
\usepackage{stackengine}
\usepackage{lipsum}
\usepackage{mathtools}
\usepackage{xspace}

\usepackage{bm}
\usepackage{lineno}
\usepackage{pdfpages}
\usepackage{enumitem,xcolor}
\newcommand*\rfrac[2]{{}^{#1}\!/_{#2}}

\def\SM{$\mathrm{ SU(3)_C \otimes SU(2)_L \otimes U(1)_Y }$ }
\newcommand{\sm}{{Standard Model }}

\definecolor{darkred}{rgb}{0.6,0,0}
\definecolor{darkpurple}{rgb}{0.5,0,0.5}
\definecolor{darkgreen}{rgb}{0, 0.6, 0}

\usepackage[pagebackref=true]{hyperref}
\hypersetup{
  colorlinks=true,
  allcolors=darkpurple,
  urlcolor=darkred,
  citecolor=darkgreen,
  pdfborder={0 0 1},
  linktocpage=false,
  bookmarks=true,
  pdftitle={Tesis}
}
\renewcommand*{\backref}[1]{}
\renewcommand*{\backrefalt}[4]{[{\tiny%
    \ifcase #1 Not cited.%
          \or Cited on page~#2.%
          \else Cited on pages #2.%
    \fi%
    }]}

\makeatletter
\def \cleardoublepage {\clearpage \if@twoside
\ifodd \c@page
\else
\null\thispagestyle{empty}\clearpage
\fi
\fi}
\makeatother

\def\vev#1{\left\langle #1\right\rangle}
\def\ket#1{\left\vert #1\right\rangle}
\def\bra#1{\left\langle #1\right\vert}
\newcommand{\braket}[2]{\ensuremath{\left\langle #1 | #2 \right\rangle}}
\newcommand{\hc}{\mathrm{h.c.}}

\begin{document}

\setlength{\unitlength}{1cm} 
\thispagestyle{empty}
\begin{center}

\begin{picture}(3.5,2.4)
\put(0,-3){\includegraphics[scale=0.06]{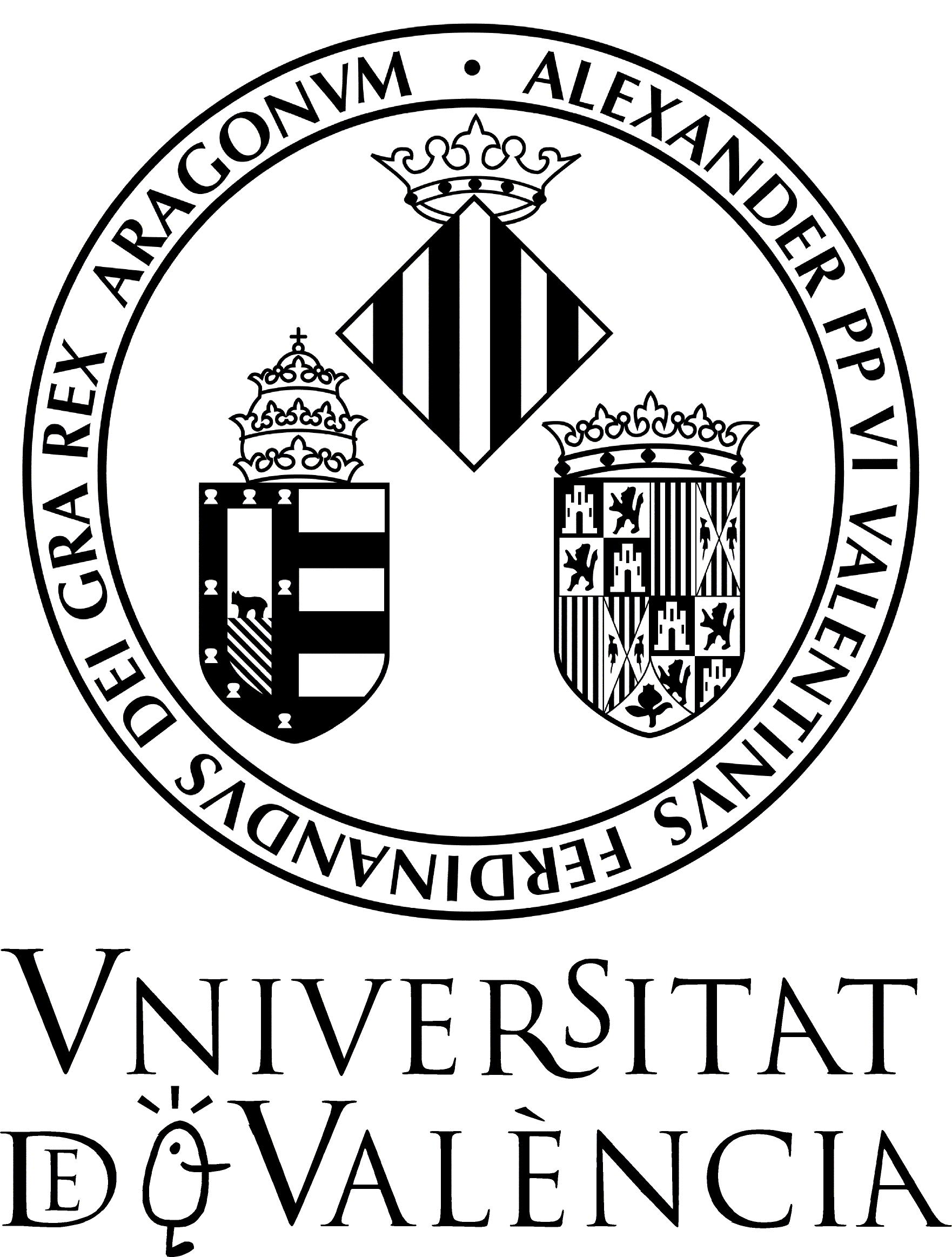}}
\end{picture}
\vspace*{4.55cm}

\textbf{\Large{Theory and phenomenology of Dirac neutrinos: Symmetry breaking patterns, flavour implications and Dark Matter}} \\[4ex]
{\Large Tesi Doctoral\\[1ex]
Programa de Doctorat en F\'isica}

\vspace{\fill}

{\LARGE \bf{Salvador Centelles Chuli\'a}} \\[4ex]

{\Large Director: Dr. Jos\'e Wagner Furtado Valle}\\
{\Large Co-Director: Dr. Rahul Srivastava}\\
{\Large Tutora: Dra. Mar\'ia Amparo T\'ortola Baixauli}\\
[8ex]

{IFIC - CSIC/Universitat de València}\\
{Department de F\'isica Teòrica}\\[2ex]

{\Large València, Junio 2021}

\end{center}

\newpage
\thispagestyle{empty}

\begin{titlepage}
\cleardoublepage
\thispagestyle{empty}
\vspace*{4cm}
\epigraph{\textit{A mis padres por ayudarme y apoyarme siempre. \\A Olaia porque contigo todo es más fácil. \\A mis hermanos aunque crean que estoy tronao.\\
A mi abuelita porque dirá que si lo explico yo se entiende todo y a mi abuelito que le hubiera encantado leerla. \\ A Ravioli por ayudarme a escribir la tesis y a Daphne por intentar sabotearla. }}{}
\end{titlepage}

\doublespacing 
\frontmatter
\thispagestyle{empty}

\noindent Dr. Jos\'e Wagner Furtado Valle,\\
\noindent Profesor de investigaci\'on del Consejo Superior de Investigaciones Cient\'ificas (CSIC),\\[2ex]
\noindent Dr. Rahul Srivastava,\\
\noindent Assistant professor, Indian Institute of Science Education and Research - Bhopal,\\[2ex]
\noindent CERTIFICAN:\\[2ex]
\noindent Que la presente memoria "Theory and phenomenology of Dirac neutrinos: Symmetry breaking patterns, flavour implications and Dark Matter" ha sido realizada bajo su direcci\'on en el Institut de F\'isica Corpuscular, centro mixto del CSIC y de la Universitat de València, por Salvador Centelles Chuli\'a y constituye su Tesis para optar al grado de Doctor en F\'isica.\\[2ex]
\noindent Y para que as\'i conste, en cumplimiento de la legislaci\'on vigente, presenta en el Departament de F\'isica Teòrica de la Universidad de Valencia la referida T\'esis Doctoral, y firma el presente certificado.

Paterna (Valencia), a 22 de Agosto de 2021.
\begin{figure}[h!]
\begin{center}
 	\includegraphics[scale=0.3]{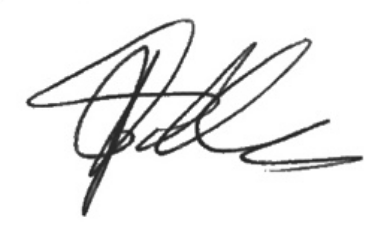}
	\caption*{Dr. Jos\'e Wagner Furtado Valle }
\end{center}
\end{figure}

Bhopal (India), a 22 de Agosto de 2021.

\begin{figure}[h!]
\begin{center}
	\includegraphics[scale=0.3]{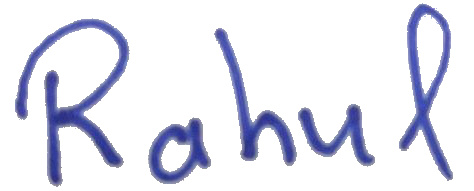}
	\caption*{Dr. Rahul Srivastava}
	\end{center}
\end{figure}


\chapter*{Acknowledgements}
\addcontentsline{toc}{chapter}{Acknowledgements} 

This thesis wouldn't have been possible without the amazing people that surrounded me in the last few years. For all of you, you have my deepest gratitude.

Thanks to Jose for guiding me and making me read Schechter-Valle 1980 when I was in physics kindergarten, always opening opportunities for me in terms of assistance to schools, conferences, summer positions and financial support even before starting my PhD, and for directing my bachelor and master's thesis. In the same line, thanks to Rahul for patiently explaining everything I needed to know everytime I bugged you with simple questions, physics or otherwise, and I am happy to call you friend. I cannot think of better PhD advisors that you both and your guidance has been paramount to my career.

Similarly, it was a pleasure to collaborate with so many great scientists and persons: Ricardo, Werner, Gui-Jun, Peng, Eduardo, Ulises, Manfred, Matthias, Björn, Christian, Avelino, C\'esar, Andreas, Pedro and Ernest. I have learned a lot from all of you but most importantly we've had a lot of fun discussing physics.

I also want to dedicate some words to my office mates over the years Pedro, Lucho, Pablo I, Ricardo, Christoph, Ivania and Jacopo, as well as the members of the AHEP group in Valencia and the MPIK in Heidelberg. All the good moments will not be forgotten, nor Martin's monday cakes.

I feel the need to remember the IFIC Secretar\'ia people who helped me navigate the bureaucracy seas: Amparo, Teresa, Luis, Pilar and Elena.

I also acknowledge the funding from the Spanish FPI fellowship BES-2016-076643.

Out of the physics realm I have to foremost thank my extended family and specially my parents. Their constant support, love and guidance have made me who I am, and their advice proved to be good even when I ignored it. I also have to thank my childhood friends for being how they are, because when we all took diverging paths, friendship continued and we still find time to converge. 

My final thank you goes to Olaia, because enduring me for all these years is a PhD in itself.

This thesis is also yours. 
So long and thanks for all the fish. 


{\hypersetup{hidelinks}
\tableofcontents
}

\mainmatter


\fancyhf{}
\fancyhead[LE,RO]{\thepage}

\chapter{List of scientific publications}

\fancyhf{}
\fancyhead[LE,RO]{\thepage}
\fancyhead[RE]{\slshape\nouppercase{\leftmark}}
\fancyhead[LO]{\slshape\nouppercase{\rightmark}}

\section{Publications included in the thesis}
This thesis is based on the following scientific publications:

\begin{enumerate}
\item \textit{Dirac Neutrinos and Dark Matter Stability from Lepton Quarticity} \cite{CentellesChulia:2016rms}

Salvador Centelles Chuliá (Valencia U., IFIC), Ernest Ma (UC, Riverside), Rahul Srivastava (Ahmedabad, Phys. Res. Lab), José W. F. Valle (Valencia U., IFIC)

e-Print: 1606.04543 [hep-ph]

DOI: 10.1016/j.physletb.2017.01.070

Published in: Phys.Lett.B 767 (2017), 209-213

\item \textit{Generalized Bottom-Tau unification, neutrino oscillations and dark matter: predictions from a lepton quarticity flavour approach} \cite{CentellesChulia:2017koy}

Salvador Centelles Chuliá (Valencia U., IFIC), Rahul Srivastava (Valencia U., IFIC), José W. F. Valle (Valencia U., IFIC)

e-Print: 1706.00210 [hep-ph]

DOI: 10.1016/j.physletb.2017.07.065

Published in: Phys.Lett.B 773 (2017), 26-33

\item \textit{Seesaw roadmap to neutrino mass and dark matter} \cite{CentellesChulia:2018gwr}

Salvador Centelles Chuliá (Valencia U., IFIC), Rahul Srivastava (Valencia U., IFIC), José W.F. Valle (Valencia U., IFIC)

e-Print: 1802.05722 [hep-ph]

DOI: 10.1016/j.physletb.2018.03.046

Published in: Phys.Lett.B 781 (2018), 122-128

\item \textit{Seesaw Dirac neutrino mass through dimension-six operators} \cite{CentellesChulia:2018bkz}

Salvador Centelles Chuliá (Valencia U., IFIC), Rahul Srivastava (Valencia U., IFIC), José W.F. Valle (Valencia U., IFIC)

e-Print: 1804.03181 [hep-ph]

DOI: 10.1103/PhysRevD.98.035009

Published in: Phys.Rev.D 98 (2018) 3, 035009

\item \textit{Dark matter stability and Dirac neutrinos using only Standard Model symmetries} \cite{Bonilla:2018ynb}

Cesar Bonilla (Munich, Tech. U.), Salvador Centelles-Chuliá (Valencia U., IFIC), Ricardo Cepedello (Valencia U., IFIC), Eduardo Peinado (Mexico U.), Rahul Srivastava (Valencia U., IFIC)

e-Print: 1812.01599 [hep-ph]

DOI: 10.1103/PhysRevD.101.033011

Published in: Phys.Rev.D 101 (2020) 3, 033011

\item \textit{The inverse seesaw family: Dirac and Majorana} \cite{CentellesChulia:2020dfh}

Salvador Centelles Chuliá (Valencia U., IFIC), Rahul Srivastava (IISER, Bhopal), Avelino Vicente (Valencia U., IFIC and Valencia U.)

e-Print: 2011.06609 [hep-ph]

DOI: 10.1007/JHEP03(2021)248

Published in: JHEP 03 (2021), 248

\end{enumerate}

\section{Other publications not included in the thesis}

\begin{enumerate}
 \item \textit{CP violation from flavour symmetry in a lepton quarticity dark matter model} \cite{CentellesChulia:2016fxr}

Salvador Centelles Chuliá (Valencia U., IFIC), Rahul Srivastava (Ahmedabad, Phys. Res. Lab and IMSc, Chennai), José W.F. Valle (Valencia U., IFIC)

e-Print: 1606.06904 [hep-ph]

DOI: 10.1016/j.physletb.2016.08.028

Published in: Phys.Lett.B 761 (2016), 431-436

\item \textit{Neutrino oscillations from warped flavour symmetry: predictions for long baseline experiments T2K, NOvA and DUNE} \cite{Pasquini:2016kwk}

Pedro Pasquini (Campinas State U. and Valencia U., IFIC), S.C. Chuliá (Valencia U., IFIC), J.W.F. Valle (Valencia U., IFIC and Valencia U.)

e-Print: 1610.05962 [hep-ph]

DOI: 10.1103/PhysRevD.95.095030

Published in: Phys.Rev.D 95 (2017) 9, 095030

\item \textit{Neutrino Predictions from Generalized CP Symmetries of Charged Leptons} \cite{Chen:2018lsv}

Peng Chen (Ocean U.), Salvador Centelles Chuliá (Valencia U., IFIC), Gui-Jun Ding (Hefei, CUST), Rahul Srivastava (Valencia U., IFIC), José W.F. Valle (Valencia U., IFIC)

e-Print: 1802.04275 [hep-ph]

DOI: 10.1007/JHEP07(2018)077

Published in: JHEP 07 (2018), 077

\item \textit{Realistic tribimaximal neutrino mixing} \cite{Chen:2018eou}

Peng Chen (Ocean U.), Salvador Centelles Chuliá (Valencia U., IFIC), Gui-Jun Ding (Hefei, CUST), Rahul Srivastava (Valencia U., IFIC), José W.F. Valle (Valencia U., IFIC)

e-Print: 1806.03367 [hep-ph]

DOI: 10.1103/PhysRevD.98.055019

Published in: Phys.Rev.D 98 (2018) 5, 055019

\item \textit{CP symmetries as guiding posts: revamping tri-bi-maximal mixing. Part I.} \cite{Chen:2018zbq}

Peng Chen (Ocean U.), Salvador Centelles Chuliá (Valencia U., IFIC), Gui-Jun Ding (Hefei, CUST), Rahul Srivastava (Valencia U., IFIC), José W.F. Valle (Valencia U., IFIC)

e-Print: 1812.04663 [hep-ph]

DOI: 10.1007/JHEP03(2019)036

Published in: JHEP 03 (2019), 036

\item \textit{Systematic classification of two loop $d$ = 4 Dirac neutrino mass models and the Diracness-dark matter stability connection} \cite{CentellesChulia:2019xky}

Salvador Centelles Chuliá (Valencia U., IFIC), Ricardo Cepedello (Valencia U., IFIC), Eduardo Peinado (Mexico U.), Rahul Srivastava (Valencia U., IFIC)

e-Print: 1907.08630 [hep-ph]

DOI: 10.1007/JHEP10(2019)093

Published in: JHEP 10 (2019), 093, JHEP1910(2019)093

\item \textit{Scotogenic dark symmetry as a residual subgroup of Standard Model symmetries} \cite{CentellesChulia:2019gic}

Salvador Centelles Chuliá (Valencia U., IFIC), Ricardo Cepedello (Valencia U., IFIC), Eduardo Peinado (UNAM, Mexico), Rahul Srivastava (Valencia U., IFIC and IISER, Bhopal)

e-Print: 1901.06402 [hep-ph]

DOI: 10.1088/1674-1137/44/8/083110

Published in: Chin.Phys.C 44 (2020) 8, 083110

\item \textit{CP symmetries as guiding posts: Revamping tribimaximal mixing. II.} \cite{Chen:2019fgb}

Peng Chen (Zhejiang Ocean U.), Salvador Centelles Chuliá (Valencia U., IFIC), Gui-Jun Ding (Hefei, CUST), Rahul Srivastava (Valencia U., IFIC), José W.F. Valle (Valencia U., IFIC)

e-Print: 1905.11997 [hep-ph]

DOI: 10.1103/PhysRevD.100.053001

Published in: Phys.Rev.D 100 (2019) 5, 053001

\item \textit{Two-Higgs-doublet models with a flavoured $\mathbb{Z}_2$ symmetry} \cite{CentellesChulia:2019ijn}

S. Centelles Chuliá (Valencia U., IFIC), W. Rodejohann (Heidelberg, Max Planck Inst.), U.J. Saldaña-Salazar (Heidelberg, Max Planck Inst.)

e-Print: 1911.06824 [hep-ph]

DOI: 10.1103/PhysRevD.101.035013

Published in: Phys.Rev.D 101 (2020) 3, 035013

\item \textit{Asymmetric tri-bi-maximal mixing and residual symmetries} \cite{CentellesChulia:2019ldn}

Salvador Centelles Chuliá (Valencia U., IFIC), Andreas Trautner (Heidelberg, Max Planck Inst.)

e-Print: 1911.12043 [hep-ph]

DOI: 10.1142/S0217732320502922

Published in: Mod.Phys.Lett.A 35 (2020) 35, 2050292

\item \textit{Natural axion model from flavour} \cite{CentellesChulia:2020bnf}

Salvador Centelles Chuliá (Valencia U., IFIC), Christian Döring (Heidelberg, Max Planck Inst.), Werner Rodejohann (Heidelberg, Max Planck Inst.), Ulises J. Saldaña-Salazar (Heidelberg, Max Planck Inst.)

e-Print: 2005.13541 [hep-ph]

DOI: 10.1007/JHEP09(2020)137

Published in: JHEP 09 (2020), 137

\item \textit{Gravitational wave induced baryon acoustic oscillations} \cite{Doring:2021gue}

Christian Döring (Heidelberg, Max Planck Inst.), Salvador Centelles Chuliá (Valencia U., IFIC), Manfred Lindner (Heidelberg, Max Planck Inst.), Björn Malte Schaefer (Heidelberg, Astron. Rechen Inst.), Matthias Bartelmann (Heidelberg U.)

e-Print: 2107.10283 [gr-qc]

Sent for publication to SciPost Physics
\end{enumerate}

\chapter{Introduction}

The building of the Standard Model (SM) of particle physics has been a tremendous joint effort between theorists and experimentalists \cite{Glashow:1961tr, Englert:1964et, Higgs:1964ia, Guralnik:1964eu, Weinberg:1967tq, Glashow:1970gm, tHooft:1972tcz}, culminating with the discovery of the Higgs boson in 2012 \cite{ATLAS:2012yve, CMS:2012qbp}. While the success of the theory cannot be denied, there are still some shortcomings and open questions in particle physics. An incomplete list includes neutrino masses, Dark Matter (DM) \cite{Planck:2018vyg}, the hierarchy problem \cite{tHooft:1979rat} and the strong CP problem \cite{tHooft:1976rip, Peccei:1977hh, Peccei:1977ur}. In this dissertation we will focus on the first, with special attention to the nature of neutrinos: they may be Dirac particles like the rest of the fermions of the Standard Model, in contrast to the canonical expectations that they are Majorana particles \cite{Majorana:1937vz, Schechter:1980gr}, meaning that the neutrino mass Eigenstates are their own self conjugate fields.

While a detailed exposition of the SM is out of the scope of this work it can be found in standard quantum field theory textbooks, reviews and textbooks (see for example \cite{Pich:2012sx}). However, since the focus of this work will be directed towards neutrino masses, we will first give a short overview on the mass mechanism in the SM, i.e. the so-called Englert-Brout-Higgs mechanism \cite{Englert:1964et, Higgs:1964ia, Guralnik:1964eu} in Sec.~\ref{sec:Higgsmech}. While the SM predicts massless neutrinos, neutrino oscillation experiments show that they are indeed massive particles. We will briefly review neutrino oscillations in Sec.~\ref{sec:oscillations}. To conclude the introduction chapter, in Sec.~\ref{sec:seesawcompletions} we will cover the seesaw completions of the SM which predict Majorana neutrino masses.

 \section{Fermion masses in the Standard Model}
 \label{sec:Higgsmech}
 
  The unbroken $SU(3)_{QCD} \times SU(2)_L \times U(1)_Y$ gauge structure of the SM can qualitatively reproduce the particle interactions seen in collider experiments. A big drawback is that it automatically predicts massless fermions and massless gauge bosons, all of which are experimentally known to be massive \cite{ParticleDataGroup:2020ssz}. The reason is that their Lagrangian mass terms are forbidden by the gauge symmetry. For example, take the fermionic mass term for the electron:

\begin{equation}
\mathcal{L}_{m_e} = m_e \bar{e}_L \, e_R \label{eq:baremass}
\end{equation}

The field $e_L$ forms part of an $SU(2)_L$ doublet, $L_e = (e_L, \nu_{L,e})^T$ (analogously, the left handed quarks form part of a different $SU(2)_L$ doublet: $Q_u = (u_L, d_L)^T$, and part of a colour triplet) and therefore transforms non-trivially under an $SU(2)_L$ local transformation. In contrast,  $e_R$ is a singlet, as well as the other right-handed fermions in the theory, $u_R$ and $d_R$. We could repeat the exercise for any fermion and reach the same conclusion. Traditionally $\nu_R$ is not added to the model as during the early days of the construction of the SM neutrinos were thought to be massless. 

Since we know that the fermions are massive, we know that the symmetry must be broken. The bare mass terms like Eq.~\ref{eq:baremass} are forbidden by $SU(2)_L \times U(1)_Y$, but do not break the $SU(3)_{QCD} \times U(1)_{QED}$ symmetry group since i.e. left and right chiralities transform in the same manner. Moreover, the associated gauge bosons of the unbroken group, the gluons and the photon, are known to be massless \cite{ParticleDataGroup:2020ssz}. These are strong hints that point towards the need of a breaking of the electroweak group as

\begin{equation}
SU(2)_L \times U(1)_Y \rightarrow U(1)_{QED} 
\end{equation}

while respecting the color group $SU(3)_{QCD}$. Finally, let us note that an explicit breaking, for example simply adding the term in \ref{eq:baremass} to the lagrangian, would spoil the desired features of the gauge framework and render the model non-renormalizable \cite{tHooft:1972tcz}. Thus, the solution is to \textbf{spontaneously} break the symmetry, meaning that the Lagrangian is invariant under a certain symmetry group $G$, but the lowest energy state, the vacuum, is only invariant under a subgroup $F \subset G$. For a more detailed exposition of this mechanism we direct the reader to the vast literature in the topic and proceed to sketch how it generates masses for the SM fermions and gauge bosons.

In order to achieve Spontaneous Symmetry Breaking (SSB), to the fermion and gauge inventory of the massless SM we now add a complex scalar $\phi$. The scalar potential will be given by 

\begin{equation}
-\mathcal{L}_{\phi} = \mu^2 \phi^\dagger \phi + \lambda (\phi^\dagger \phi)^2 
\end{equation}

These two terms will \textit{always} be allowed by the gauge symmetry irrespective of the charges that we assign to $\phi$ and since its hypercharge is non-zero they will be the \textit{only} allowed terms. Let us now study a simplified case in which $\phi$ is an $SU(2)_L$ singlet. If the mass parameter $\mu^2$ is negative (i.e. $\mu$ is purely imaginary) then we can see that the minimum of the Hamiltonian in the classical theory is not found in $\phi = 0$ but in $|\phi_{\text{min}}| = \sqrt{\frac{\mu^2}{2 \lambda}} \equiv \frac{v}{\sqrt{2}}$. 
This is the lowest energy state of the theory and perturbations must be made around this point (and not around $\phi = 0$). We can thus rewrite $\phi$ as $\frac{1}{\sqrt{2}} (v + \phi_1 + i \phi_2)$ and immediately see that $\phi_2$ is a massless field. This is a direct consequence of the Goldstone theorem \cite{Nambu:1960tm, Goldstone:1961eq, Goldstone:1962es}.

Since we want to break $SU(2)_L \times U(1)_Y \rightarrow U(1)_{QED}$ we must give the scalar $\phi$ a non-trivial charge under this group, while it must be a singlet of the color group in order to respect $SU(3)_{QCD}$. Moreover, we must ensure that the fields which couple to the photon do not get a vacuum expectation value (vev) in order to respect the subgroup $U(1)_{QED}$. If $\phi$ is a doublet of $SU(2)_L$ with hypercharge $Y = 1/2$, then we can write

\begin{equation}
 \phi = \left( \begin{matrix} \phi^+ \\ \phi^0 \end{matrix}\right)
\end{equation}

Where the names $\phi^+$ and $\phi^0$ are given according to their electromagnetic interactions: $\phi^0$ does not couple with the photon at tree level and the superscripts represent the electric charge. Since only $\phi^0$ gets a vev and not $\phi^+$ and therefore the photon remains massless. If we rewrite

\begin{equation}
\phi = \frac{1}{\sqrt{2}} e^{\frac{i}{2} \sigma_i \theta_i} \left(\begin{matrix} 0 \\ v + h \end{matrix}\right),
\end{equation}

where $\sigma_i$ are the three Pauli matrices, now the 4 degrees of freedom are in the three real fields $\theta_i$ and $h$ instead of in the two coplex fields $\phi^+$ and $\phi^0$. Taking advantage of the $SU(2)_L$ local symmetry of the lagrangian we can rotate away any dependence on $\theta_i$. This is the so-called unitary gauge.

We turn our focus to the following Yukawa terms, which are allowed by the gauge symmetry

\begin{equation}
 \mathcal{L}_Y = Y_e \, \bar{L} \,\phi\, e_R \, + \, Y_ u\, \bar{Q}\, \phi^c \,u_R \, + \,  Y_d \,\bar{Q} \,\phi d_R + h.c.
\end{equation}

Where the sum over flavour indices is understood, $\tilde{\phi} = i \sigma_2 \phi^*$. Now expanding the $SU(2)_L$ indices in the unitary gauge we find

\begin{equation}
  \mathcal{L}_Y = \frac{m_e}{v} \, (v + h)\, \bar{e}_L \, e_R \, + \,  \frac{m_u}{v} \, (v + h) \, \bar{u}_L\, \,u_R \, + \,   \frac{m_d}{v} \, (v + h) \,\bar{d}_L \, d_R + h.c.
\end{equation}

Where we have explicitly rewritten the Yukawa matrices into mass matrices $m_x = \frac{Y_x v}{\sqrt{2}}$. We can now see that the fermions are massive thanks to the interaction with the Higgs boson. Additionally, an important prediction arises: the Yukawa interaction of a given fermion with the physical scalar $h$ will be proportional to the fermion mass.

The last step is to find the fermion mass Eigenstates. If we perform a unitary transformation on flavour space for each fermion type and chirality we have

\begin{equation}
 f_L \rightarrow U^f_L f_L^{(m)}, \hspace{2cm} f_R \rightarrow U^f_R f_R^{(m)}, \hspace{2cm} f \in \{e, u, d\},
\end{equation}

where $U^f_{L/R}$ are unitary matrices which we call mixing matrices and $f_{L/R}^{(m)}$ are chiral mass Eigenstates. The right-handed mixing matrices are unphysical, but the left-handed quark mixing matrices enter in the $W^\pm$ interactions in the combination

\begin{equation}
 U_{CKM} \equiv U_L^u \left(U_L^d\right)^\dagger
\end{equation}

$U_{CKM}$ has been measured with great precision \cite{ParticleDataGroup:2020ssz}. Analogously, in the lepton sector of a SM extension where  neutrinos are massive we have the lepton mixing matrix given by

\begin{equation}
 U_{lep} \equiv \left(U_L^l\right)^\dagger U_L^\nu,
\end{equation}

which has been measured in neutrino oscillations \cite{deSalas:2020pgw}, see Sec.~\ref{sec:oscillations}.

The Higgs mechanism also gives mass to the gauge bosons, this time through the covariant derivative $D^\mu \phi = \left( \partial^\mu + i g \tilde{W}^\mu + i g' y_\phi B^\mu\right )\phi$. Since our focus is on fermion masses we will not expand on this further.

 \section{The status of neutrino oscillations}
 \label{sec:oscillations}

  We can now note that neutrinos are massless in the SM simply by construction: since the right handed neutrino $\nu_R$ is missing from the model, the Higgs mechanism that gives mass to the other fermions in the SM cannot work for neutrinos. This was a natural choice at the time. In the 60 and the 70s, when the SM was being built, there was no experimental indication that neutrinos where massive and only upper limits could be set on the neutrino mass. Today we know that at least two neutrinos are massive thanks to the discovery of neutrino oscillations \cite{Kajita:2016cak,McDonald:2016ixn}. Irrespective of the underlying mass generation mechanism, if at least one neutrino is massive and non-degenerate and there is a mixing between gauge states to form the massive Eigenstates then there will be neutrino oscillations, which can and have been measured with great precision \cite{deSalas:2020pgw}. What is more, the $L/E$ dependence of the oscillations cannot be explained, to the best of our knowledge, by any other alternative mechanism \footnote{If one only has access to a single measurement, for example Davis experiment \cite{Davis:1968cp} in which a neutrino deficit is observed, possible alternative explanations to neutrino oscillations could be, for example, non-standard interactions. However these alternative explanations cannot explain the detailed $L/E$ profile we have measured today and have been ruled out by other observations.}. We now briefly lay out the neutrino oscillations mechanism in terms of simple quantum mechanics. For a more complete, quantum field theory derivation see for example \cite{Grimus:1996av, Akhmedov:2010ms, Grimus:2019hlq}, yielding the same result.
  
  We start by assuming that some underlying mechanism gives mass to neutrino fields. Moreover, the gauge states $\ket{\nu_\alpha}$, $\alpha \in \{e, \mu, \tau\}$ mix to give the mass eigenstates $\ket{\nu_i}$, $i \in \{1, 2, 3\}$ as:
  
  \begin{equation}
 \ket{\nu_\alpha} = U_{\alpha i} \ket{\nu_i}
  \end{equation}
  
  In a neutrino beam of fixed trimomenta $\vec{p}$ the Hamiltonian Eigenstates will be simply the mass Eigenstates. Their time evolution is, therefore, trivial
  
  \begin{equation}
   \ket{v_i}(t) = \ket{v_i}(t=0) e^{-i E_i t}
  \end{equation}
  
  And the energy $E_i$ is given in the ultra relativistic approximation as $E_i = \sqrt{m_i^2 + \vec{p}^2} \approx |\vec{p}| + \frac{1}{2} \frac{m_i^2}{E}$. Then, the energy difference between two mass Eigenstates will be
  
  \begin{equation}
   E_j-E_i  = \frac{1}{2} \frac{m_j^2 - m_i^2}{|\vec{p}|} \equiv \frac{1}{2} \frac{\Delta m_{ij}^2 }{|\vec{p}|} \equiv \Delta E_{ij}
  \end{equation}
  
  Now let's assume that a neutrino of a given flavour $\alpha$ is produced, for example in a proton-proton fusion (which is the main production mechanism of electron neutrinos in the sun). The time evolution of such a state will be given by
  
  \begin{equation}
   \ket{\psi}(t=0) = \ket{\nu_\alpha}, \hspace{1cm} \ket{\psi}(t) = U_{\alpha i}  e^{-i E_i t} \ket{\nu_i}
  \end{equation}

And the probability of detecting a given flavour $\beta$ is then

\begin{equation}
 P_{\alpha\rightarrow \beta} = |\braket{\nu_\beta}{\psi}|^2 = U^*_{\beta j} U_{\alpha j} U_{\beta k} U_{\alpha, k}^* e^{-i \Delta E_{kj} t}
\end{equation}

Since neutrinos are ultrarelativistic we can replace $t \rightarrow L$, $|\vec{p}| \rightarrow E$ and the final expression becomes

\begin{equation}
 P_{\alpha\rightarrow \beta} = U^*_{\beta j} U_{\alpha j} U_{\beta i} U_{\alpha, i}^* e^{-i \frac{1}{2}\Delta m_{ij}^2 \frac{L}{E}}
\end{equation}

Two important features can be immediately extracted. First, the aforementioned dependence of the oscillations on $L/E$. Second, the dependence not on the absolute mass of neutrinos but on the square mass differences.

It is common to parametrize the mixing matrix $U$ in a convenient way. Let's start with some mathematical considerations. An $n\times n$ general complex matrix has $2 n ^2$ independent real parameters ($n^2$ moduli and $n^2$ phases). If the matrix is unitary then the equation $U^\dagger U = I$ imply $n^2$ conditions and therefore the number of independent parameters is $n^2$. Note that the equation $U U^\dagger = I$ does not impose any new independent constraint. Coming to physics and assuming $3$ generations of light neutrinos we get $9$ parameters, of which $3$ can be rotated away thanks to the the $U(1)$ symmetry of the lepton doublets. Then we are left with $3$ angles and $3$ phases. $2$ of these phases are irrelevant to neutrino oscillations and are actually unphysical if neutrinos are Dirac particles. Let us define the matrices

\begin{align}
& U_{12}(\theta, \delta) \equiv \left(\begin{matrix}
\cos \theta & e^{-i\delta} \sin \theta  & 0 \\ 
- e^{i\delta}  \sin \theta & \cos \theta & 0\\
0 & 0 & 1
\end{matrix} \right), \hspace{5mm} 
U_{13}(\theta, \delta) \equiv \left(\begin{matrix}
\cos \theta &0& e^{-i\delta} \sin \theta  \\ 
0 & 1 & 0\\
- e^{i\delta}  \sin \theta & 0& \cos \theta
\end{matrix} \right)\\
& U_{23}(\theta, \delta) \equiv \left(\begin{matrix}
1 & 0 & 0\\
0 & \cos \theta & e^{-i\delta} \sin \theta\\ 
0 & - e^{i\delta}  \sin \theta & \cos \theta
\end{matrix} \right), \hspace{5mm}
P(\delta_1,\delta_2,\delta_3) \equiv \left(\begin{matrix}
e^{i\delta_1} & 0 & 0\\
0 &  e^{i\delta_2}&0\\ 
0 & 0 & e^{i\delta_3}
\end{matrix} \right) 
\end{align}

The parametrization used by the Particle Data Group (PDG) \cite{ParticleDataGroup:2020ssz} is given by

\begin{equation}
U_{PDG} = U_{23}(\theta_{12}, 0) U_{13}(\theta_{13}, \delta_{CP}) U_{12}(\theta_{12}, 0) P(\alpha, \beta, 0) 
\end{equation}

Where the parameters $\theta_{ij}$ and $\delta_{CP}$ are relevant for neutrino oscillations. The phases $\alpha$ and $\beta$ are unphysical in the Dirac neutrino case and irrelevant for neutrino oscillations in any case, but appear in neutrinoless double beta decay calculations. Alternatively one could use the so-called symmetric parametrization \cite{Schechter:1980gr, Rodejohann:2011vc} which is more transparent from a theoretical and model building point of view:

\begin{equation}
 U_\text{symmetric} = U_{23}(\theta_{23}, \delta_{23}) U_{13}(\theta_{13}, \delta_{13}) U_{12} (\theta_{12}, \delta_{12})
 \label{eq:symparam}
\end{equation}

Which is related to $U_{PDG}$ by $\delta_{CP} = \delta_{13} - \delta_{12} - \delta_{23}$.

\begin{figure}
\begin{center}
	\includegraphics[scale=0.5]{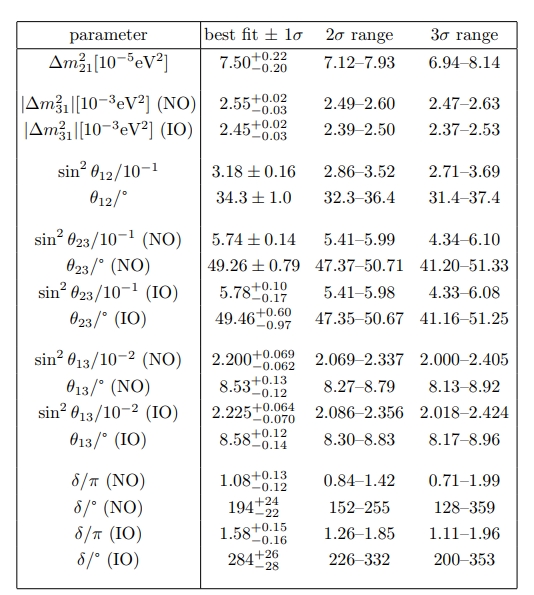}, 
	\includegraphics[scale=0.3]{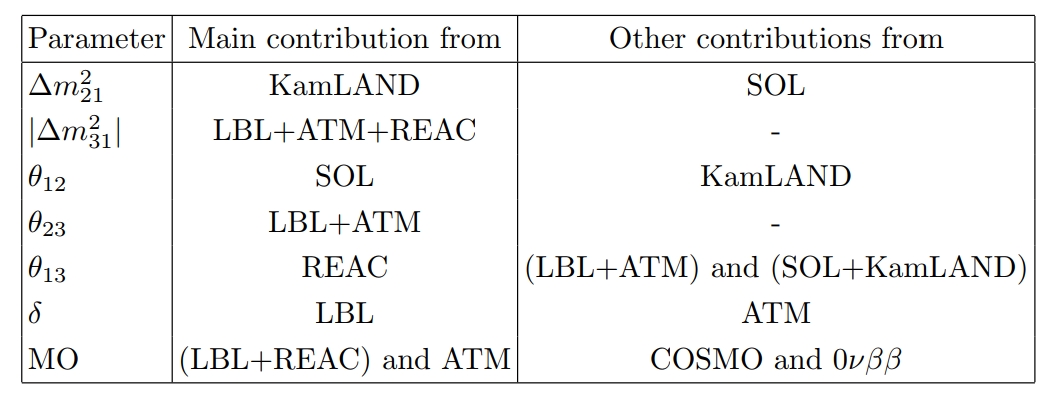},
	\includegraphics[scale=0.7]{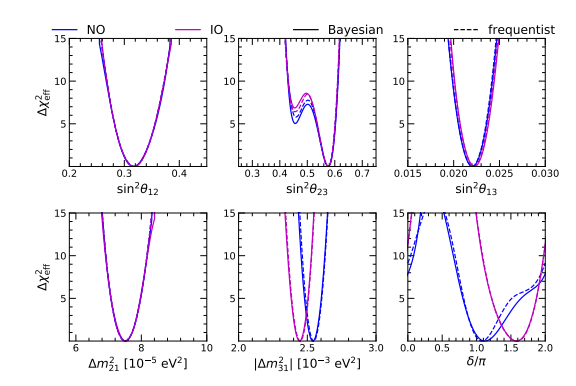},
	\caption{\textbf{Up:} Global fit to the oscillation parameters. \textbf{Middle:} Sensitivity of the different oscillation experiments to each parameter. \textbf{Down:} $\chi^2$ profiles for the $6$ oscillation parameters. All tables taken from \cite{deSalas:2020pgw} with their kind permission.}
	\label{fig:oscillationparameters}
\end{center}
\end{figure}

These oscillations parameters are now measured with great precision by many complementary experiments. A global fit to all the oscillation parameters can be found in \cite{deSalas:2020pgw}. The relevant experiments for each of the parameters, the best-fit values and the $\chi^2$ profiles are shown in Fig.~\ref{fig:oscillationparameters}

 \section{Majorana seesaw completions}
 \label{sec:seesawcompletions}
 
We know that neutrinos are massive, as discussed in Sec.~\ref{sec:oscillations}, and that the SM predicts massless neutrinos simply by construction. We can then naively think of adding $\nu_R$ and generate a Yukawa term of the form
 
 \begin{equation}
  \mathcal{L}_{Y\nu} = Y_\nu \, \bar{L} \phi^c \nu_R + h.c.\rightarrow \frac{m_D}{v} \, (v + h)\, \bar{\nu}_L \, \nu_R  + h.c.
 \end{equation}

 Where $m_D = \frac{Y_\nu v}{\sqrt{2}}$ and the subindex $D$ stands for `Dirac'. Note however that if we don't extend the symmetry inventory of the SM the following term is also allowed by the gauge symmetries
 
 \begin{equation}
  \mathcal{L}_M =\frac{1}{2} M \bar{\nu}_R^c \nu_R \, + h.c.
 \end{equation}
 
\begin{figure}[!h]
 \centering
  \includegraphics[width=0.3\textwidth]{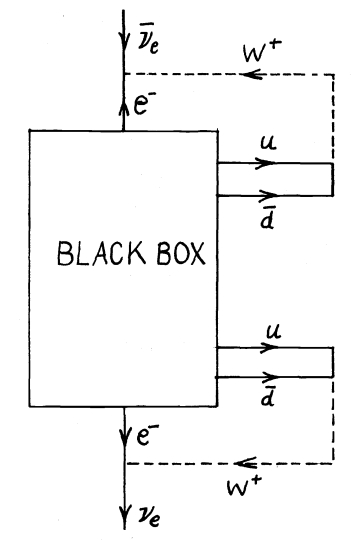}~~~~~~
    \caption{Diagram representing the black box theorem. If neutrinoless double beta decay exists in nature (even if it is not mediated by neutrinos) then the `black box' mechanism generating it can be `dressed' with $W$ bosons, becoming a Majorana mass for neutrinos. Original figure taken from \cite{Schechter:1981bd} with their kind permission.}
    \label{fig:neutrinoless}
  \end{figure}

 Where we have defined the conjugate of a field as $\psi^c = \mathcal{C} \bar{\psi}^T$, $\mathcal{C}$ is the charge conjugation matrix given by
 
 \begin{equation}
  \mathcal{C} = \left(\begin{matrix} - i \sigma_2 & 0 \\ 0 & i \sigma_2\end{matrix}\right)
 \end{equation}

 and $M$ is a symmetric complex matrix. Note that this term violates lepton number symmetry, which is accidentally conserved in the SM \footnote{An accidental symmetry is one which is not explicitly imposed upon the construction of the model but is conserved anyway.}. The presence of this term implies that the neutrino mass Eigenstates will be Majorana particles and it is thus conceptually extremely important. The existence of such lepton number violating terms may imply the existence of $\Delta L = 2$ low energy processes. There are several ongoing and planned experiments searching for this phenomena, such as neutrinoless double beta decay \cite{KamLAND-Zen:2016pfg, GERDA:2018zzh, Agostini:2017iyd, Alduino:2017ehq, Arnold:2016qyg, Albert:2014awa} or $\mu^- \rightarrow e^+$ conversion in muonic atoms \cite{SINDRUMII:1998mwd} as well as high energy processes that could be detected in colliders \cite{ATLAS:2018dcj,Scutti:2020ulw, Aaboud:2018spl}. However, the most conceptually important implication is the reverse, and comes from the Schechter-Valle black box theorem \cite{Schechter:1981bd}. This theorem states that if neutrinoless double beta decay is present in nature then neutrinos must be Majorana particles, see Fig.\ref{fig:neutrinoless} for a diagramatic proof.

 Note however that the black box theorem cannot provide a quantitative prediction for the `Majorananess' of neutrino masses \cite{Duerr:2011zd}. As an example scenario, let's assume that neutrinos are Dirac at tree level. Lepton number is broken in the Lagrangian through a leptoquark which mediates the neutrinoless double beta decay process, see \cite{Hirsch:1996ye} as an example. Then, the black box theorem ensures that there will be a loop induced Majorana mass through the black-box loop' depicted in Fig.\ref{fig:neutrinoless}, but this contribution will necessarily be small: at least 4-loop. Neutrinos will thus be quasi-Dirac \cite{Valle:1982yw}. For an updated study of quasi-Dirac neutrinos phenomenology see for example \cite{Anamiati:2016uxp, Anamiati:2017rxw}. An extension of the black-box theorem to other lepton number violating processes is given in \cite{Hirsch:2006yk}.
 
 Coming back to our discussion, adding the two neutrino mass Lagrangians and writing in matrix form we get after SSB
 
 \begin{equation}
  \mathcal{L}_{m_\nu} = \mathcal{L}_{Y\nu} + \mathcal{L}_M \rightarrow \text{SSB} \rightarrow 
  \left(\begin{matrix}
                                                                                        \bar{\nu}_L & \bar{\nu}_R^c
                                                                                       \end{matrix}\right)
                                                                                       \left(\begin{matrix}
                                                                                        0 & m_D \\
                                                                                        m_D^\dagger & M
                                                                                       \end{matrix} \right)
\left(\begin{matrix}
                                                                                        \nu_L^c\\
                                                                                        \nu_R                                                                                     \end{matrix}\right) + h.c. \label{eq:massmatrix}
 \end{equation}

Where we have ignored the interaction with the physical Higgs. Now we perform some transformations and definitions

\begin{align}
 & \nu \equiv \left(\begin{matrix}  \nu_L^c\\  \nu_R  \end{matrix}\right), \hspace{2cm} M_\nu =  \left(\begin{matrix}
                                                                                        0 & m_D \\
                                                                                        m_D^T & M
                                                                                       \end{matrix} \right)\\
 &   \nu\rightarrow U \nu_m, \hspace{1cm} \text{Relation between gauge and mass Eigenstates} 
\end{align}
\begin{equation}
   \bar{\nu}^c M_\nu \nu \rightarrow \bar{\nu}_m^c U^T M_\nu U \nu_m 
\end{equation}

Where we impose the condition that $U^T M_\nu U$ is diagonal, real and positive. In \cite{Schechter:1980gr} it is beautifully shown that this $U$ will always exist. The diagonal entries will be the neutral mass Eigenvalues.

In this discussion we are avoiding writing expliticly the flavour indices. Therefore we are not commited to a certain number of generations. Let us now, in the spirit of \cite{Schechter:1980gr}, define $n$ as the number of generations of `active' neutrinos, i.e. those coming from the lepton doublet $L$, and $m$ as the number of `sterile' neutrinos, i.e. those coming from the gauge singlet $\nu_R$. While it may be useful to perform some conceptual calculations in the $n=m=1$ limit, realistic models need $n=3$ since we have measured the number of active neutrinos through the $Z$ boson decay \cite{ALEPH:1989kcj} and $m\geq 2$, since we know from oscillations that at least $2$ neutrinos are massive \cite{deSalas:2017kay, Whitehead:2016xud, Abe:2017uxa, Decowski:2016axc, deSalas:2020pgw}, see Sec.~\ref{sec:oscillations}. The matrix sub blocks $m_D$ and $M$ are of dimensions $n \times m$ and $m \times m$ respectively, while the symmetric mass matrix $M_\nu$ and the unitary transformation matrix $U$ are both $(n+m) \times (n+m)$.

The beauty of the mass matrix in \ref{eq:massmatrix} is that it can explain the smallness of neutrino masses without the need to impose extremely small values for the Yukawa couplings. If we assume the hierarchy $M \gg m_D$, then we can approximately diagonalize the matrix $M_\nu$ in a perturbative way. Following \cite{Schechter:1981cv} we start with the Ansatz

\begin{align}
  &U=e^{iH}V \\
H = \left( \begin{array}{cc}
0 & S\\
S^\dagger & 0 \end{array} \right ),& \hspace{5mm} V = \left( \begin{array}{cc}
V_\text{light} & 0\\
0 & V_\text{heavy} \end{array} \right )
\end{align}

The matrix $e^{iH}$ block diagonalizes the matrix $ M_\nu$:

\begin{equation} \large  \label{Blockdiag}
{e^{iH}}^T\, M_\nu \, e^{iH} = \left( \begin{array}{cc}
m_\text{light} & 0\\
0 & m_\text{heavy} \end{array} \right )
\end{equation} 

where $m_\text{light}$ and $m_\text{heavy}$ are general symmetric complex matrices of dimensions $n\times n$ and $m\times m$, respectively, which are diagonalized by 

 \begin{equation} \label{Vdiag}
V_i^T m_i V_i = \text{real, positive, diagonal}, \hspace{0.5cm} i\in\{\text{light}, \text{heavy}\}
\end{equation}

If we define the `parameter order' $\epsilon \sim O\left(\frac{m_D}{M}\right)$. This is more an intuitive definition than a rigurous one, since the order of magnitude of a matrix is not well defined. However its meaning is clear. If $\epsilon <<1$ we can assume $S \sim O(\epsilon)$ and we can make a perturbative expansion in the parameter $\epsilon$ keeping only the terms of order $\epsilon ^2$ or lower and, after some algebra, find

\begin{equation}
 S = - i m_D^* M^{-1}, \hspace{2cm} m_\text{light} = - m_D M^{-1} m_D^T
\end{equation}

Particularizing in the one generation limit we obtain

\begin{equation}
 m_\nu \propto Y^2 \frac{v^2}{M}
 \label{eq:Majoranaseesaw}
\end{equation}
 
The celebrated `seesaw formula'. While we have showed the method to block diagonalize the type I seesaw mass matrix, it can be easily adapted to different seesaw scenarios \cite{Minkowski:1977sc,GellMann:1980vs,Yanagida:1979as,mohapatra:1980ia,Schechter:1980gr,Schechter:1982cv}. As a benchmark scenario of order of magnitude values we can consider $m_\nu\sim 0.1 eV$, $Y \sim 0.1$, $v\sim 100 GeV$ we get $M \sim 10^{12} GeV$. We can compare this scenario with the one obtained taking $M=0$. In this case neutrinos will be purely Dirac particles, lepton number is conserved and neutrino masses come solely from the Higgs mechanism $m_\nu \sim Y_\nu v \rightarrow Y_\nu \sim 10^{-12}$. While this choice is not unnatural in the 't Hooft sense \cite{tHooft:1979rat}, because the radiative corrections of $Y_\nu$ are proportional to themselves, it is considered unaesthetical. However, it is indeed unnatural in the Dirac sense \cite{Dirac:1938mt}.

The drawbacks of both limiting scenarios are immediate: there are no phenomenological predictions that can be realistically tested. In the Majorana case, the reason is the suppresion of the phenomenological signatures by the high scale $M \sim 10^{12} GeV$. In the Dirac case, the suppresion comes from the small Yukawas $Y_\nu \sim 10^{-12}$. These two cases are not the only possibilities in order to give neutrino masses. In what follows we will give a quick overview on the Majorana mass generation mechanisms. Before starting the discussion, let us define what we mean by mass models and mass mechanisms. \\ \\
\textbf{Neutrino Mass Models:} A proper neutrino mass model
should be capable of generating neutrino masses and should be
renormalizable. It is also highly desirable, though not essential,
that the mass model also provides an ``explanation'' for the non-zero
yet so tiny masses of neutrinos when compared to masses of all the
other fermions in the SM.  \\
\textbf{Neutrino Mass Generation Mechanisms:} A mechanism is a class
of models which generates the neutrino masses in the same or very closely
related ways. For example, various variants of the canonical type-I
seesaw model can be classified together as type-I seesaw mechanism.
\\ \\
The model-building opportunities for neutrino masses are vast and the quantity and quality of the bibliography on the subject in the past few decades gives proof. We refer the interested reader to the cited references and the references within for more details on neutrino model building.

From a model-independent perspective, if only
SM fields are present, there is a unique dimension-5 operator that
gives rise to neutrino masses, the well-known Weinberg
operator~\cite{weinberg:1980bf},
\begin{equation}
\frac{C}{\Lambda}  \, L \phi \phi L
\label{weinberg} 
\end{equation}

where $\Lambda$ represents the cutoff scale. 
Above the cutoff scale the Ultra-Violet (UV) complete theory is at
play, involving new `messenger' fields, whose masses lie close to
the scale $\Lambda$.

\begin{figure}[!h]
 \centering
  \includegraphics[width=0.6\textwidth]{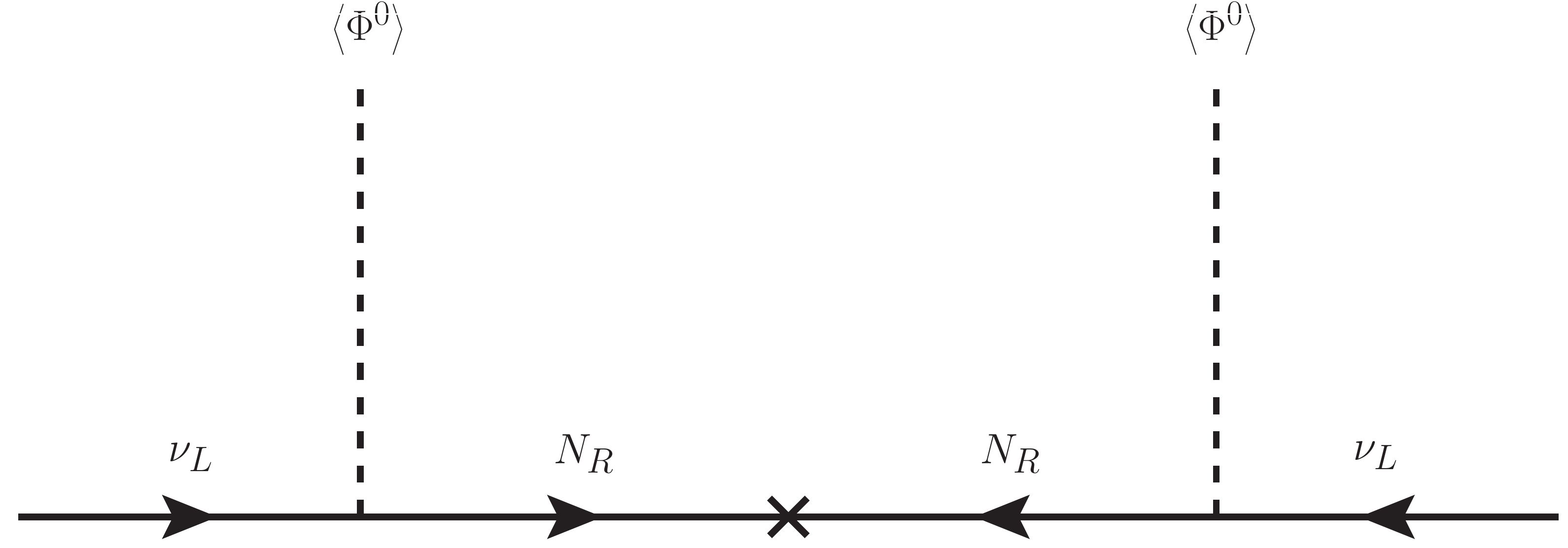}~~~~~~
 \includegraphics[width=0.35\textwidth]{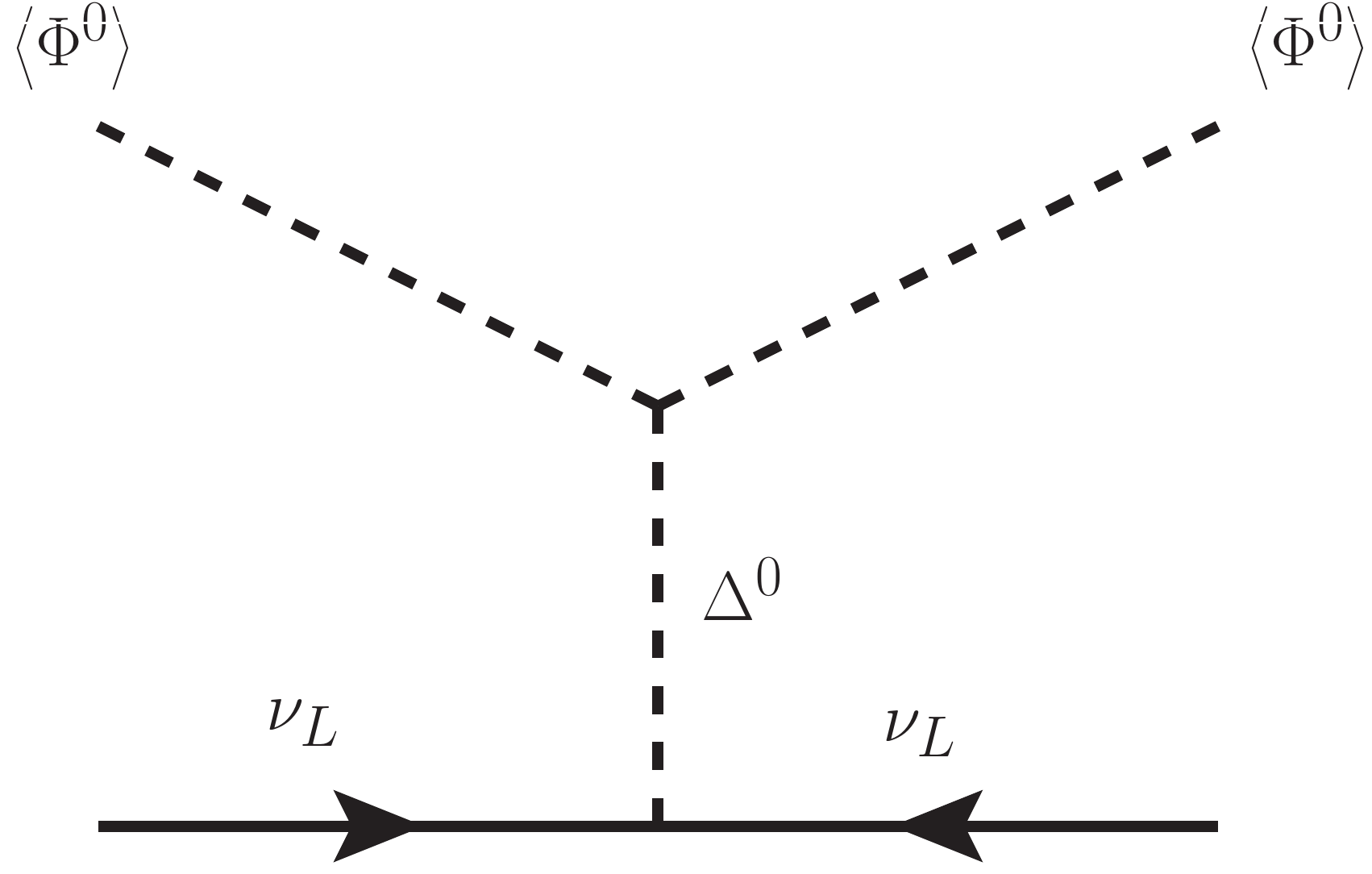} \\
 \includegraphics[width=0.6\textwidth]{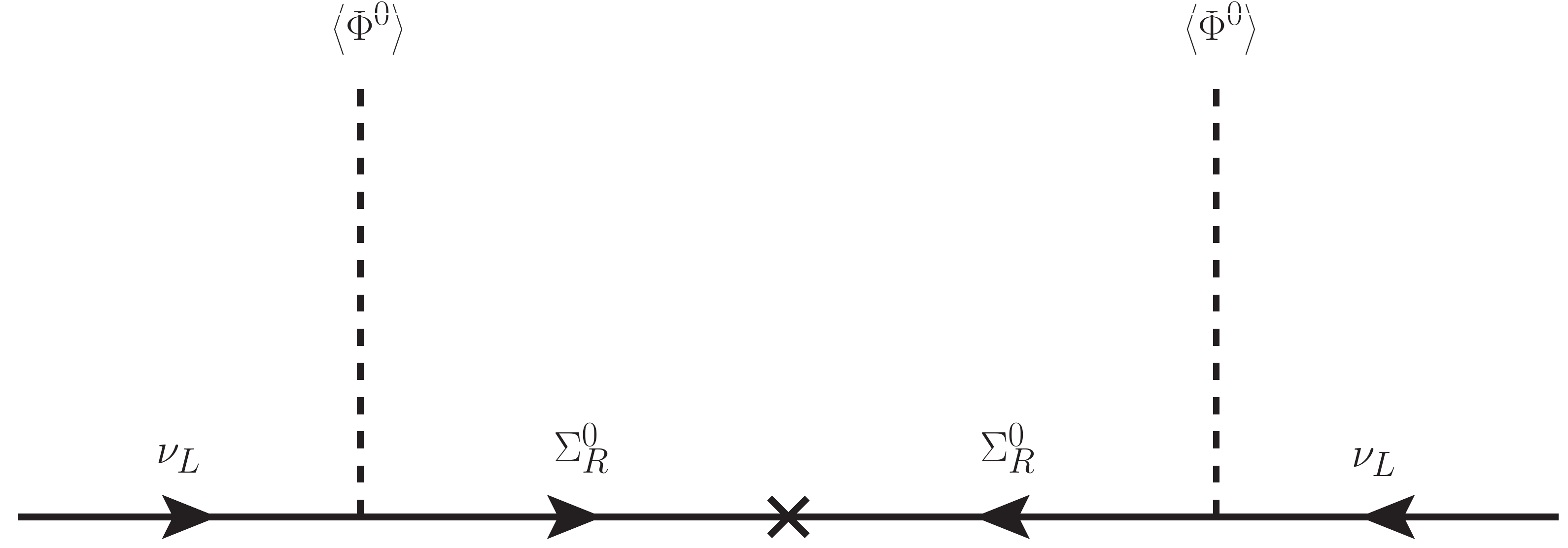}
    \caption{Feynman diagram generating Majorana masses in Type I, II and III seesaw mechanism.}
    \label{MT1}
  \end{figure}

There are three different ways of contracting the relevant fields at tree level and the UV completions of these three different contraction possibilities correspond to the so-called type I, II and III seesaw mechanism:
\begin{equation}
\underbrace{\underbrace{\bar{L}^c \otimes \phi}_1 \otimes \underbrace{\phi \otimes L}_{1}}_{ \textnormal{Type I }} \hspace{0.2cm} , \hspace{0.5cm}
\underbrace{\underbrace{\bar{L}^c \otimes L}_3 \otimes \underbrace{\phi \otimes \phi}_{3}}_{\textnormal{Type II }} \hspace{0.2cm} , \hspace{0.5cm}
\underbrace{\underbrace{\bar{L}^c \otimes \phi}_3 \otimes \underbrace{\phi \otimes L}_{3}}_{\textnormal{Type III }} \hspace{0.2cm}
\label{wopcont}
\end{equation}
In \eqref{wopcont} the underbrace denotes a $SU(2)_L$ contraction of
the fields involved. The number under the brace denotes the $SU(2)_L$
transformation of the contracted fields. Although not explicitly
written, the global contraction should always be a singlet of the full SM gauge group. The UV complete
realization of these contractions results into the three well known
seesaw variants as shown in a diagramatical way in Figure \ref{MT1}.
 
  Figure \ref{MT1} illustrates three UV--complete seesaw realizations
  of the same Weinberg operator of \eqref{wopcont}, differing from
  each other in the nature of the messenger fields involved.  
  In the left--panel diagram of Fig.~\ref{MT1}, corresponding to
  the type-I seesaw explained before, the field $N_R$ is a heavy fermion which transforms
  as a singlet under $SU(2)_L$ and carries no $SU(3)_C$ or $U(1)_Y$
  charge. 
 In the middle  diagram corresponding to type-II seesaw, the field
  $\Delta^0$ is the neutral component of a heavy scalar multiplet
  transforming as triplet under $SU(2)_L$. 
  In the right--panel diagram corresponding to type-III seesaw,
  $\Sigma_R^0$ is the neutral component of the heavy fermion multiplet
  transforming as triplet under $SU(2)_L$ symmetry.

  Notice that the three possible messenger fields and their $SU(2)_L$
  transformation properties arise from the different possibilities of
  field contractions of the Weinberg operator as shown in
  \eqref{wopcont}. 
  
  The possibility where both $L L$ and
  $\phi \phi$ contract to a singlet is forbidden, since
  $\phi \phi$ is symmetric, while the singlet contraction is
  antisymmetric, and therefore vanishes. Even in the presence of
  another Higgs doublet, the singlet contraction would vanish due to
  electric charge conservation.

  Weinberg's dimension-5 operator is the lowest one which can generate
  Majorana neutrino masses. In general, using only \sm
  fields, i.e.  a single Higgs doublet $\phi$, it is easy to show that
  the $SU(2)_L$ symmetry implies that Majorana masses can only be
  generated by odd-dimensional operators. This means that they can
  only arise from operators involving even number of Higgs
  doublets. The general operator allowed by \SM symmetry is
\begin{equation}
 \frac{1}{\Lambda^{2n+1}}\,  \bar{L}^c \phi^2 \left(\phi^\dagger \phi\right)^n L , \hspace{0.2cm} n \in \{0,1,2,3...\}
\end{equation}

\begin{figure}[!h]
 \centering
  \includegraphics[width=0.7\textwidth]{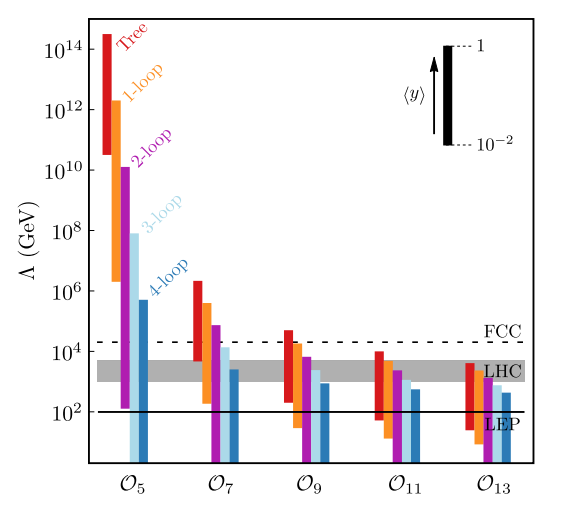}~~~~~~
    \caption{Typical new physics scale that reproduce the correct neutrino mass for different operator dimension and dominant loop level. Figure taken from \cite{Anamiati:2018cuq} with their kind permission.}
    \label{fig:scales}
  \end{figure}

A systematic classification of all the possibilities for operators of dimension $5$ up to three loops, dimension $7$ up to one loop and dimensions $9$, $11$ and $13$ at tree-level have been discussed \cite{Ma:1998dn, Babu:2001ex, Bonnet:2009ej, Bonnet:2012kz, Farzan:2012ev, Sierra:2014rxa, Cepedello:2017eqf, Cai:2017jrq, Anamiati:2018cuq, Cepedello:2018rfh, Klein:2019iws}, see Fig.~\ref{fig:scales}. Moreover, a plethora of UV complete models can be found in the literature. We will not develop this further and move on to the traditionally understudied possibility that neutrinos are Dirac particles.

\chapter{Dirac neutrino mass models}
\label{chap:Dirac}

 We now turn our focus to the possibility that neutrinos are Dirac particles. Traditionally this option has received less attention than the Majorana option from a theoretical point of view. Several seesaw \cite{Minkowski:1977sc, Yanagida:1979as, GellMann:1980vs, Mohapatra:1979ia, Schechter:1980gr, Schechter:1981cv, Foot:1988aq} and loop \cite{Zee:1980ai, Zee:1985id, Babu:1988ki, Ma:2006km} mass generation mechanism for Majorana neutrinos have been known for a long time. Furthermore, a systematic classification of all Majorana neutrinos mass mechanisms at a given operator dimensionality and up to certain number of loops also exist in the literature \cite{Ma:1998dn, Babu:2001ex, Bonnet:2009ej, Bonnet:2012kz, Farzan:2012ev, Sierra:2014rxa, Cepedello:2017eqf, Cai:2017jrq, Anamiati:2018cuq, Cepedello:2018rfh, Klein:2019iws}. The reason is twofold: first, Dirac neutrinos necessarily need a right-handed partner of the neutrinos, $\nu_R$, to be added to the SM. If it is added without further symmetry conditions then the Majorana mass term is automatically allowed too, as explained in Sec.~\ref{sec:seesawcompletions}. A protecting symmetry that forbids Majorana mass terms is needed if neutrinos are to be Dirac, see Sec.~\ref{sec:symmetryrequirement}. While the obvious choice is lepton number, already present in the SM itself as a global accidental symmetry, this raises the second concern: why is the neutrino mass so low compared to the electroweak scale? If neutrinos acquire a mass via the SM Higgs mechanism then Yukawas must be `ad-hoc' of order $10^{-12}$. This model is consistent and natural in 't Hooft's sense, but utterly unaesthetical and has no distinguishing phenomenology. These concerns can be addressed with the addition of new fields and symmetries. Then, we conclude that the simplest Majorana mass model, the type I seesaw \cite{Schechter:1980gr}, is simpler in terms of new fields and symmetries with respect to the Standard Model than the simplest Dirac model with naturally small masses, the Dirac type I seesaw \cite{CentellesChulia:2016rms}. However in more involved scenarios the Dirac option may be as appealing as the Majorana one, as we will argue in what follows.
 
 Most importantly, from an observational point of view the nature of neutrinos is still a major unknown. As explained in Sec.~\ref{sec:seesawcompletions}, the chief process for distinguishing Dirac and Majorana neutrinos is neutrinoless double beta decay \cite{KamLAND-Zen:2016pfg, GERDA:2018zzh, Agostini:2017iyd, Alduino:2017ehq, Arnold:2016qyg, Albert:2014awa} thanks to the black-box theorem~\cite{Schechter:1981bd,Duerr:2011zd}. Other related lepton number violation processes are also being attracting experimental attention \cite{SINDRUMII:1998mwd, ATLAS:2018dcj,Scutti:2020ulw, Aaboud:2018spl}.
On the other hand, the fact that the weak interaction is V-A turns
this quest into a major
challenge~\cite{Avignone:2007fu,Barabash:2004pu}. Indeed, the `Dirac-Majorana confusion theorem' \cite{Schechter:1980gk, Kayser:1981nw, Kayser:1982br} states that the difference between Dirac and Majorana neutrinos must vanish in the massless limit. Still, some works have tried to find alternative ways to distinguish the neutrino nature, see \cite{Nieves:1985ir, Chhabra:1992be,Rodejohann:2017vup, Berryman:2018qxn, Kim:2021dyj} and the references within.
As of now the nature of neutrinos remains as mysterious as the
mechanism responsible for generating their small masses. Little is
known regarding the nature of its associated messenger particles, the
underlying mass scale or its flavour structure~\cite{Valle:2015pba},
currently probed only in neutrino oscillation
experiments~\cite{deSalas:2020pgw}.


For these reasons, in the last few years there has been a renewed interest in looking at natural, elegant and predictive mass models for Dirac neutrinos. In this direction, several seesaw mechanisms have been proposed \cite{Ma:2014qra, Ma:2015mjd, Ma:2015raa, Valle:2016kyz, CentellesChulia:2016rms, CentellesChulia:2016fxr, Reig:2016ewy, CentellesChulia:2017koy, CentellesChulia:2017sgj, Borah:2017leo, Bonilla:2017ekt, Borah:2017dmk, Borah:2018nvu, Ma:2018bow,  Borah:2019bdi, Alvarado:2021fbw, Bernal:2021ppq}, as well as loop models for Dirac neutrinos \cite{Farzan:2012sa, Okada:2014vla, Bonilla:2016diq, Wang:2016lve, Ma:2017kgb, Wang:2017mcy, Helo:2018bgb, Reig:2018mdk, Han:2018zcn, Kang:2018lyy, Bonilla:2018ynb, Calle:2018ovc, Carvajal:2018ohk, Ma:2019yfo, Bolton:2019bou, Saad:2019bqf, Bonilla:2019hfb, Dasgupta:2019rmf, Jana:2019mez, Enomoto:2019mzl, Ma:2019byo,Restrepo:2019soi, CentellesChulia:2020dfh}. Other mechanisms for light Dirac neutrinos can be seen in models with extra dimensions~\cite{Chen:2015jta,Addazi:2016xuh, Pasquini:2016kwk}, where the $\nu_R$ states are required for the consistent high energy completion of the theory, GUT theories \cite{Ma:2021att, Ma:2021jvk}, or natural explanations of the SM fermion mass hierarchy \cite{CentellesChulia:2020bnf}. Other Dirac neutrino works without an explicit mass mechanism are, for example, \cite{Heeck:2013rpa,Aranda:2013gga,Abbas:2013uqh, Correia:2019vbn}. Regarding the symmetries utilized to protect the Dirac nature of neutrinos, in the literature we can find several $U(1)$ lepton number symmetries \cite{Farzan:2012sa, Ma:2014qra,Ma:2015mjd} or some of its discrete $Z_n$ subgroups~\cite{Ma:2014qra, Ma:2015mjd, Ma:2015raa,Bonilla:2016zef, Bonilla:2016diq, CentellesChulia:2016rms, CentellesChulia:2016fxr, CentellesChulia:2017koy, Hirsch:2017col, Fonseca:2018ehk, CentellesChulia:2018gwr, Bonilla:2018ynb, Calle:2018ovc, Bonilla:2019hfb, Dasgupta:2019rmf}, but also by invoking the presence of additional new symmetries \cite{Ma:2019yfo}, including non-abelian flavour groups \cite{Aranda:2013gga}.

Finally, Dirac neutrino masses can be found in different new physics scenarios which have the need of extra symmetries. Examples are Peccei-Quinn mechanisms with Dirac neutrinos \cite{Baek:2019wdn, Peinado:2019mrn, delaVega:2020jcp, CentellesChulia:2020bnf,Dias:2020kbj}, which require a 'PQ symmetry', Dirac neutrinos and dark matter models \cite{CentellesChulia:2016rms, CentellesChulia:2016fxr, CentellesChulia:2017koy, Bonilla:2018ynb, Bonilla:2019hfb, CentellesChulia:2019xky}, which require a 'dark matter symmetry', Dirac neutrinos arising from Grand Unification Theories \cite{Ma:2021att, Ma:2021jvk} etc.

Due to the increasing interest in Dirac neutrino mass models, a classification of such tree-level and one-loop models at dimension 4~\cite{Ma:2016mwh}, dimension 5~\cite{Yao:2018ekp, CentellesChulia:2018gwr}, dimension 6~\cite{Yao:2017vtm, CentellesChulia:2018bkz} and two-loops \cite{CentellesChulia:2019xky} have also been considered. Moreover, in some recent works it has been argued that in a full theory with the weak gravity conjecture neutrinos are expected to be Dirac fermions \cite{Ibanez:2017kvh}.

 \section{Symmetry requirements}
 \label{sec:symmetryrequirement}
 
 Before beginning our discussion on possible Dirac neutrino mass models, let us mention certain generic
conditions which must be satisfied in order to have Dirac neutrinos.
By definition, a Dirac fermion can be viewed as two chiral fermions,
one left handed and other right handed, having exactly degenerate
masses~\cite{Schechter:1980gr}. 
Thus in order to have massive Dirac neutrinos one must extend the \sm
particle content by adding the right handed partners of the
known neutrinos, $\nu_R$, being singlets under \SM gauge
symmetry. 
Second, owing to the color and electric charge neutrality of
neutrinos, an additional exactly conserved symmetry beyond the \sm
gauge symmetries is required to protect the Diracness of neutrinos by forbidding the Majorana masses of the neutrino states - and any neutral fermion that mixes with them.
While not the only option, from a theoretical point of view, the issue of the Dirac/Majorana nature of neutrinos is intimately connected with the $U(1)_{B-L}$ symmetry of the SM and its possible breaking pattern \cite{Heeck:2013rpa, Hirsch:2017col}. 
If the $U(1)_{B-L}$ symmetry is conserved in nature, like it is in the Standard Model, then the neutrinos will be Dirac fermions. We do not know if physics Beyond the Standard Model (BSM) will break the lepton or baryon number symmetries or, if they are broken, if some non-trivial residual subgroups will survive the spontaneous symmetry breaking.
However, if it is broken to a residual $\mathcal{Z}_m$ subgroup with $m\in \mathbb{Z}^+$ and $m \geq 2$, with $\mathbb{Z}^+$ being the set of all positive integers, then the Dirac/Majorana nature will depend on the residual $\mathcal{Z}_m$ symmetry provided that the SM lepton doublets $L_i = (\nu_{L_i}, l_{L_i})^T$  do not transform trivially under it. Thus, we have
\begin{eqnarray}
\label{eq:oddzn}
U(1)_{B-L}   & \, \to  \, &   \mathcal{Z}_m \equiv \mathcal{Z}_{2n+1} \, \text{with} \,  n \in \mathbb{Z}^+    \nonumber \\
& \, \Rightarrow \, & \text{neutrinos are Dirac particles}      \nonumber \\
U(1)_{B-L}  & \, \to  \,  & \mathcal{Z}_m \equiv \mathcal{Z}_{2n} \, \text{with} \,  n \in \mathbb{Z}^+  \\
& \, \Rightarrow \, & \text{neutrinos can be Dirac or Majorana } \nonumber
\end{eqnarray}
If the $U(1)_{B-L}$ is broken to a $\mathcal{Z}_{2n}$ subgroup, then one can make a further classification depending on how the $L_i$ transform,
\begin{eqnarray}
 L_i \left\{ \begin{array}{ll}
          \nsim  \omega^{n} \ \ \text{under $\mathcal{Z}_{2n}$} &  \Rightarrow\text{Dirac neutrinos}\\
          \sim  \omega^{n}\ \ \text{under $\mathcal{Z}_{2n}$} &  \Rightarrow\text{Majorana neutrinos}\end{array} \right. 
\label{evenzndir}
\end{eqnarray}
where $\omega^{2n}=1$. 

Irrespective of the choice of the symmetry that will forbid the Majorana mass terms for the neutrinos at all orders, throughout this work we will call such a symmetry the `Diracness symmetry'. Any Dirac neutrino mass model must feature this (unbroken) symmetry protection. \\

Additionally, it is clear that the renormalizable dimension $4$ operator $Y \bar{L} \phi^c \nu_R$ is allowed by the gauge symmetries of the Standard Model, giving rise to a Dirac neutrino mass at tree level through the Higgs mechanism \ref{sec:Higgsmech}. If this tree level Lagrangian term is present it will give a contribution to Dirac neutrino masses of $m_\nu = Y v$. Therefore, except for the case of extremely fine tuned cancellations with other new physics sector, if neutrinos are Dirac and the bare mass term is present then the Yukawa coupling $Y$ should be very small, of order $10^{-12}$ to reproduce the observed neutrino masses. Therefore in order to have 'natural' (in the Dirac sense) small neutrino masses we must ensure that this Lagrangian term is forbidden by some symmetry, which we call `seesaw symmetry'. The simplest choice is a simple $Z_2$ \cite{CentellesChulia:2016rms} under which the Standard model transform trivially but $\nu_R\sim -1$, forbidding the tree level mass term. Another option is to choose chiral lepton number charges: If the left and right chiralities transform differently under lepton number, the tree level is automatically forbidden \cite{Ma:2014qra, Ma:2015mjd, Ma:2015raa, Bonilla:2018ynb, Bonilla:2019hfb}. Furthermore, one could also choose a non-abelian symmetry \cite{CentellesChulia:2016fxr, CentellesChulia:2017koy}, leading to flavour predictions. See Sec.~\ref{sec:flavourpred} for more details.

Dirac neutrino masses can also be found in different new physics scenarios which have the need of extra symmetries. Some examples are Peccei-Quinn symmetry, dark matter stability, GUT models etc.

Note, however, that it is not necessary to have a different symmetry group for each of these purposes. In many of the cited works one symmetry group is fulfilling two or more requirements at the same time. Following this minimalistic approach, in \cite{Bonilla:2018ynb} and in Sec.~\ref{sec:radanddark} we develop an elegant mechanism with a single symmetry group, a $U(1)_{B-L}$ with chiral charges for the $\nu_R$ which plays the role of the `Diracness symmetry', `seesaw symmetry' and 'dark matter symmetry' at the same time. This $U(1)_{B-L}$ is also anomaly free and thus can be gauged, leading to a richer phenomenology.

 \section{The Dirac seesaw formula}
 
\begin{figure}[H]
\centering 
\includegraphics[scale=0.3]{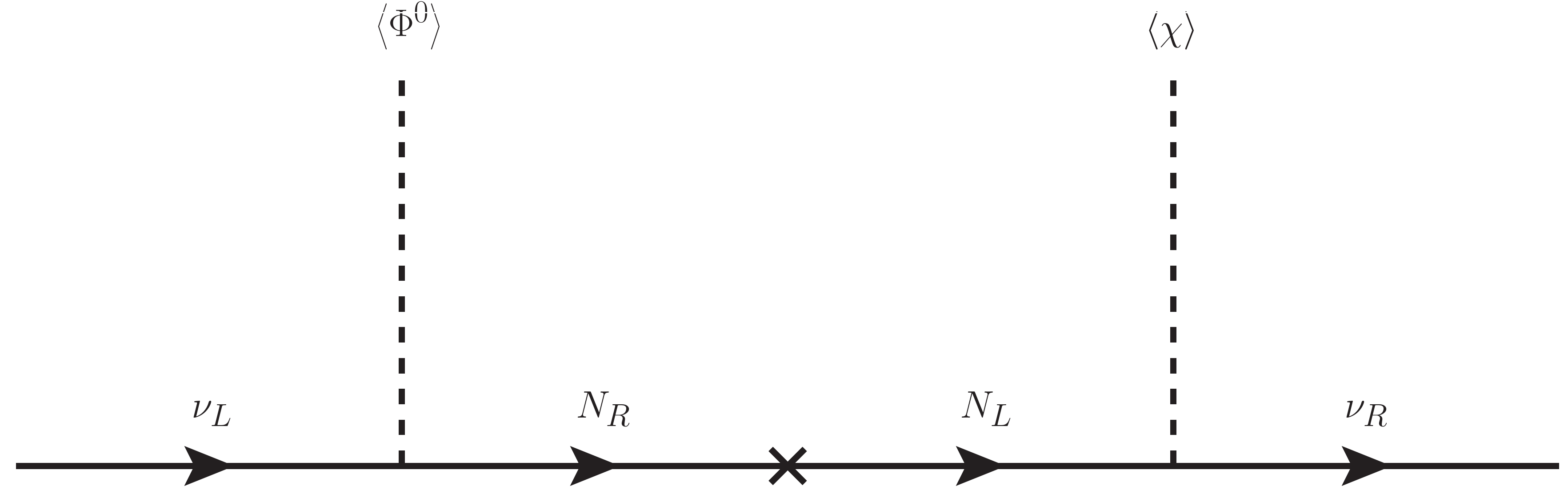}
    \caption{Feynman diagram representing the Dirac Type I seesaw.}
    \label{fig:diractypeI}
  \end{figure}

 We will now develop the analogue of the Majorana seesaw formula given in Eq.~\ref{eq:Majoranaseesaw}. As explained in Sec.~\ref{sec:symmetryrequirement}, the tree level Yukawa term $\bar{L} \phi^c \nu_R$ must be forbidden by some `seesaw symmetry' in order to have naturally small neutrino masses and Majorana mass terms are forbidden for neutrino fields by some `Diracness symmetry'. As the simplest seesaw scenario, consider the Standard Model with the addition of $n \ge 2$ gauge singlets $\nu_R$ and $m$ vector-like (VL) pairs\footnote{'Vector-like fermions' here means simply that the left and right chiralities transform in the same way under the symmetries considered, as opposed to the term 'chiral fermions' where the symmetry transformations are chiral-dependent.}, which are also gauge singlets and we call them $N_L$ and $N_R$. Since they are vector-like pairs, their number is unconstrained by theory in the sense that they won't contribute to gauge anomalies \cite{Schechter:1980gr}. The neutrino mass Lagrangian is then given by
 
 \begin{equation}
  \mathcal{L}_{m_\nu} = Y_\phi \bar{L} \phi^c N_R \, + \, Y_\chi \bar{N}_L \chi \nu_R \, + \, M N_L N_R \, + \, h.c. 
 \end{equation}
 
 where $\chi$ is a SM singlet carrying a vev $\bra{0} \chi \ket{0} = u$. The mass mechanism can be diagramatically portrayed as Fig.~\ref{fig:diractypeI}. 
 
 After spontaneous symmetry breaking the Lagrangian becomes
 
 \begin{equation}
  \mathcal{L}_{m_\nu} \rightarrow \text{SSB} \rightarrow 
  \left(\begin{matrix}
                                                                                        \bar{\nu}_L & \bar{N}_L
                                                                                       \end{matrix}\right)
                                                                                       \left(\begin{matrix}
                                                                                        0 & Y_\phi v \\
                                                                                        Y_\chi u & M
                                                                                       \end{matrix} \right)
\left(\begin{matrix}
                                                                                        \nu_R\\
                                                                                        N_R                                                                                     \end{matrix}\right) + h.c. \label{eq:DiracmassmatrixtypeI}
 \end{equation}
 
Where we have ignored the interaction with the physical Higgses. Similar to the Majorana case, we define

 \begin{align}
 & n_L \equiv \left(\begin{matrix}  \nu_L\\  N_L  \end{matrix}\right), \hspace{1cm} n_R \equiv \left(\begin{matrix}  \nu_R\\  N_R  \end{matrix}\right), \hspace{1cm} M_\nu =  \left(\begin{matrix}
                                                                                        0 & Y_\phi v \\
                                                                                        Y_\chi u & M
                                                                                       \end{matrix} \right)\\
 &   n_L \rightarrow U_L n_L^{(m)}, \hspace{1cm} n_L \rightarrow U_R n_R^{(m)}, \hspace{1cm} n_{L/R}^{(m)} \text{ are the mass Eigenstates} 
\end{align}

\begin{equation}
   \bar{n}_L M_\nu n_R \rightarrow \bar{n}_L^{(m)} U_L^\dagger M_\nu U_R n_R^{(m)}
\end{equation}

Where we impose the condition that $U_L^\dagger M_\nu U_R $ is diagonal, real and positive. Under these conditions the unitary matrices $U_L$ and $U_R$ are unique modulo rearranging the mass Eigenvalues in the diagonal matrix \footnote{Note that failing to impose the reality and positivity of the diagonal entries breaks the uniqueness of $U_L$ and $U_R$}. The diagonal entries will be the masses of the neutral states. Now me make the Ansatz

\begin{align}
  &U_L=e^{iH_L}V_L, \hspace{1cm} U_R=e^{iH_R}V_R \\
H_L = \left( \begin{array}{cc}
0 & S_L\\
S_L^\dagger & 0 \end{array} \right ),& \hspace{5mm} V_L = \left( \begin{array}{cc}
V_{\text{light}L} & 0\\
0 & V_{\text{heavy}L} \end{array} \right ) \\
H_R = \left( \begin{array}{cc}
0 & S_R\\
S_R^\dagger & 0 \end{array} \right ),& \hspace{5mm} V_R = \left( \begin{array}{cc}
V_{\text{light}R} & 0\\
0 & V_{\text{heavy}R} \end{array} \right ) \\
\end{align}

We impose that the matrices $e^{iH_L}$ and $e^{i H_R}$ block diagonalize the matrix $ M_\nu$:

\begin{equation}   \label{Blockdiag}
{e^{iH_L}}^\dagger\, M_\nu \, e^{iH_R} = \left( \begin{array}{cc}
m_\text{light} & 0\\
0 & m_\text{heavy} \end{array} \right )
\end{equation} 

where $m_\text{light}$ and $m_\text{heavy}$ are general complex matrices of dimensions $n\times n$ and $m\times m$, respectively, which are diagonalized by 

 \begin{equation}  \label{Vdiag}
V_{iL}^\dagger m_i V_{iR} = \text{real, positive, diagonal}, \hspace{0.5cm} i\in\{\text{light}, \text{heavy}\}
\end{equation}

And once again we define the 'parameters order' $\epsilon_1 \sim O\left(\frac{Y_\phi v}{M}\right)$ and $\epsilon_2 \sim O\left(\frac{Y_\chi u}{M}\right)$. If $\epsilon_1, \epsilon_2 << M$ and if we assume $S \sim O(\epsilon)$ we can make a perturbative expansion in the $\epsilon$ parameters keeping only the terms of order $\epsilon^2$ or lower and, after some algebra, find

\begin{equation}
 S_L = - i v Y_\phi M^{-1}, \hspace{1cm} S_R = - i u Y_\chi M^{-1}, \hspace{1cm} m_\text{light} = -Y_\phi v M^{-1} Y_\chi u 
\end{equation}

At first order this implies for the left handed mixing matrix

\begin{align}
 U_L = e^{i H_L} V_L \approx (I_{n+m} + i H_L) V_L = \left(\begin{matrix}
                              I_{n} & v Y_\phi M^{-1} \\
                              -v (M^{-1})^\dagger Y_\phi^\dagger & I_{m}
                             \end{matrix}\right)
                             \left(\begin{matrix}
                              V_{\text{light}L} & 0\\
                              0 & V_{\text{heavy}L}
                             \end{matrix}\right) = \\
                          \left(\begin{matrix}
                              V_{\text{light}L} & v Y_\phi M^{-1} V_{\text{heavy}L} \\
                              -v (M^{-1})^\dagger Y_\phi ^\dagger V_{\text{light}L} & V_{\text{heavy}L}
                             \end{matrix}\right)
\end{align}

Where $I_{k}$ is the identity matrix of dimension $k \times k$. We can now note two relevant features. The matrix $V_{\text{light}L}$ will diagonalize the light neutrino mass matrix from the left, and can thus be identified with the lepton mixing matrix of Eq.~\ref{eq:symparam} (modulo charged lepton contributions to the mixing). $V_{\text{light}L}$ is a unitary matrix and the $U_L$ as written above is unitary to first order. We can conclude that non-unitarity effects on the lepton mixing matrix which affect neutrino oscillations come at order $\sim (v/M)^2$. Non-unitarity bounds can then be translated into lower limits in the mass scale $M$. Moreover, the mixing between the $\nu$ and $N$ states comes at order $\sim v/M$. This mixing will generate interactions between SM particles and the heavy neutrinos which can be tested for example in Charged Lepton Flavour Violation (LFV) processes, see Sec.~\ref{sec:LFV}. As a side note, the right-handed neutrino mixing will instead come at $u/M$ order, but $U_R$ is unphysical.

In the one generation limit we find for the light eigenstates

\begin{equation}
 m_\nu = Y_\phi Y_\chi \frac{v \, u}{M},
 \label{eq:Diracseesaw}
\end{equation}
in clear analogy to Eq.~\ref{eq:Majoranaseesaw}, where now the vev of the singlet $u$ appears multiplying a single power of the SM vev $v$. Again, like in the Majorana case this formula applies to more involved escenarios than the simple one depicted here, as will be explained in Sec.~\ref{sec:Diraczoo}. The $SU(2)_L \times U(1)_Y$ charges can be generalized and textures can be applied to the mass matrices.

 \section{The Dirac seesaw zoo}
 \label{sec:Diraczoo}
 
 We will now generalize the previous result to a class of models which can elegantly accomodate naturally small Dirac neutrino masses in the seesaw spirit. Many of these models are analogous to some popular Majorana seesaw models, but some are completely new. Let us start with the simplest case, the dimension 4 operator

\begin{equation}
y_\nu \,  \bar{L} \Phi^c \nu_R
 \label{nyuk}
\end{equation}
where $y_\nu$ is the Yukawa coupling constant. 
Although this possibility is allowed on theory grounds, it leaves the
smallness of neutrino masses unexplained, implying the need for a tiny
Yukawa coupling for neutrinos ($ y_\nu \sim \mathcal{O}(10^{-13})$). Therefore, this operator is assumed to be forbidden by the `seesaw symmetry' discussed in Sec.~\ref{sec:symmetryrequirement}. Moreover, the Weinberg operator and any other operator for Majorana neutrino masses is assumed to be forbidden by the `Diracness symmetry'.
A more attractive possibility would be to obtain naturally small neutrino masses
through generalized Weinberg operators or their higher dimensional
counterparts.

In general, using only the \sm Higgs doublet $\Phi$, the Standard Model gauge 
symmetry implies that the only allowed dimensions for the operators
that can induce Dirac neutrino masses are even, i.e.  operators
involving odd number of Higgs doublets, namely
\begin{equation}
\frac{1}{\Lambda^{2n}} \bar{L} \Phi^c \left(\Phi^\dagger \Phi \right)^n \nu_R , \hspace{0.2cm} n \in \{0,1,2,3,4...\}
\label{eq:dimnSM}
\end{equation} 
Therefore, after the dimension 4 Yukawa term of \eqref{nyuk}, the next
allowed operator involving only \sm Higgs would be of dimension
6. 
We will first study the dimension 5 operators, which necesarily imply the existence of new scalar fields.

\subsection{Generalized Weinberg operators}
\label{sec:genWein}

We will now build the possibilities for the dimension-5 operators in the Weinberg spirit. We start by constructing the generalized dimension 5 Weinberg operator for Dirac neutrinos, given by
\begin{equation}
\frac{1}{\Lambda} \, \bar{L} \otimes X \otimes Y \otimes \nu_R
\label{dgenw}
\end{equation}
where $X$ and $Y$ are scalar fields transforming as some $n$-plets of
$SU(2)_L$ with appropriate $U(1)_Y$ charges. 

Invariance of \eqref{dgenw} under $SU(2)_L$ symmetry implies that, if
$X$ transforms as a $n$-plet under $SU(2)_L$, then $Y$ must transform
either as a $n+1$-plet, or a $n-1$-plet under $SU(2)_L$ symmetry. 
For example, if we take $X$ to be a singlet then $Y$ should be a
doublet. If we take $X$ to be a doublet then $Y$ can only be a singlet
(equivalent to the previous case) or a triplet. 

\begin{figure}[H]
    \centering
\includegraphics[scale=0.3]{Figures/Dirac-type-I.pdf} \\ \vspace{1 cm}
 \includegraphics[scale=0.4]{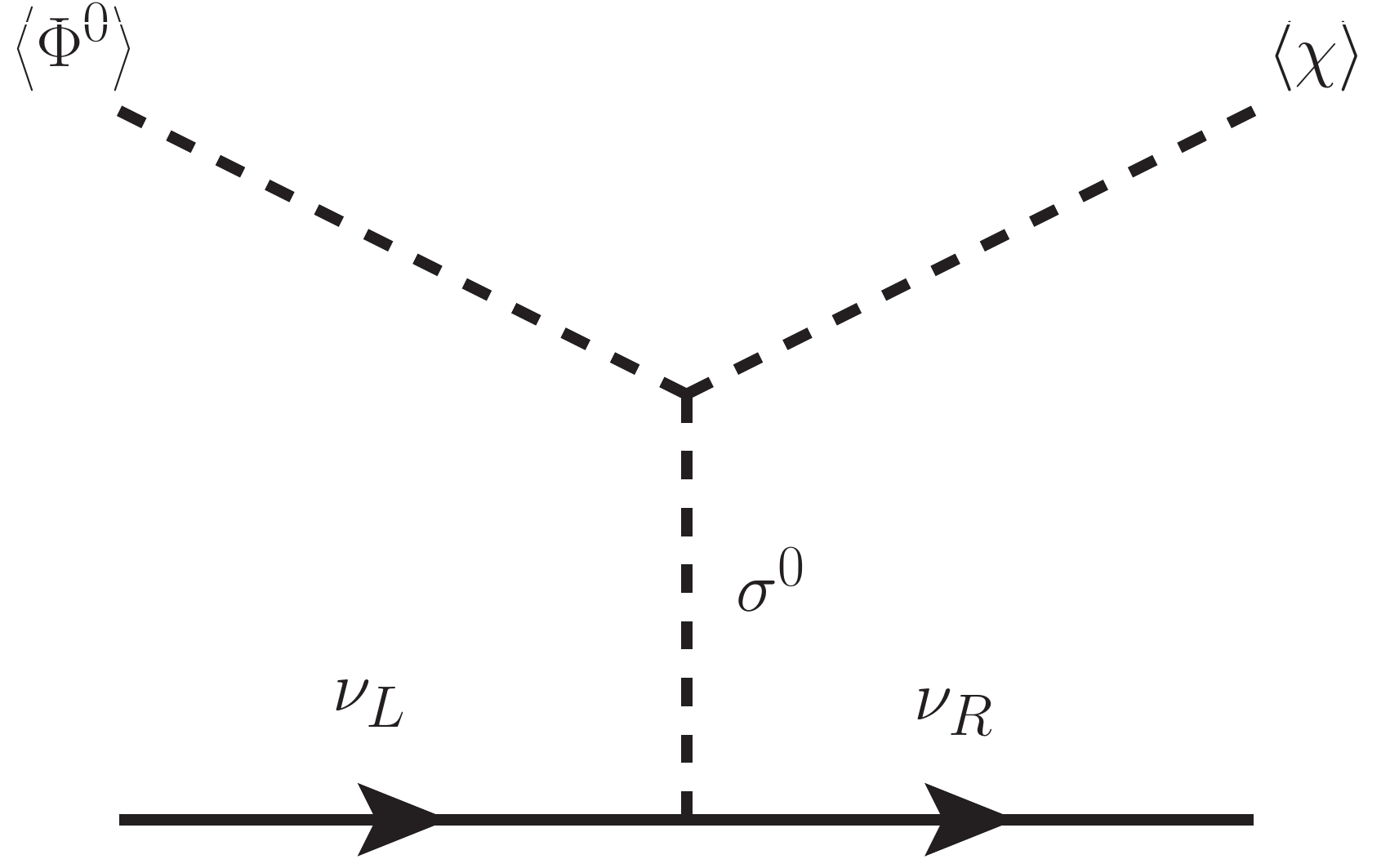} \\ \vspace{1 cm}
 \includegraphics[scale=0.3]{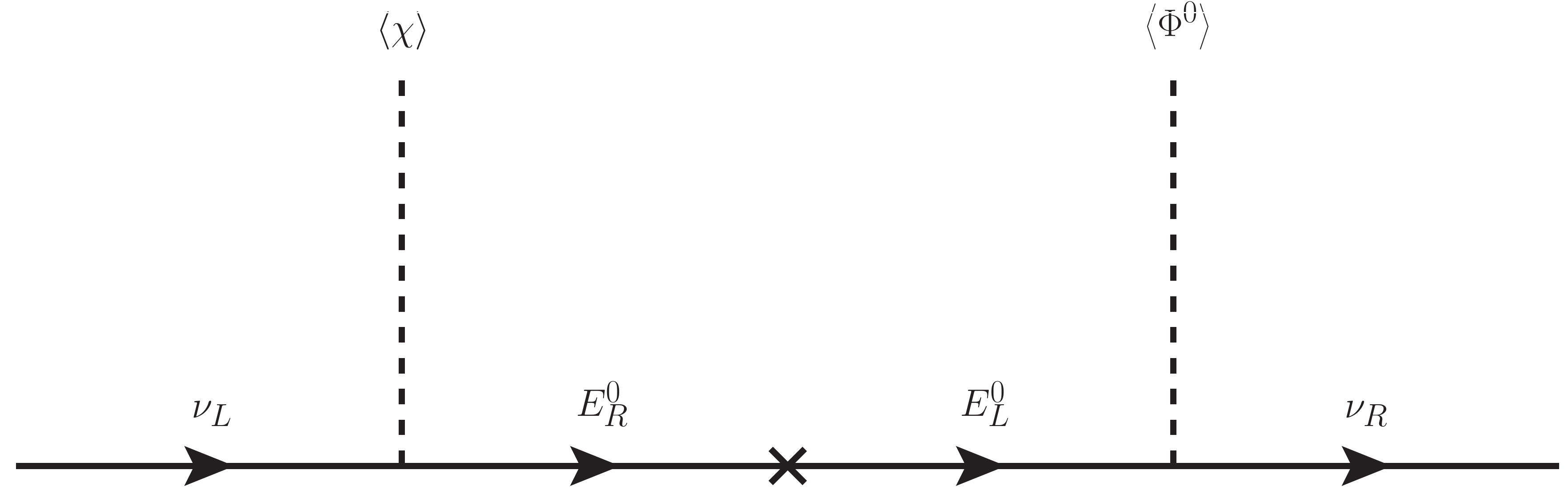}
    \caption{Feynman diagrams representing the Dirac Type I, II and III seesaw analogues.}
    \label{D1}
  \end{figure}

Each of these cases leads to different $SU(2)_L$ contractions which,
as we will see shortly, will lead to different seesaw UV completions.
For example, for the case $X = \chi$ i.e. a singlet under $SU(2)_L$
symmetry and $Y = \Phi$ i.e. a doublet under $SU(2)_L$ symmetry, we
have the following possible contractions, which can be viewed as Dirac
analogues of the type I, II and III Majorana seesaw mechanism,
\begin{equation}
\underbrace{\underbrace{\bar{L} \otimes \Phi^c}_1 \otimes \underbrace{\chi \otimes \nu_R}_{1}}_{\textnormal{Type I analogue}} \hspace{0.2cm} , \hspace{0.5cm}
\underbrace{\underbrace{\bar{L} \otimes \nu_R}_2 \otimes \underbrace{\Phi^c \otimes \chi}_{2}}_{\textnormal{Type II analogue}} \hspace{0.2cm} , \hspace{0.5cm}
\underbrace{\underbrace{\bar{L} \otimes \chi}_2 \otimes \underbrace{\Phi^c \otimes \nu_R}_{2}}_{\textnormal{Type III analogue}} \hspace{0.2cm}, 
\end{equation}
where again the underbrace denotes a contraction of the fields
involved and the number under the brace denotes the $n$-plet
contraction of $SU(2)_L$ to which the fields contract. Note that invariance under $U(1)_Y$ requires that $\Phi^c$ should appear in this operator. The global contraction should be a singlet in order that the operator is allowed
by $SU(2)_L$. The three possible seesaw completions of this operator are shown in figure
\ref{D1}.
  Here the fields $N_L$ and $N_R$ are heavy
  fermions transforming as singlets under $SU(2)_L$, $\sigma^0$ is the
  neutral component of a $SU(2)_L$ doublet scalar, and $E_L^0$ and
  $E_R^0$ are the neutral components of the heavy vector like fermions
  transforming as doublets under $SU(2)_L$.
Note that the new $SU(2)_L$ doublet scalar $\sigma$ in the Type-II case must be a new
scalar, and cannot be identified with the Standard model Higgs $\Phi$,
as such an identification will also imply presence of a tree-level
Dirac neutrino mass term we assumed to be forbidden by symmetries. 

Some of the corresponding UV--complete theories have been discussed in
the literature, while others have not. For example, explicit models
employing type I Dirac seesaw have already been realized in
\cite{Ma:2014qra, CentellesChulia:2016fxr, CentellesChulia:2016rms,
  CentellesChulia:2017koy}, while explicit models for type II
  Dirac seesaw where considered in \cite{Valle:2016kyz,
    Bonilla:2016zef, Reig:2016ewy, Bonilla:2017ekt}. 
  In contrast, to the best of our knowledge, a full--fledged
  UV--complete theory using the Dirac type III seesaw has so far not
  been explicitly developed.\\[-.2cm] 

\begin{figure}[H]
  \centering
  \includegraphics[scale=0.4]{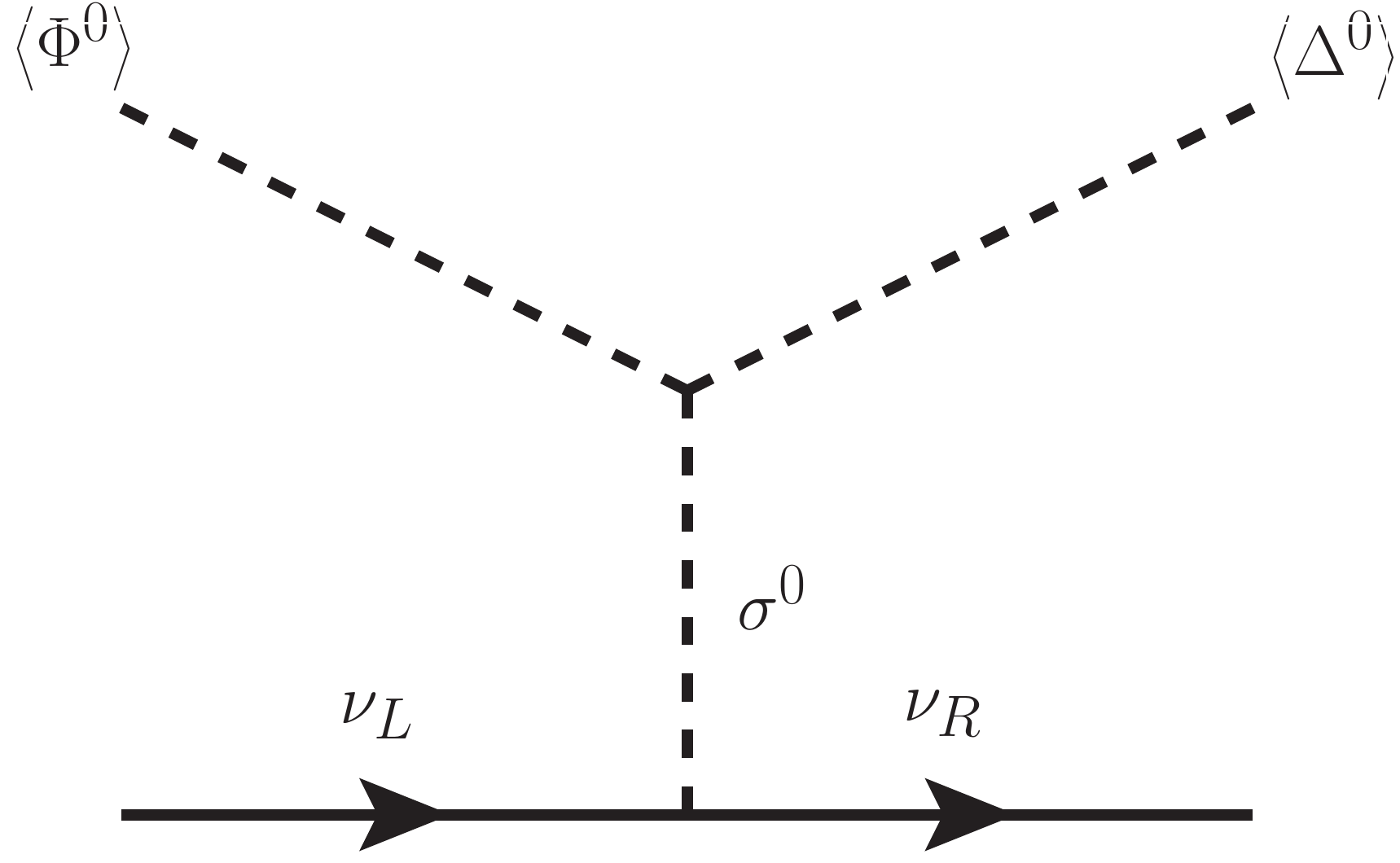}\\ \vspace{1 cm}
  \includegraphics[scale=0.3]{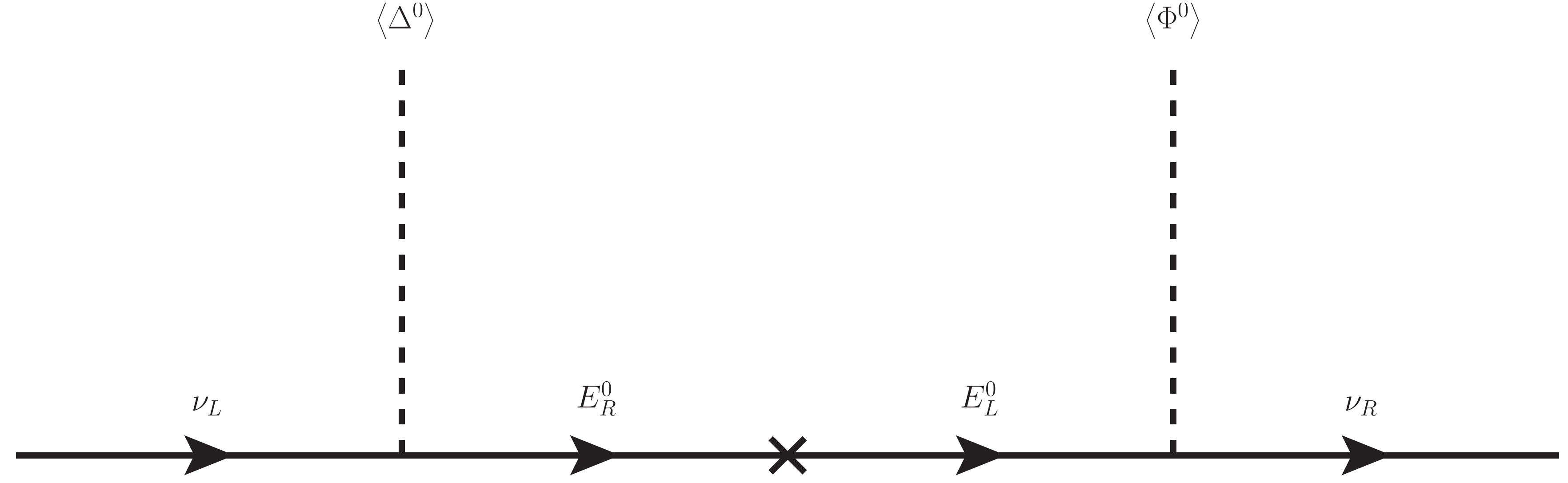}\\ \vspace{1 cm}
\includegraphics[scale=0.3]{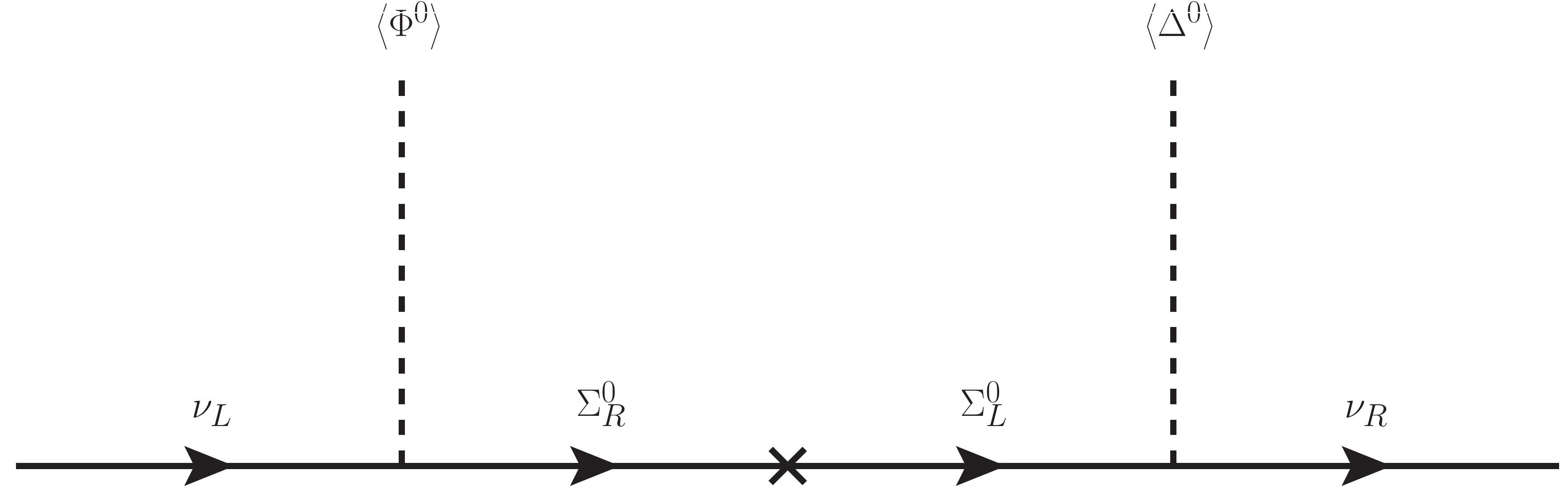}\\ \vspace{1 cm}
\caption{Feynman diagrams for Type-II and Type-III Dirac seesaw mechanism.}
    \label{D2}
  \end{figure}

  Going beyond singlets and doublets opens up still more
  possibilities. For example, taking $X = \Phi$ and $Y = \Delta$, where $\Delta$ is an $SU(2)_L$ triplet \footnote{$SU(2)_L$ multiplets of order higher than $2$ face stringent phenomenological constraints. In particular, their vev must be much smaller than the EW vev and there must not exist a massless Goldstone boson associated to it. Both problems can be solved by inducing the vev and thus generating a dimension 6 operator for neutrino mass, see Sec.~\ref{sec:dim6}.},
  yields  
\begin{equation}
\underbrace{\underbrace{\bar{L} \otimes \nu_R}_2 \otimes \underbrace{\Phi \otimes \Delta}_{2}}_{\textnormal{Type II like}} \hspace{0.2cm}, 
\hspace{0.5cm}
\underbrace{\underbrace{\bar{L} \otimes \Delta}_2 \otimes \underbrace{\Phi \otimes \nu_R}_{2}}_{\textnormal{Type III like}} \hspace{0.2cm} , \hspace{0.5cm}
\underbrace{\underbrace{\bar{L} \otimes \Phi}_3 \otimes \underbrace{\Delta \otimes \nu_R}_{3}}_{\textnormal{Type III like}} \hspace{0.2cm}
\label{dtrip}
\end{equation}
The underbrace denotes, as before, field contraction, and the number
under the brace denotes the $n$-plet contraction of $SU(2)_L$ to which
the fields reduce.
The global contraction should be an $SU(2)_L$ singlet. Note that for this operator we have two possibilities for $U(1)_Y$ charge of $\Delta$.
Apart from the operator in \eqref{dtrip} (which has $\Delta$ with $U(1)_Y = -2$) another operator namely $\bar{L} \Phi^c \Delta_0 \nu_R$ with $\Delta_0$ carrying $U(1)_Y = 0$ is also possible. The diagrams for this case will be identical to those discussed here but the hypercharges of the intermediate fields will be different.
  Note that one can always induce the vevs of either $\chi$ or
  $\Delta$ with the coupling to a pair of $\Phi$'s. Such operators will
  have dimension 6 and will be discussed in Sec.~\ref{sec:dim6}. The
  diagrams leading to the seesaw completion of \eqref{dtrip} are
  shown in Figure \ref{D2}.
  where, as before, the field $\Delta^0$ is the neutral component of
  the $SU(2)_L$ triplet, $\Sigma_L^0$ and $\Sigma_R^0$ are the neutral
  components of the heavy $SU(2)_L$ triplet fermions, $\sigma^0$ is
  the neutral component of an $SU(2)_L$ doublet scalar $\sigma$, while
  $E_L^0$ and $E_R^0$ are the neutral components of the heavy
  vector--like fermions transforming as $SU(2)_L$ doublets. 
  As before, owing to the symmetry requirements, $\sigma$ must be a
  new $SU(2)_L$ doublet, distinct from the Standard model Higgs $\Phi$.
  There have been so far no dedicated study of UV--complete theories
  in literature corresponding to these new possibilities.
  
  Going yet to higher multiplets of $SU(2)_L$ will open up novel ways
  to generate Dirac neutrino mass at the dimension 5 level.  These can
  easily be realized following our procedure in a straightforward way,
  so here we skip the details.
  Furthermore, as already discussed, Dirac neutrino masses can also be
  generated at higher dimensions. The general operator for such a
  scenario involves several scalar fields $X_i$; $i = 1, \cdots , n$
  which can be different $SU(2)_L$ multiplets, carrying appropriate
  $U(1)_Y$ charges, as follows
\begin{equation}
\frac{1}{\Lambda^{n-1}} \, \bar{L} \otimes X_1 \otimes \cdots \otimes X_n \otimes \nu_R
\label{dhwop}
\end{equation}
Where some of the $X_i$ may coincide and the overall combination $X_1 \otimes \cdots \otimes X_n$ must transform as an $SU(2)$ doublet of hypercharge $-1/2$.
\subsection{Dimension 6 operators}
\label{sec:dim6}
 We now turn our discussion to operators involving only \sm scalar
doublet $\Phi$ and discuss the various possible UV-complete models for
this case. 
As has been argued in \cite{CentellesChulia:2018gwr} and in Eq.~\ref{eq:dimnSM}, for Dirac
neutrinos, after the Yukawa term, the lowest dimensional operator
involving only the \sm scalar doublet appears at dimension-6 and is
given by
\begin{equation}
 \frac{1}{\Lambda^2} \bar{L}\otimes \Phi^c \otimes \Phi^c \otimes \Phi \otimes \nu_R
 \label{op-sm}
\end{equation}
where $L$ and $\Phi$ denote the lepton and Higgs doublets, $\nu_R$ is
the right-handed neutrino field and $\Lambda$ represents the cutoff
scale.
Above $\Lambda$ the UV complete theory is at play,
involving new ``messenger'' fields, whose masses lie close to the
scale $\Lambda$. Recently, this operator has also been studied in
\cite{Yao:2017vtm} and our results agree.

\begin{figure}[!h]
 \centering
  \includegraphics[scale=0.35]{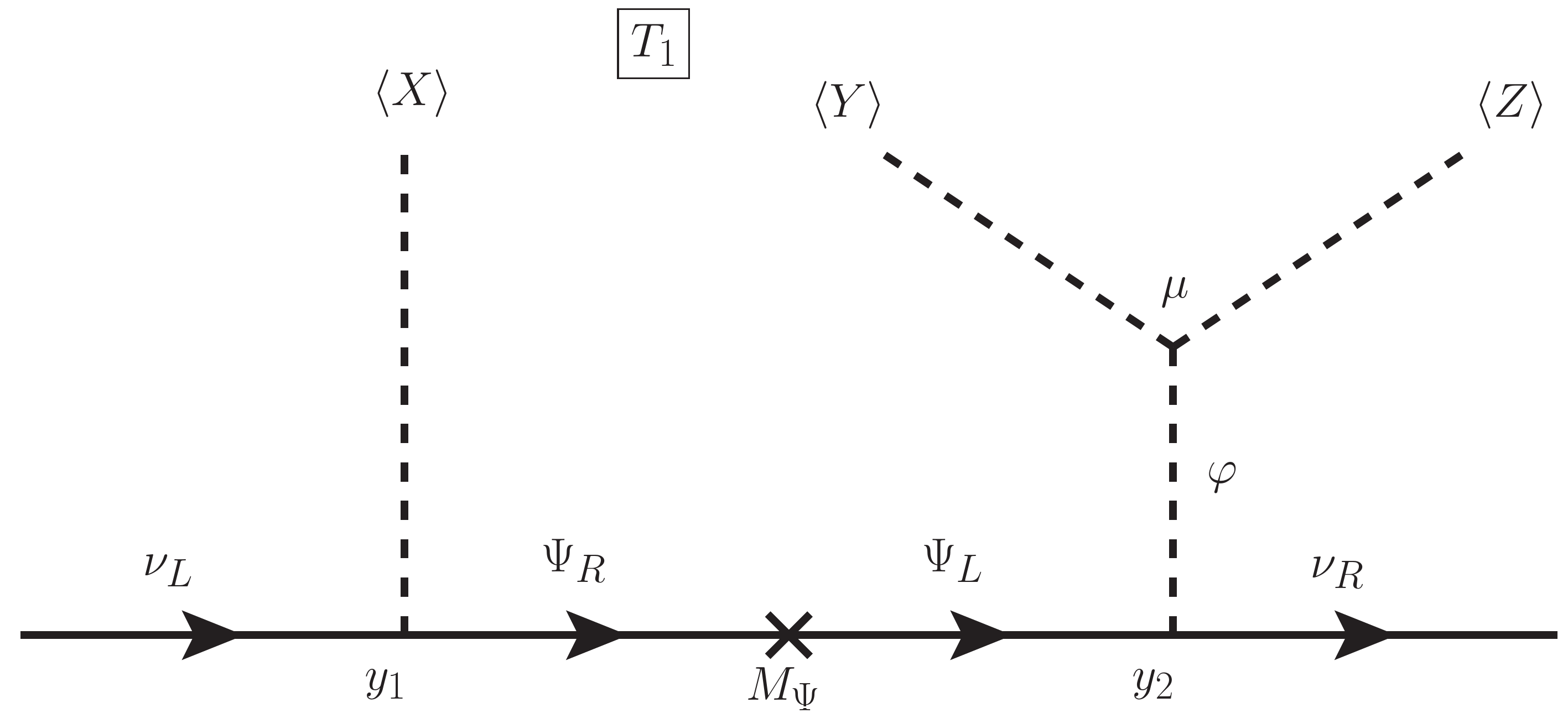} \\ \vspace{0.5cm}
   \includegraphics[scale=0.35]{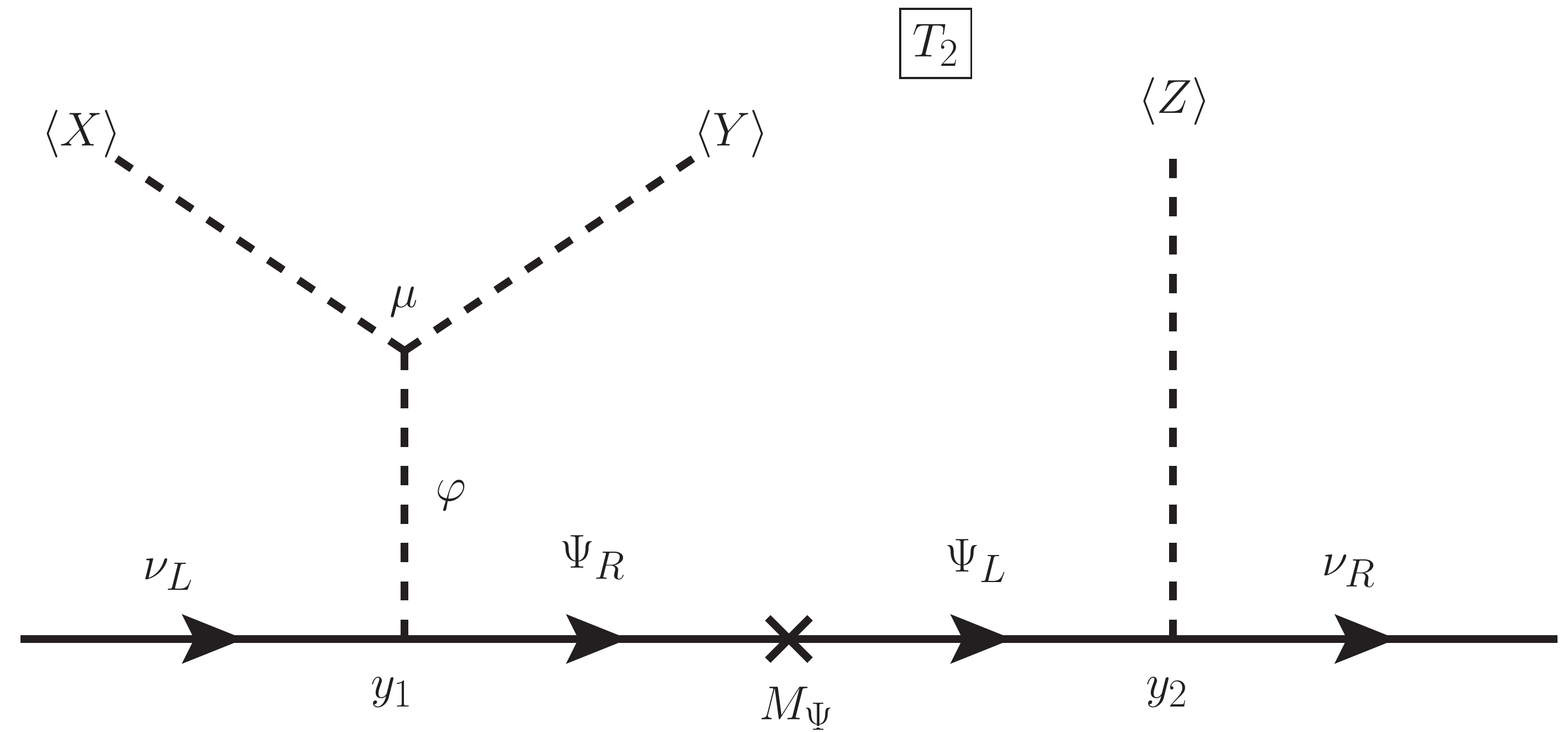} \\ \vspace{0.5cm}
    \includegraphics[scale=0.35]{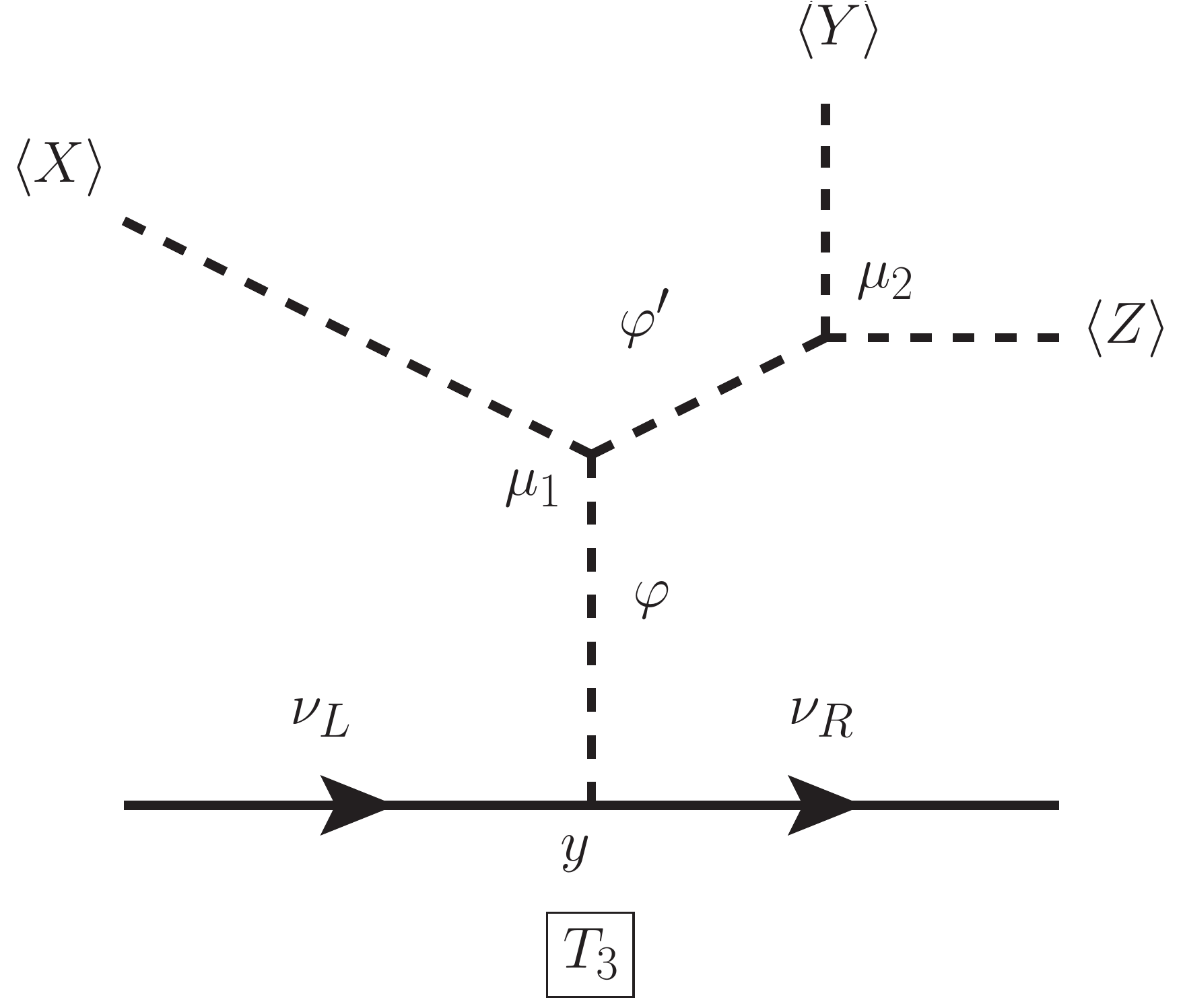}
     \includegraphics[scale=0.35]{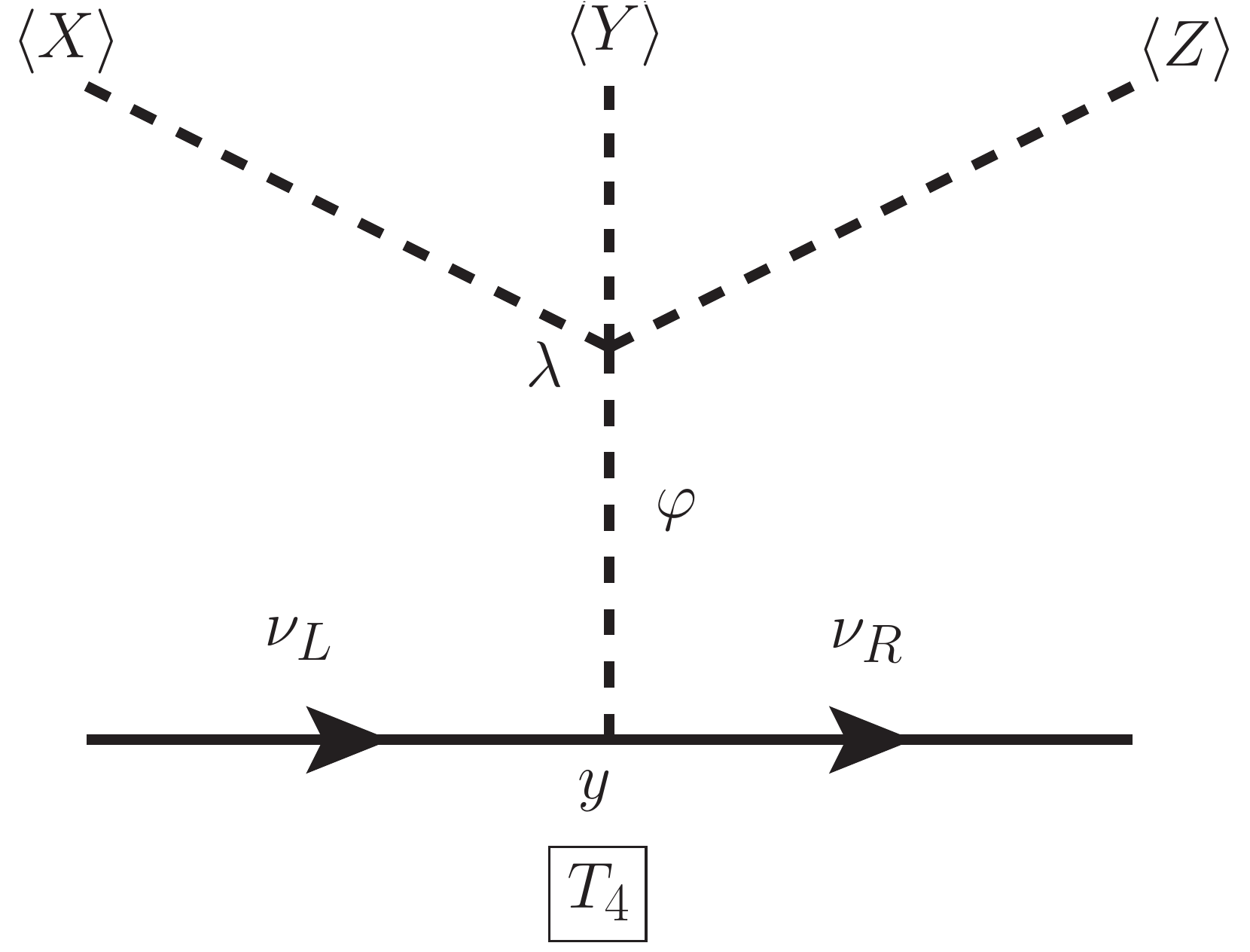} \\ \vspace{0.5cm}
      \includegraphics[scale=0.35]{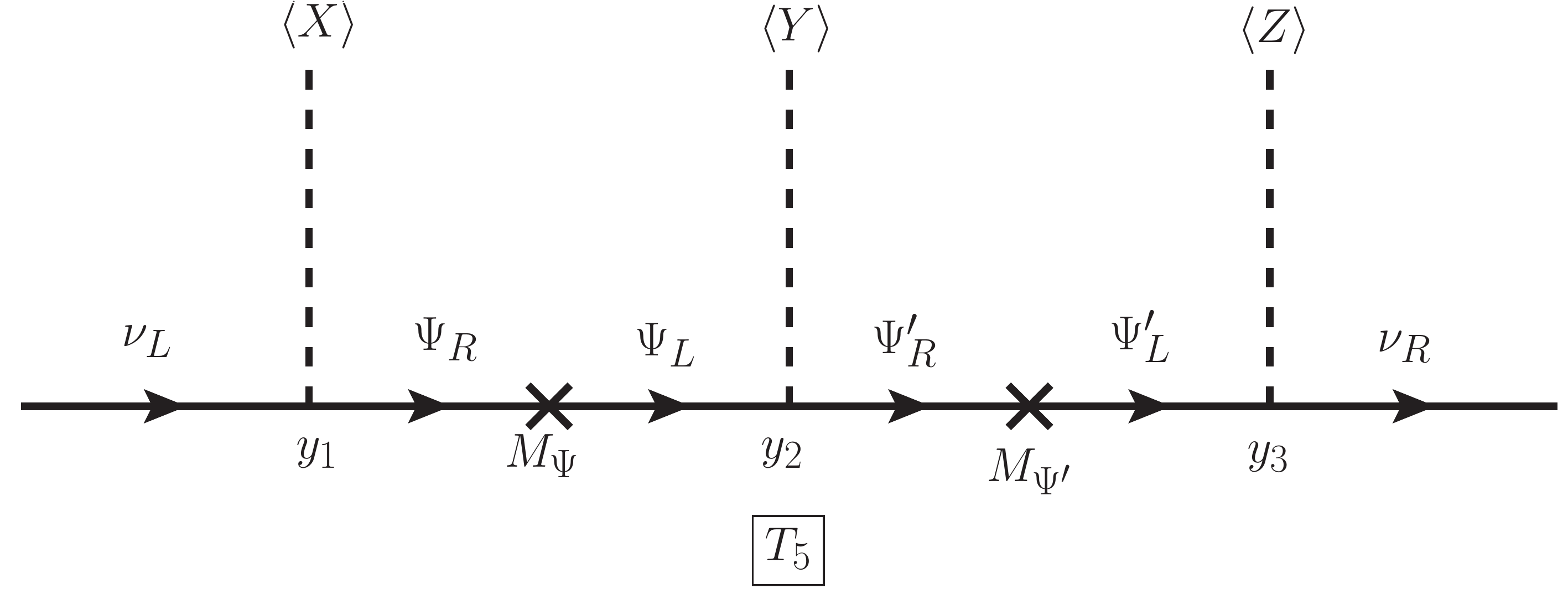}
       \caption{Feynman diagrams representing the five different topologies, $T_1$, $T_2$, $T_3$, $T_4$, $T_5$,  respectively. }
        \label{fig:topologies}
\end{figure}

Before starting our systematic classification of the UV-complete
seesaw models emerging from this operator, we stress again that, in order
for this operator to give the leading contribution to naturally small Dirac neutrino
masses, the lower dimensional Yukawa term $\bar{L} \Phi^c \nu_R$
should be forbidden by the `seesaw symmetry'. 
The operator in \eqref{op-sm} can lead to several different
UV-complete seesaw models, depending on the field contractions
involved, which can be arranged into five distinct topologies for
the Feynman diagrams of neutrino mass generation.  For lack of better
names, we are calling these five topologies as $T_i$,
$i \in \{1, 2, 3, 4, 5\}$. These topologies are shown in Fig.~\ref{fig:topologies}. 

Each topology involves new ``messenger fields'' which can be either
new scalars ($\varphi$), new fermions ($\Psi$) or both. The masses of
these heavy messenger fields lying typically at or above the cutoff
scale $\Lambda$ of the dimension-6 operators.
The first topology $T_1$ involves two messenger fields, a scalar
($\varphi$) and a Dirac fermion ($\Psi$). The scalar $\varphi$ gets a
small induced vev through its trilinear coupling with $Y$ and $Z$
scalars. 
Topology $T_2$ is very similar to $T_1$ and also involves two
messenger fields, a scalar $\varphi$ and a Dirac fermion
$\Psi$. However, the small vev of $\varphi$ in $T_2$ is induced by its
trilinear coupling with $X$ and $Y$ scalars.
The third topology $T_3$ is distinct from the first two and only
involves scalar messengers $\varphi$ and $\varphi^\prime$. The scalar
$\varphi^\prime$ gets a small induced vev through its trilinear coupling
with $Y$ and $Z$ scalars. The other scalar $\varphi$ subsequently
gets a ``doubly-induced vev'' through its trilinear coupling with
$\varphi^\prime$ and $X$ scalars. 
The fourth topology $T_4$ only involves a single messenger scalar
$\varphi$ which gets an induced vev through its quartic coupling with
scalars $X$, $Y$ and $Z$. 
The final fifth topology involves only fermionic Dirac messengers
$\Psi$ and $\Psi'$, as shown in Fig.~\ref{fig:topologies}. 
The $SU(2)_L \otimes U(1)_Y$ charges of the messenger fields in all
topologies will depend on the details of the operator under
consideration and the contractions involved. We will discuss all such
possibilities in the following sections.
Each topology leads to different estimates for the associated light
neutrino mass generated in each case. Neglecting the three generation
structure of the various Yukawa coupling matrices in family space, the resulting formulas for the neutrino masses are listed in Table
\ref{Tab:mass-formula}. 
\begin{table}[ht]
\begin{center}
\begin{tabular}{c c c}
  \hline \hline  
Topology  \hspace{1cm} &   Messenger Fields   \hspace{1cm}   & Neutrino Mass Estimate                  \\
\hline \hline 
T1        \hspace{1cm} &  $\Psi$, $\varphi$   \hspace{1cm}   &  $ \frac{\mu y_1 y_2 \, v_X v_Y v_Z}{M_\psi M^2_\varphi}$   \vspace{2mm}   \\
T2        \hspace{1cm} &  $\Psi$, $\varphi$   \hspace{1cm}   &  $ \frac{\mu y_1 y_2 \, v_X v_Y v_Z}{M_\psi M^2_\varphi}$   \vspace{2mm}    \\
T3        \hspace{1cm} &  $\varphi$, $\varphi^\prime$ \hspace{1cm} &  $ \frac{\mu_1 \mu_2 y \, v_X v_Y v_Z}{M^2_{\varphi} M^2_{\varphi^\prime}}$ \vspace{2mm}  \\
T4       \hspace{1cm}  & $\varphi$            \hspace{1cm}   &  $ \frac{ y \lambda \, v_X v_Y v_Z}{M^2_\varphi}$              \vspace{2mm}  \\
T5       \hspace{1cm}  & $\Psi$, $\Psi'$      \hspace{1cm}   &  $ \frac{ y_1 y_2 y_3 \, v_X v_Y v_Z}{M_{\psi} M_{\psi'}}$        \vspace{2mm}                \\
  \hline
  \end{tabular}
\end{center}
\caption{Possible topologies and messengers leading to light neutrino
  masses, and the associated estimates for each topology.  }
 \label{Tab:mass-formula}
\end{table}

There are fifteen inequivalent ways of contracting the operator in Eq.~\ref{op-sm}, each
of which will require different types of messenger fields for
UV-completion.
Out of these $15$, $9$ are very similar to the 'pure' type-I ($1$), type-II ($3$) and type-III ($5$) Dirac seesaws described in Sec.~\ref{sec:genWein}, with the particularity that the singlet or triplet external scalars will now acquire an 'induced' vev, which in turn can provide a further suppression to neutrino mass. Moreover, the cubic coupling between the scalar $S$ (be it singlet or triplet) with two Higgs doublets will also remove the massless Goldstone boson associated to the pseudo scalar neutral component of $S$. This is not mandatory in the singlet case, since a singlet Goldstone can scape the experimental constraints such as stellar cooling in many appealing scenarios, see Sec.~\ref{sec:inverseseesaw}, but it is key in the triplet case, since a Goldstone triplet is ruled out by collider signatures \cite{Mandal:2021acg}. We will study these $9$ induced versions of the type I, II and III Dirac seesaws in Sec.~\ref{sec:dim6induced}. The remaining $6$ diagrams generate new models and will be shown in Sec.~\ref{sec:dim6new}

\subsubsection{Type I, II and III Dirac seesaw mechanisms with induced vev}
\label{sec:dim6induced}

\begin{figure}[!h]
 \centering
  \includegraphics[scale=0.35]{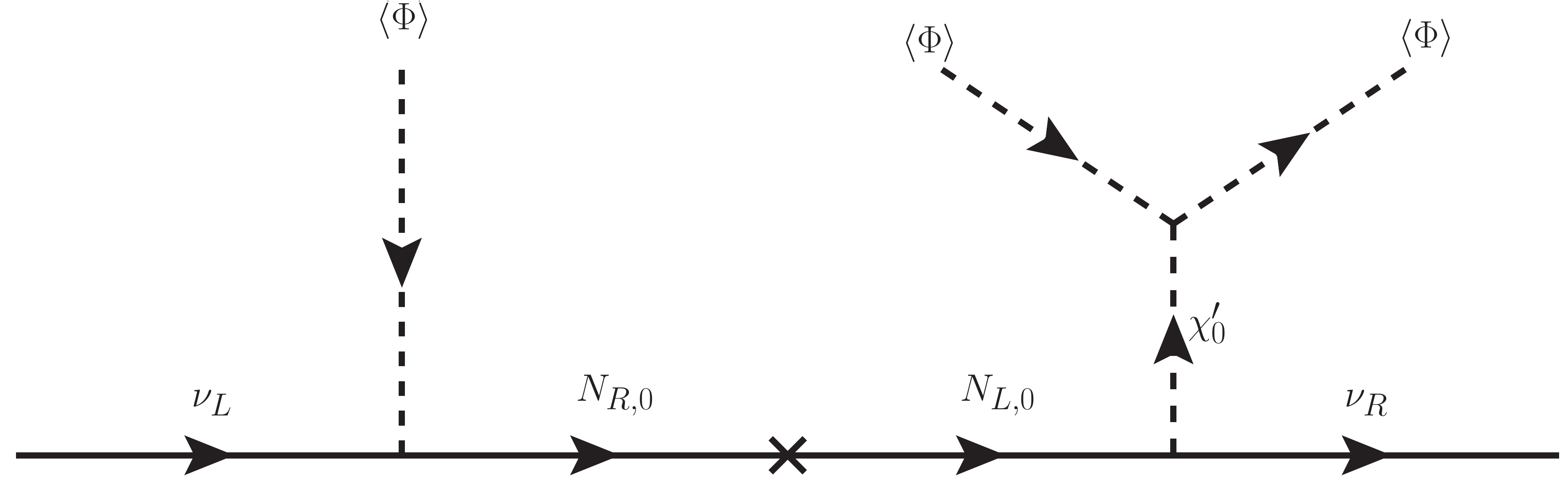}
  \caption{Feynman diagram representing the Dirac Type-I seesaw with
    an induced vev for $\chi^\prime_0$.}
    \label{DT1}
 \end{figure}

We start by laying out the simplest contraction of the dimension-6 operator of
\eqref{op-sm}, which is as follows:
\begin{equation}
\underbrace{\underbrace{\underbrace{\bar{L} \otimes \Phi^c}_1 \otimes \underbrace{\Phi^c \otimes \Phi}_1}_{1} \otimes \underbrace{\nu_R}_{1}}_{\textnormal{Type I with induced vev  Fig.~\ref{DT1}}}
\label{DT1-op}
\end{equation}

Like in previous sections, in ~\eqref{DT1-op} the under-brace denotes a $SU(2)_L$ contraction of the fields involved, whereas the number given under it denotes the
transformation of the contracted fields under $SU(2)_L$ (note that the
other possible contraction in which $\Phi^c \otimes \Phi^c$
goes to a singlet is simply $0$).
Although not made explicit, we take it for granted that the global
contraction leading to a UV-complete model where the neutrino mass is
generated by the diagram shown in Fig.~\ref{DT1} should always be an
$SU(2)_L$ singlet.
 
 The diagram in Fig.~\ref{DT1} involves two messenger fields, a vector-like neutral fermion $N_0$ and a scalar $\chi^\prime_0$, both singlets under the \SM gauge
 group. 
 As listed in Table \ref{Tab:mass-formula}, the light neutrino mass is
 doubly suppressed first by the mass of the fermion $N_0$ and also by
 the small induced vev for $\chi^\prime_0$. 
 In contrast to the type I Dirac seesaw diagram of Fig.~\ref{D1}, here the messenger field $\chi^\prime_0$
 required for the UV-completion gets a small induced vev via its cubic
 coupling with the Standard Model Higgs doublet, also removing the massless Goldstone boson associated to it.

\begin{figure}[!h]
 \centering
  \includegraphics[scale=0.35]{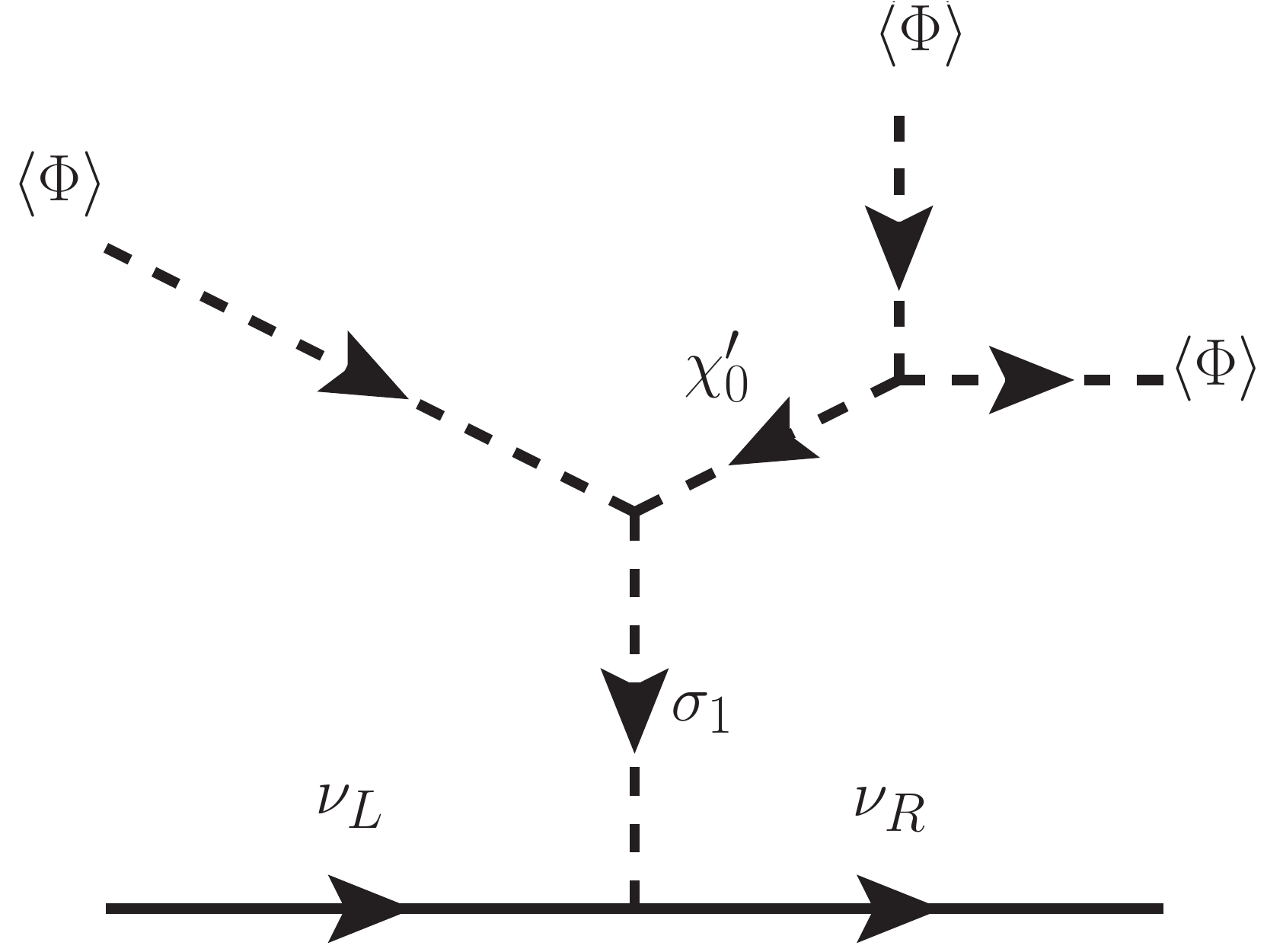}, \hspace{2mm}
   \includegraphics[scale=0.35]{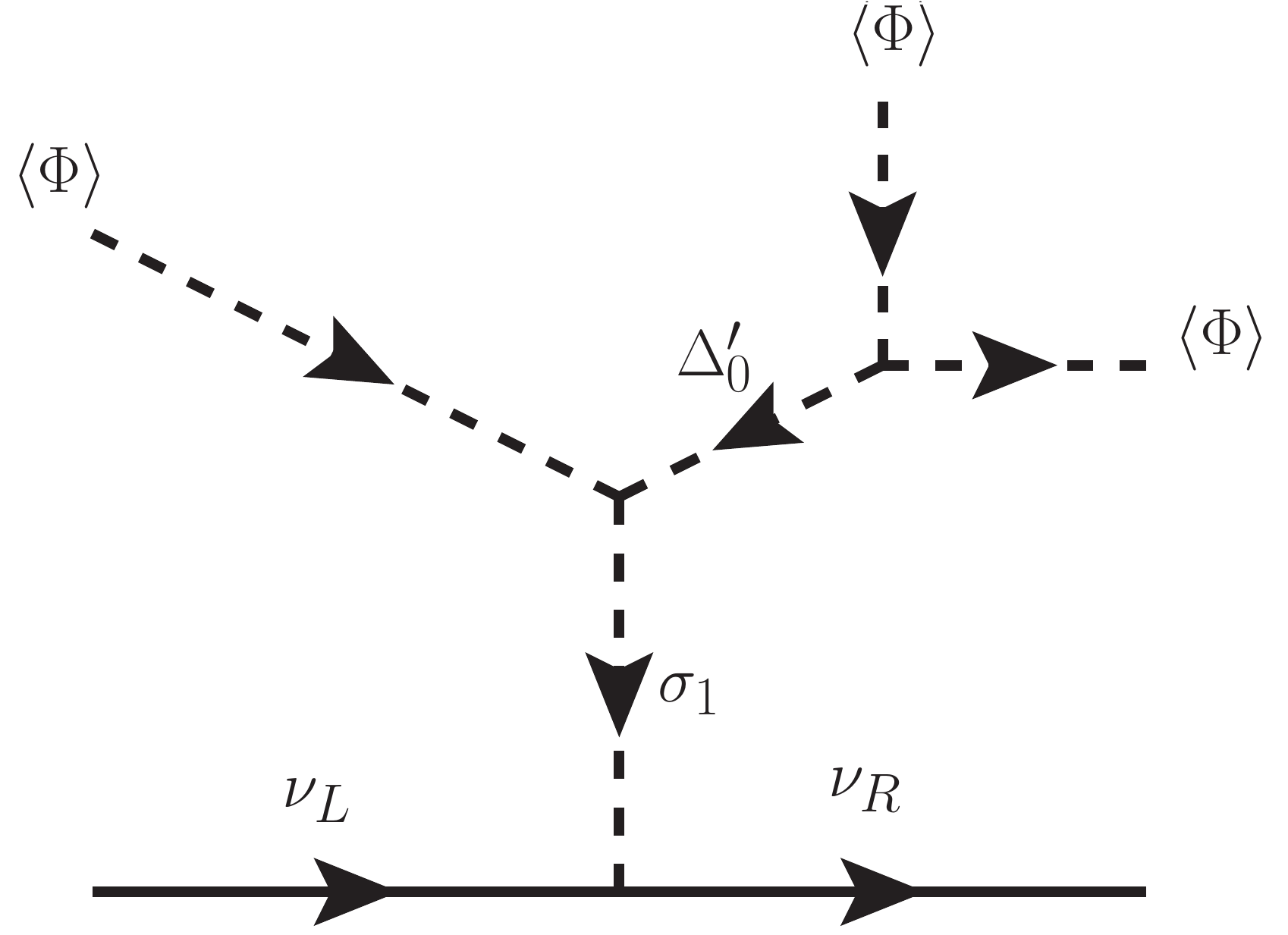}, \hspace{2mm}
    \includegraphics[scale=0.35]{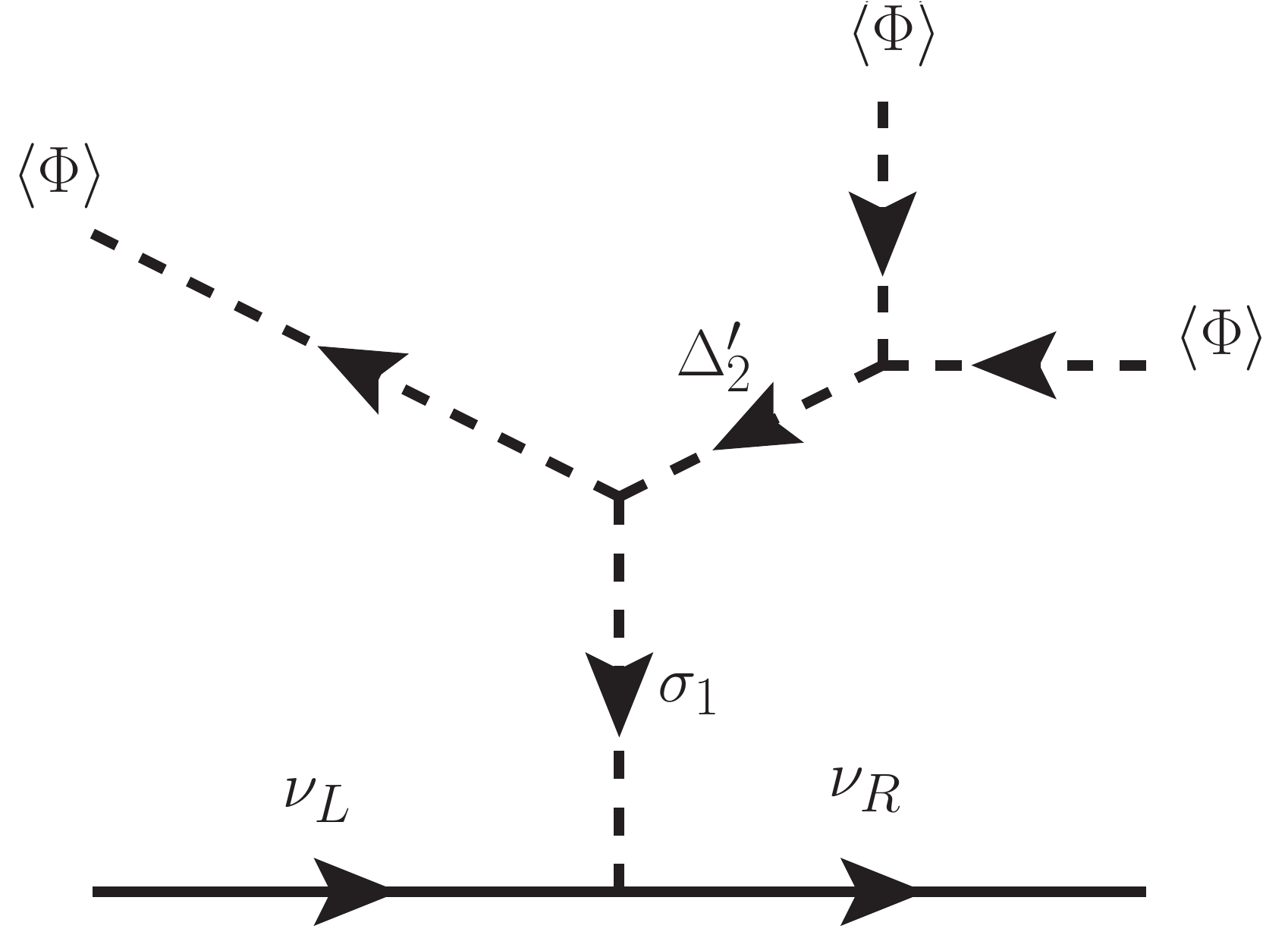}, \hspace{2mm}
     \caption{Feynman diagrams representing the three realizations of the Dirac Type-II seesaw with an induced vev for $\chi^\prime_0$ or $\Delta^\prime_i$.}
    \label{DT2}
 \end{figure}

The three possibilities for this case are shown in \eqref{DT2-op}. 
\begin{equation}
\underbrace{\underbrace{\underbrace{\bar{L} \otimes \nu_R}_2 \otimes \underbrace{\Phi^c \otimes \Phi}_1}_{2} \otimes \underbrace{\Phi^c}_{2}}_{\textnormal{Type II induced  Fig.~\ref{DT2}}}, \hspace{0.5cm}
\underbrace{\underbrace{\underbrace{\bar{L} \otimes \nu_R}_2 \otimes \underbrace{\Phi^c \otimes \Phi}_3}_{2} \otimes \underbrace{\Phi^c}_{2}}_{\textnormal{Type II induced   Fig.~\ref{DT2}}} , \hspace{0.5cm}
\underbrace{\underbrace{\underbrace{\bar{L} \otimes \nu_R}_2 \otimes \underbrace{\Phi^c \otimes  \Phi^c_3}_{2}} \otimes \underbrace{\Phi}_{2}}_{\textnormal{Type II induced Fig.~\ref{DT2}}}
\label{DT2-op}
\end{equation}

These three contraction possibilities lead to three different
UV-completions, as illustrated in Fig.~\ref{DT2}.
 All diagrams in Fig.~\ref{DT2} belong to the $T_3$ topology, and
 require two scalar messengers.  The diagram on the left requires a
 $SU(2)_L$ singlet $\chi^\prime_0$ and a new doublet $\sigma_1$ (different from
 the \sm Higgs doublet) with $U(1)_Y = 1$.
 The middle one requires an $SU(2)_L$ triplet $\Delta^\prime_0$ (with
 $U(1)_Y = 0$) and an $SU(2)_L$ doublet $\sigma_1$ (with $U(1)_Y = 1$)
 scalar messengers.  
 The third diagram is identical to the second, exchanging
 $\Phi \leftrightarrow \Phi^c$ in two external legs. Note that the
 hypercharges of the intermediate fields $\Delta^\prime_{0}$ and $\Delta^\prime_{2}$
 are different so that, although the UV-completions share the same
 topology, the underlying models are different. 
 The associated light neutrino mass estimate is given in Table
 \ref{Tab:mass-formula}. 

\begin{figure}[H]
 \centering
  \includegraphics[scale=0.3]{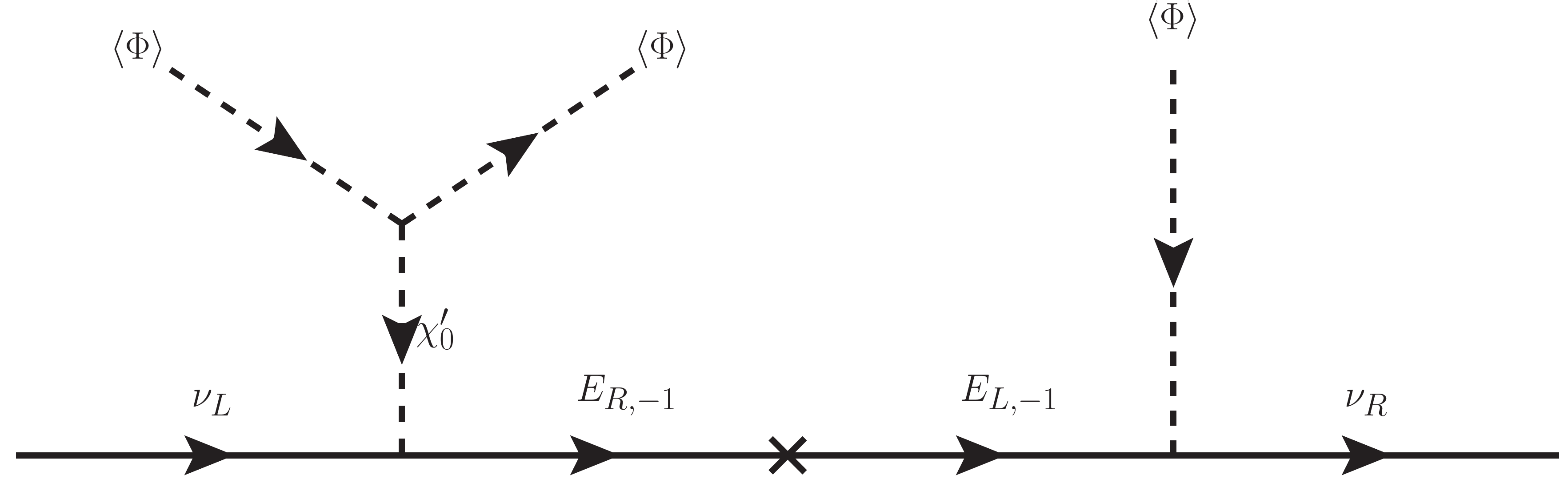}, \hspace{2mm}
   \includegraphics[scale=0.3]{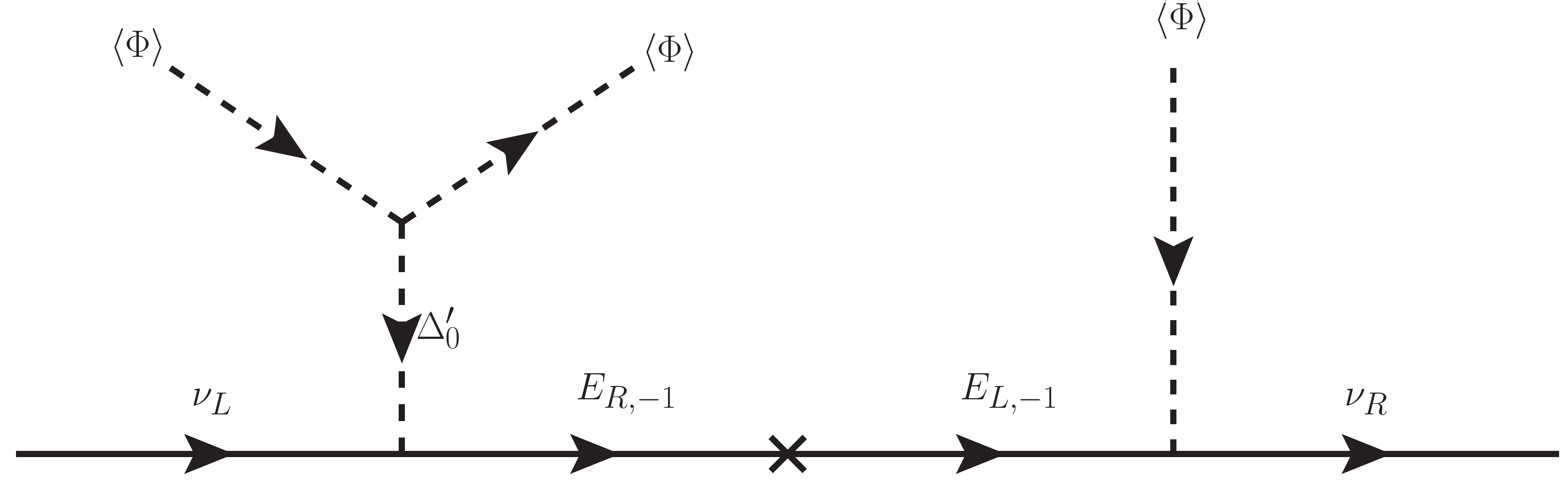}, \hspace{2mm}
    \includegraphics[scale=0.3]{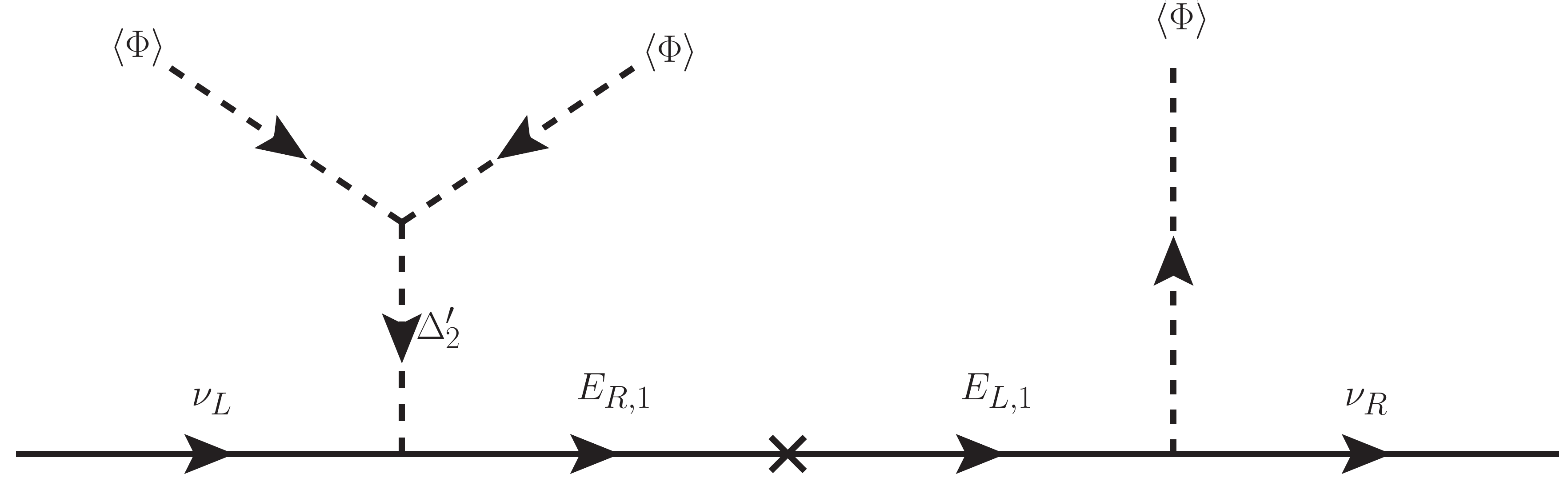}, \hspace{2mm}
     \includegraphics[scale=0.3]{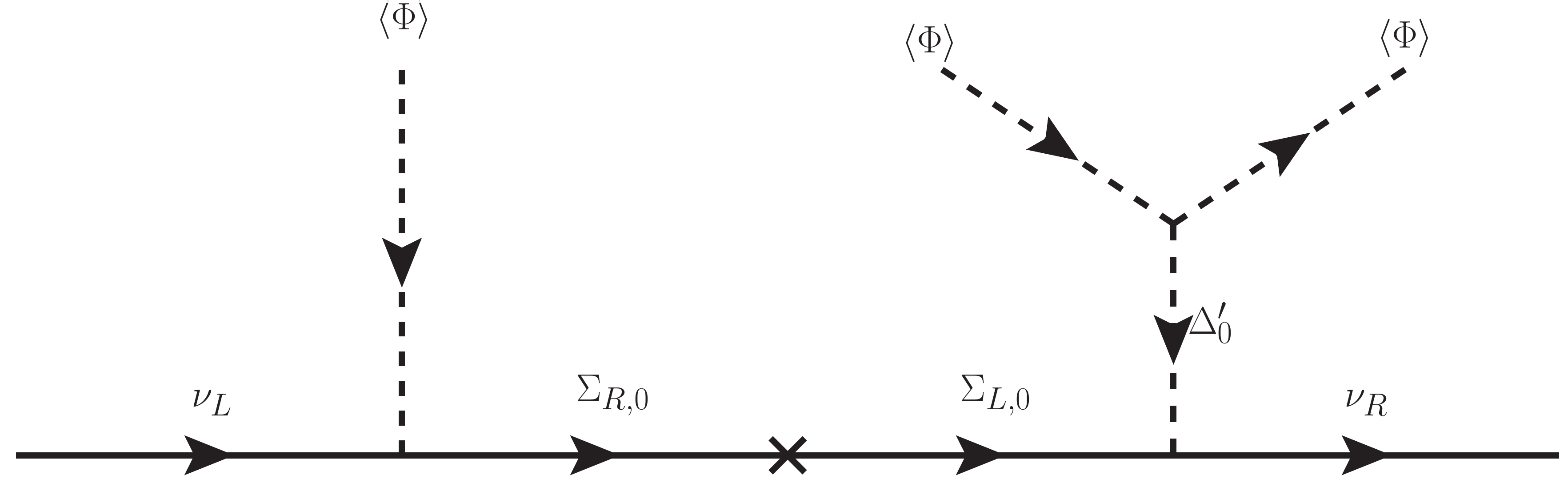}, \hspace{2mm}
      \includegraphics[scale=0.3]{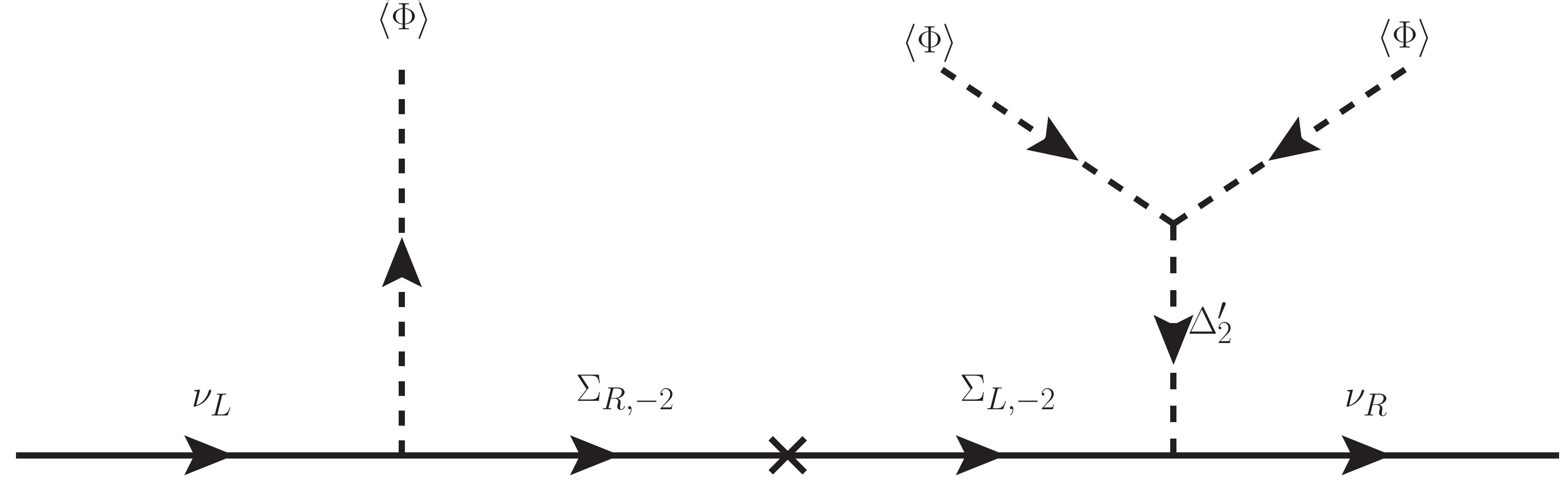}
       \caption{Feynman diagram representing the five realizations of the Dirac Type-III seesaw with an induced vev for $\chi^\prime_0$ or $\Delta^\prime_i$.}
    \label{DT3}
 \end{figure}

The operator of \eqref{op-sm} also leads to five distinct type-III
like seesaw possibilities with induced vevs~\footnote{We denote all
  diagrams with $T_1$ or $T_2$ topologies as type-III seesaw-like if
  they involve fermions transforming non-trivially under $SU(2)_L$.}.
The various possible contractions leading to such possibilities as
shown in \eqref{DT32-op}. 
\begin{equation}
\underbrace{\underbrace{\underbrace{\bar{L}}_2 \otimes \underbrace{\Phi^c \otimes \Phi}_1}_2 \otimes \underbrace{\Phi^c \otimes \nu_R}_{2}}_{\textnormal{Type III induced Fig.~\ref{DT3}}} , \hspace{0.5cm}
\underbrace{\underbrace{\underbrace{\bar{L}}_2 \otimes \underbrace{\Phi^c \otimes \Phi}_3}_2 \otimes \underbrace{\Phi^c \otimes \nu_R}_{2}}_{\textnormal{Type III induced Fig.~\ref{DT3}}} , \hspace{0.5cm}
\underbrace{\underbrace{\underbrace{\bar{L}}_2 \otimes \underbrace{\Phi^c \otimes \Phi^c}_3}_2 \otimes \underbrace{{\Phi} \otimes \nu_R}_{2}}_{\textnormal{Type III induced Fig.~\ref{DT3}}} \nonumber
\label{DT3-op}
\end{equation}
\begin{equation}
\underbrace{\underbrace{\underbrace{\bar{L} \otimes \Phi^c}_3 \otimes \underbrace{\Phi^c \otimes \Phi}_3}_{1} \otimes \underbrace{\nu_R}_{1}}_{\textnormal{Type III induced Fig.~\ref{DT3}}}, \hspace{0.5cm}
\underbrace{\underbrace{\underbrace{\bar{L} \otimes {\Phi}}_3 \otimes \underbrace{\Phi^c \otimes \Phi^c}_3}_{1} \otimes \underbrace{\nu_R}_{1}}_{\textnormal{Type III induced Fig.~\ref{DT3}}}
\label{DT32-op}
\end{equation}

The UV-completions of each of these possible contractions involve
different messenger fields, leading to five inequivalent models. The
neutrino mass generation in these models is shown diagrammatically in
Figure \ref{DT3}. 
 The first diagram in Figure \ref{DT3} involves, as messenger fields,
 scalar singlet $\chi^\prime_0$ and vector-like fermions $E_{-1}$
 transforming as an $SU(2)_L$ doublet with $U(1)_Y = -1$. 
 The second diagram involves hypercharge-less $SU(2)_L$ triplet
 scalars $\Delta^\prime_0$ and the same $SU(2)_L$ doublet vector-like
 fermion $E_{-1}$ as messenger fields. 
 The third diagram is identical to the second one, but with exchange
 $\Phi \leftrightarrow \Phi^c$ in the external legs. This leads to
 a different hypercharge $U(1)_Y = 2$ for the intermediate scalar
 triplet $\Delta^\prime_{2}$, as well as for the $SU(2)_L$ doublet vector
 fermion $E_1$, with $U(1)_Y = 1$. 
 The fourth and fifth diagrams again are related to each other by
 exchanging $\Phi \leftrightarrow \Phi^c$ in two external
 legs. They involve, as messenger fields, $SU(2)_L$ triplet scalars
 $\Delta^\prime_i$; $i = 0, 2$ together with vector-like $SU(2)_L$ triplet
 fermions $\Sigma_i$; $i = 0, -2$.  The hypercharges of $\Delta_i^\prime$ are
 $U(1)_Y = 0, 2$ and of $\Sigma_i$ are $U(1)_Y = 0, -2$
 respectively. 
 The first three diagrams belong to $T_2$ topology, while the fourth
 and fifth diagrams have the topology $T_1$ and the associated light
 neutrino masses for $T_1$ and $T_2$ are given in Table
 \ref{Tab:mass-formula}. 
 Notice that, in contrast to the type III like Dirac seesaw diagrams
 discussed in Fig.~\ref{D1}, here the $\chi^\prime_0$ and
 $\Delta^\prime_i$ both get induced vevs from their cubic interaction terms
 with the Standard Model Higgs doublet.
 
\subsubsection{New diagrams}
\label{sec:dim6new}

\begin{figure}[!h]
 \centering
  \includegraphics[scale=0.4]{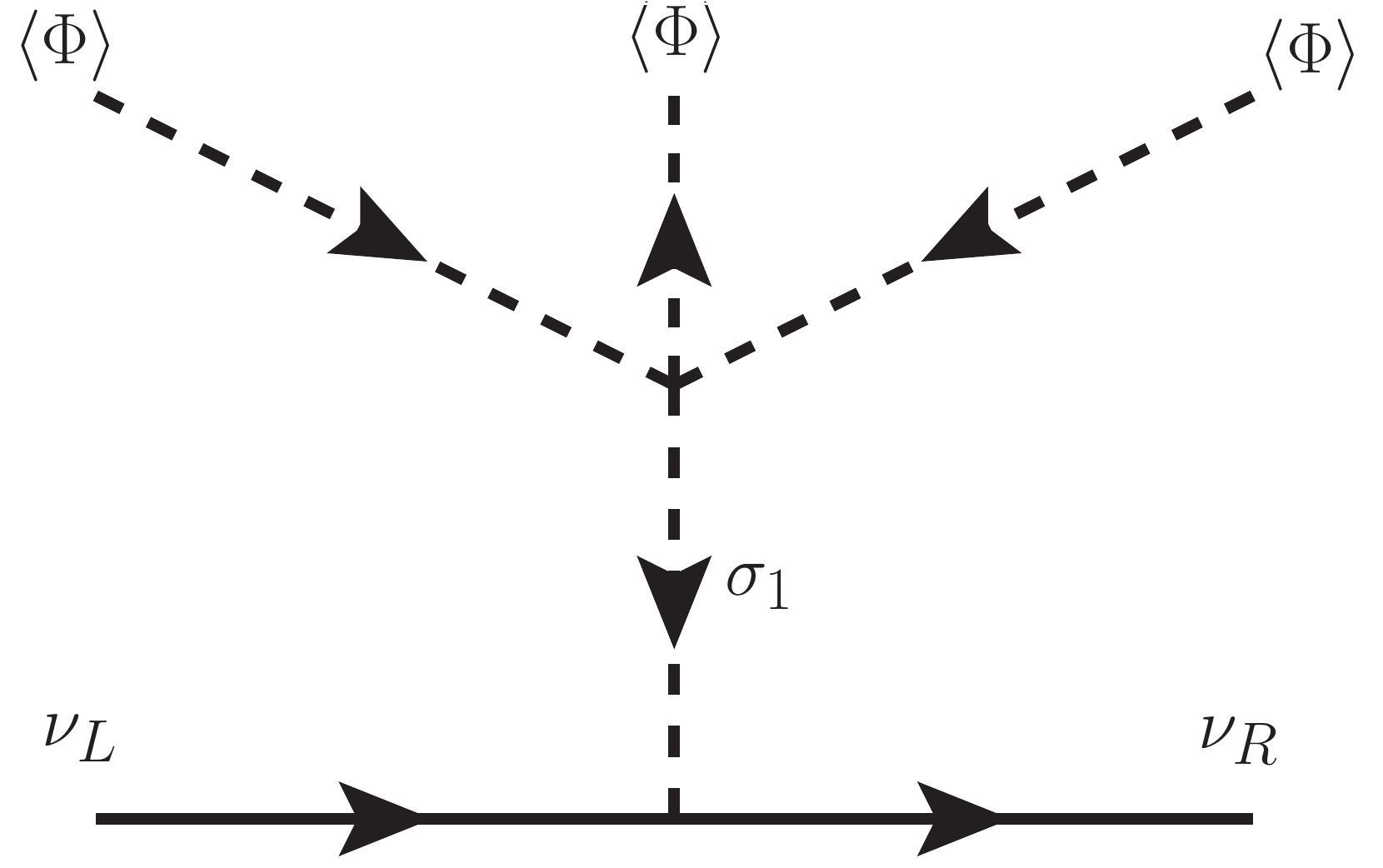}
  \caption{Feynman diagram representing the new possible UV completion
    belonging to topology $T_4$.}
    \label{DT2new}
 \end{figure}

Apart from the above diagrams, there are also six new ones which have
no dimension-5 analogues listed in Sec.~\ref{sec:genWein}.
The first of these possibilities arise from the field contraction shown in \eqref{DT2new-op}.   

\begin{equation}
\underbrace{\underbrace{\bar{L} \otimes \nu_R}_2 \otimes \underbrace{\Phi^c \otimes \Phi\otimes \Phi^c}_2}_{\textnormal{Type II as in \ref{DT2new}}}, \hspace{0.5cm}
\label{DT2new-op}
\end{equation}
This particular contraction of the operators leads to a UV-complete
model where the neutrino mass arises from the Feynman diagram shown in
Fig.~\ref{DT2new} involving a single scalar messenger field $\sigma_1$
transforming as  $SU(2)_L$ doublet with $U(1)_Y = 1$.
 The field $\sigma_1$ gets a small induced vev through its quartic
 coupling with the \sm Higgs doublet. 
 This diagram belongs to the $T_4$ topology and the resulting light
 neutrino mass estimate is given in Table \ref{Tab:mass-formula}.

\begin{figure}[!h]
 \centering
  \includegraphics[scale=0.35]{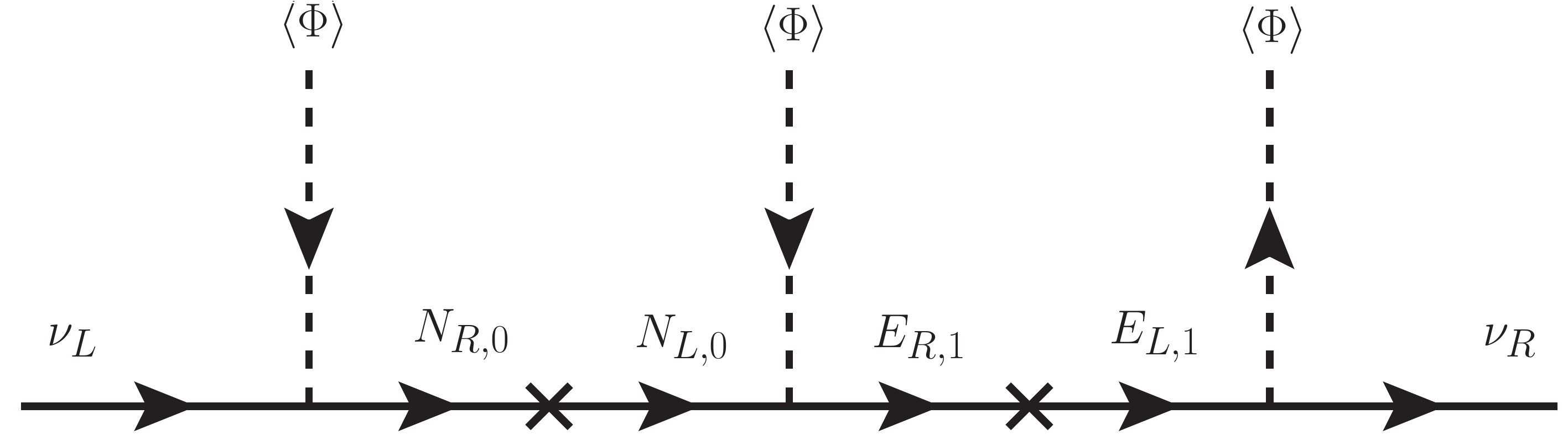}, \hspace{2mm}
  \includegraphics[scale=0.35]{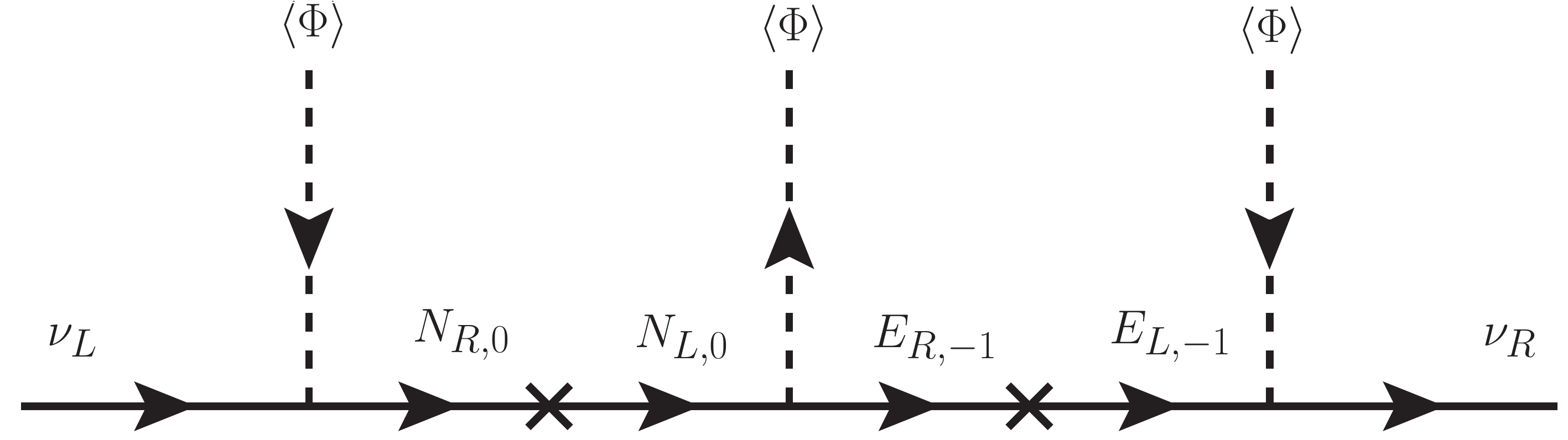}, \\
      \includegraphics[scale=0.35]{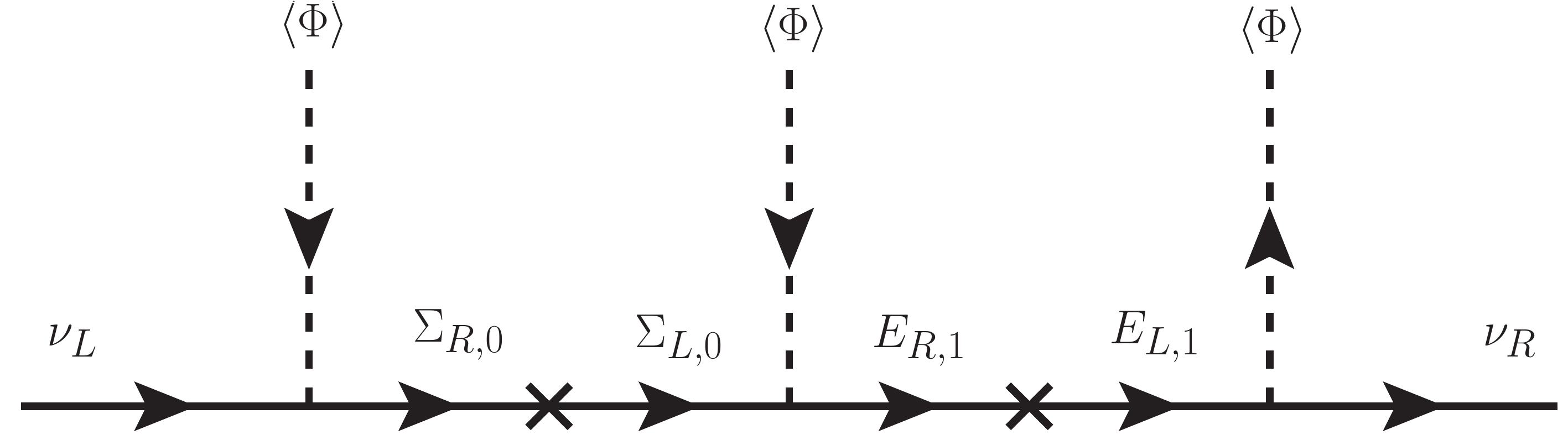}, \hspace{2mm}
      \includegraphics[scale=0.35]{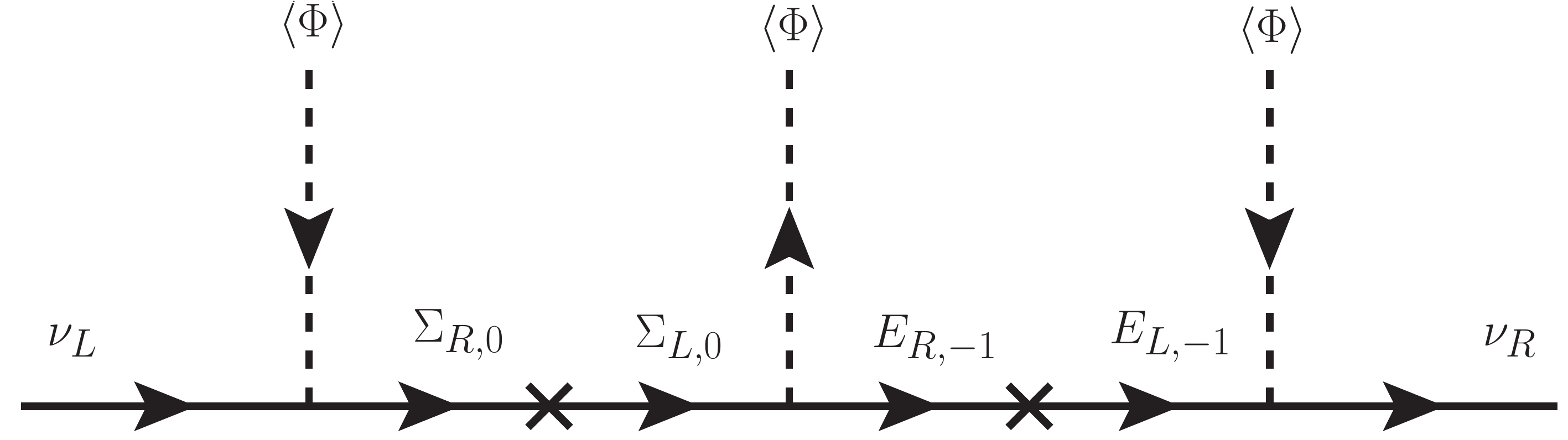}, \hspace{2mm}
      \includegraphics[scale=0.35]{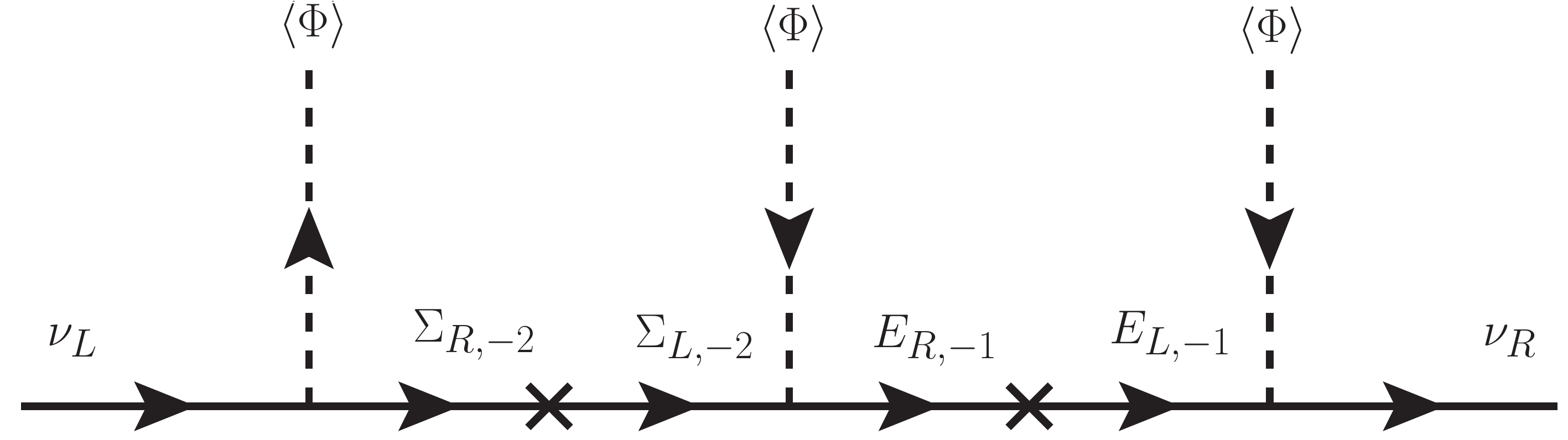}
      \caption{Feynman diagram representing the five realizations of
        the topology $T_5$ diagrams.}
    \label{DN1}
 \end{figure}

 Finally, there are five other field contractions of the dimension-6
 operator, as shown in \eqref{DN1-op} and \eqref{DN1-op2}. 
\begin{equation}
\underbrace{\underbrace{\underbrace{\bar{L} \otimes \Phi^c}_1 \otimes \underbrace{\Phi^c}_2}_{2} \otimes \underbrace{\Phi \otimes \nu_R}_{2}}_{\textnormal{Double Dirac seesaw \ref{DN1}}}, \hspace{0.5cm}
\underbrace{\underbrace{\underbrace{\bar{L} \otimes \Phi^c}_1 \otimes \underbrace{\Phi}_2}_{2} \otimes \underbrace{\Phi^c \otimes \nu_R}_{2}}_{\textnormal{Double Dirac seesaw \ref{DN1}}}
\label{DN1-op}
\end{equation}
\begin{equation}
\underbrace{\underbrace{\underbrace{\bar{L} \otimes \Phi^c}_3 \otimes \underbrace{\Phi^c}_2}_{2} \otimes \underbrace{\Phi \otimes \nu_R}_{2}}_{\textnormal{Double Dirac seesaw \ref{DN1}}}, \hspace{0.5cm}
\underbrace{\underbrace{\underbrace{\bar{L} \otimes \Phi^c}_3 \otimes \underbrace{\Phi}_2}_{2} \otimes \underbrace{\Phi^c \otimes \nu_R}_{2}}_{\textnormal{Double Dirac seesaw \ref{DN1}}}, \hspace{0.5cm}
\underbrace{\underbrace{\underbrace{\bar{L} \otimes \Phi}_3 \otimes \underbrace{\Phi^c}_2}_{2} \otimes \underbrace{\Phi^c \otimes \nu_R}_{2}}_{\textnormal{Double Dirac seesaw \ref{DN1}}}
\label{DN1-op2}
\end{equation}
Notice that the UV-completions of these five field contractions lead
to the neutrino mass generation through topology $T_5$~\footnote{This is a 'double' Dirac seesaw: neutrino masses are doubly suppressed by the two fermion propagators.}, as shown diagrammatically in
Fig.~\ref{DN1}.

All these UV complete models only involve fermionic messengers. 
One sees that two different types of fermionic messengers are
needed. 
In the first and second diagrams the $N_0$ is a vector-like gauge singlet
fermion, whereas the vector fermions $E_1$ or $E_{-1}$ are $SU(2)_L$
doublets. 
In the first diagram the field $E_1$ carries a hypercharge $U(1)_Y = 1$
while in the second diagram $E_{-1}$ carries a hypercharge $U(1)_Y = -1$.
The last three diagrams in Fig.~\ref{DN1} also involve two types of
fermionic messengers the vector-like fermions $E_{1}$ or $E_{-1}$ are
$SU(2)_L$ doublets, while the vector-like fermions $\Sigma_ i$ transform
as triplet under the $SU(2)_L$ symmetry. 
In the third diagram $\Sigma_0$ carries no hypercharge while $E_{1}$ has
$U(1)_Y = 1$. In the fourth diagram $\Sigma_0$ carries no hypercharge,
but $E_{-1}$ has $U(1)_Y = -1$. 
In the fifth diagram $\Sigma_{-2}$ has hypercharge $U(1)_Y = -2$,
while $E_{-1}$ again has $U(1)_Y = -1$.
The light neutrino mass expected for all these diagrams is the same as
that given in Table \ref{Tab:mass-formula} for $T_5$ topology.

 \subsection{Dirac inverse seesaws}
 \label{sec:inverseseesaw}

 The inverse seesaw is a popular approach for the generation of
neutrino masses with the mediator masses potentially close to
the electroweak scale. It is characterized by the presence of a small
mass parameter, generally denoted by $\mu$, which follows the
hierarchy of scales
\begin{equation} \label{eq:hierar}
 \mu \ll v \ll \Lambda \, ,  
\end{equation}
with $v$ the Higgs VEV that sets the electroweak scale and $\Lambda$
the neutrino mass generation scale, determined by the masses of the
seesaw mediators. The $\mu$-parameter suppresses neutrino masses as
$m_\nu \propto \mu$, allowing one to reproduce the observed neutrino
masses and mixing angles with large Yukawa couplings and light seesaw
mediators. This usually leads to a richer phenomenology compared to
the standard high-energy seesaw scenario.

In this section we will explore the various types of inverse seesaws
possible for Dirac neutrinos. We first define the criterion to determine which models
can be classified as belonging to the inverse seesaw mechanism:
\begin{enumerate}
\item \textbf{Presence of a Small Symmetry Breaking Parameter :} The
  first and foremost condition for a model to be classified as an
  inverse seesaw model is the requirement of a ``small'' symmetry
  breaking ``$\mu$-parameter''.  The $\mu$-parameter has to be such
  that the limit $\mu \to 0$ enhances the symmetry of the
  Lagrangian. This crucial feature implies that in the absence of the
  $\mu$-parameter, the model would have a conserved symmetry group
  $\mathcal{G}$, which gets broken by $\mu \ne 0$ as
\begin{equation}
\mathcal{G} \, \xrightarrow{\hspace*{0.4cm} \mu \hspace*{0.4cm}} \, \mathcal{G}^\prime  \, .
\label{eq:mu-term}
\end{equation}
Here $\mathcal{G}^\prime \supset \mathcal{G}$ is a residual
symmetry.\footnote{It can happen that the $\mu$-parameter completely
  breaks the symmetry group $\mathcal{G}$. In such a case
  $\mathcal{G}' \equiv \mathcal{I}$ i.e. the trivial Identity Group.}
Therefore, the limit $\mu \to 0$ enhances the symmetry of the model,
making it natural in the sense of 't Hooft~\cite{tHooft:1979rat} and
protecting the small value of $\mu$ from quantum corrections. Note
that here smallness\footnote{We leave the ``How small should be
  considered small?'' question to the model creator's taste.} of the
$\mu$-parameter is with respect to other parameters in the model under
consideration, for example the electro-weak scale.

\item \textbf{$\boldsymbol{\mu}$-parameter from Explicit/Spontaneous
  Symmetry Breaking:} The $\mu$-parameter can either be an explicit
  symmetry breaking term or a spontaneously induced symmetry breaking
  term. However, to classify as a genuine inverse seesaw, the $\mu$-parameter 
  should be a ``soft term''. In particular, this means that
  if the $\mu$-parameter is an explicit symmetry breaking term, then
  it should have a positive power mass dimension.

\item \textbf{Neutrino Mass Dependence on
  $\boldsymbol{\mu}$-parameter:} The neutrino mass at leading order
  must be directly proportional to the $\mu$-parameter.

\item \textbf{Extended Fermionic Sector:} A genuine inverse seesaw
  model should always have an extended fermionic sector directly
  participating in the neutrino mass mechanism. This means fermions
  beyond the fermionic content of the SM should be involved in
  neutrino mass generation.

\item \textbf{The $\boldsymbol{\mu}$-parameter need not be unique:} In
  cases where there are different $\mu_i$-parameters, all should be soft
  and in the limit of $\mu_i \to 0\, \, \forall \, \, i $, the symmetry
  of Lagrangian should be enhanced. Also, at least one $\mu_i$-parameter
  should be directly involved in the neutrino mass generation
  mechanism. Furthermore, all the $\mu_i$-parameters directly involved in
  neutrino mass generation should satisfy all the other conditions
  listed above.

\end{enumerate}

An example of a model which satisfies all these features is the
canonical Majorana inverse seesaw model~\cite{Mohapatra:1986bd}. The
SM field inventory is extended to include a Vector Like (VL) fermion
transforming as a singlet under the gauge group. The explicit Majorana
mass term (a soft term) for this new fermion will break lepton number
in two units explicitly and thus its smallness is protected by a
symmetry. In this notation, $\mathcal{G} = \rm{U(1)_L}$ while
$\mathcal{G}^\prime = \mathbb{Z}_2$ and $\mu$ is the explicit
Majorana mass term. However, here we will concentrate in the Dirac variants.

Let us emphasize again that the $\mu$-parameter can be explicitly
introduced in the Lagrangian, as a symmetry-breaking mass term, or
spontaneously generated by the VEV of a scalar. In the rest of the
section we will concentrate on the latter case. This is particularly
convenient for our discussion, since the identification of the broken
symmetry becomes more transparent. Furthermore, the smallness of the
$\mu$-parameter can be more easily justified in extended models that
generate it spontaneously. We note, however, that scenarios with an
explicit $\mu$-term would lead to analogous conclusions, just
replacing a VEV by a bare mass term.~\footnote{Note that this analogy
  is only true if the $\mu$-term does not break any of the SM gauge
  symmetries. Otherwise, explicit violation is forbidden while
  scenarios with spontaneous violation are in principle allowed,
  provided $\mu \ll v$, as generally assumed in the inverse seesaw
  setup. Therefore, the spontaneously broken scenario is in a sense
  more general than the explicitly broken one. Of course, electric
  charge and color should remain as conserved charges in either case.}
  
  While these remarks also apply to Majorana generalizations of the original model~\cite{Mohapatra:1986bd}, see \cite{CentellesChulia:2020dfh} for a complete discussion, here we will focus on the Dirac neutrino case. As explained in Sec.~\ref{sec:symmetryrequirement}, we need a `Diracness symmetry' and a `seesaw symmetry' in order to have consistent naturally small Dirac neutrino masses. In the particular case of the inverse seesaw, we also have the ìnverse seesaw symmetries' of Eq.~\ref{eq:mu-term}. These three symmetry requirements need not lead to three different symmetry groups and, in particular, in this Section we will stick to an elegant solution which satisfies all the requirements with a single symmetry group.
  
  We will take $\mathcal{G} = \rm U(1)_{B-L}$ and $\mathcal{G}'$ can be any of its $\mathbb{Z}_n$; $n >
2$ subgroups~\cite{Hirsch:2017col}. In this section we take the
simplest possibility of $\mathcal{G}' = \mathbb{Z}_3$. This $\rm U(1)_{B-L}$ symmetry and
its residual $\mathbb{Z}_3$ subgroup are enough to play all these
roles if the right-handed neutrinos are chiral. Moreover, the $U(1)_{B-L}$ group is anomaly free in the so-called \textit{445 chiral
  solution}~\cite{Montero:2007cd,Ma:2014qra,Ma:2015mjd}. In this case,
the right-handed neutrinos carry $(\nu_{R1},\nu_{R2},\nu_{R3}) \sim (-4, -4, 5)$ charges under $\rm
U(1)_{B-L}$. Being anomaly free, the $\rm U(1)_{B-L}$ symmetry can
also be gauged, leading to a richer phenomenology. While this is not the only
possible symmetry solution~\cite{Bonilla:2018ynb} it is just a particularly elegant one. This chiral $\rm U(1)_{B-L}$ symmetry is playing the roles of the `Diracness symmetry' (forbidding the Majorana mass terms), the `seesaw symmetry' (forbidding the tree level neutrino mass term) and the ìnverse seesaw symmetry' by protecting the smallness of the $\mu$ parameters from higher order corrections. Usually, the minimal versions of these type of models will lead to just two massive neutrinos, since $\nu_{R3} \sim 5$ will decouple from the rest of the particle content. More involved scenarios can give mass to the third neutrino and potentially explain the atmospheric and solar mass splitting scales.

\subsubsection{The simplest Dirac inverse seesaw}
\label{sec:simplestdirac}

\begin{table} [htb!]
\centering
\begin{tabular}{| c || c | c | c |}
  \hline 
&   \hspace{0.5cm} Fields          \hspace{0.5cm}  &     \hspace{0.5cm} $\rm SU(2)_L \otimes U(1)_Y$\hspace{0.5cm} &\hspace{0.5cm} $\rm U(1)_{B-L}$ \hspace{.05cm}$\to$\hspace{.05cm}  $\mathbb{Z}_3$ \hspace{.5cm}  \\
\hline \hline
\multirow{4}{*}{ \begin{turn}{90} Fermions \end{turn} }
&  \rule{0pt}{3ex} $L_i$        	  &   ($\mathbf{2}, {-1/2}$)       &    $-1  \,\, \to \,\,  \omega^2$  \\
&   $\nu_R$      	  &   ($\mathbf{1}, {0}$)          &    $(-4, -4, 5)  \,\, \to \,\,  \omega^2$ \\	
&   $N_L$        	  &   ($\mathbf{1}, {0}$)          &    $-1  \,\, \to \,\,  \omega^2$ \\
&   $N_R$        	  &   ($\mathbf{1}, {0}$)          &    $-1  \,\, \to \,\,  \omega^2$ \rule[-2ex]{0pt}{0pt}    \\	
\hline \hline 
\multirow{4}{*}{ \begin{turn}{90} \hspace{1.55cm}  Scalars \end{turn} }
&  \rule{0pt}{4ex} $H$     &   ($\mathbf{2}, {1/2}$)       &    $0  \,\, \to \,\,  \omega^0$   	\\      
&   $\chi$                 &   ($\mathbf{1}, {0}$)         &    $3  \,\, \to \,\,  \omega^0$  \rule[-3ex]{0pt}{0pt}   	       \\	
    \hline
  \end{tabular}
\caption{Particle content of the minimal model implementing the Dirac
  inverse seesaw.  All quarks transform as $1/3$ ($\omega^1$) under
  $\rm U(1)_{B-L}$ ($\mathbb{Z}_3$), while their \SM \, charges are
  identical to those in the SM. Here $\omega = e^{2\pi i/3}$ is the
  cube root of unity with $\omega^3 = 1$. Moreover, with this choice
  of charges the $\rm U(1)_{B-L}$ symmetry is anomaly free.
 \label{tab:minimaldirac}}
\end{table}

We begin the discussion on Dirac versions of the inverse seesaw
mechanism with a very minimal realization.  In this simple case, the
leading effective operator for neutrino masses is $\bar{L}
H^c \chi \nu_R$, where $\chi$ is a scalar gauge singlet. 
In the full UV complete theory, the particle content and
symmetry transformations are shown in Tab.~\ref{tab:minimaldirac},
while the relevant Lagrangian terms for the generation of neutrino
masses are given by
\begin{equation}
 \mathcal{L}_{\rm Min} = Y \, \bar{L} \widetilde{H} N_R \, + \, \lambda \, \bar{N}_L \chi \nu_R \, + \, M \, \bar{N}_L N_R \, + \hc \, .
\end{equation}
The scalar acquire VEVs  
\begin{equation}
\langle H \rangle = v \quad , \quad \langle \chi \rangle = u \, .
\end{equation}
and break the electroweak and $\rm U(1)_{B-L}$ symmetries,
respectively.  The VEV of $\chi$ induces the small symmetry breaking
$\mu$-term.
\begin{equation}
\mu = \lambda \, u \, .
\end{equation}
Also, note that the VEV of the singlet scalar $\chi$ breaks $\rm
U(1)_{B-L}$ in three units, leaving a residual
$\mathbb{Z}_3$ symmetry under which all scalars transform trivially
while all fermions (except quarks) transform as $\omega^2$, with
$\omega = e^{2 i \pi/3};\, \omega^3 = 1$ being the cube root of
unity. This symmetry forbids all Majorana terms and therefore protects
the Diracness of light neutrinos. The $H$ and $\chi$ VEVs also induce
\textit{Dirac} masses proportional to the $Y$ and $\lambda$ Yukawa
couplings.
These, in the Dirac basis $\left(\bar{\nu}_L \, \bar{N}_L \right)$ and
$\left(\nu_R \, N_R \right)^T$, gives rise to the mass matrix
\begin{equation}
 \mathcal{M}  =  
 \left ( \begin{matrix}
          0          & Y \, v\\
          \mu   & M
          \end{matrix}
 \right ) \, 
 \label{eq:min-dirac-mass}
\end{equation}
where, as mentioned before, $\mu = \lambda\, u$ will be naturally small as
it is the $\rm U(1)_{B-L}$ symmetry breaking term.

\begin{figure}[t!]
\centering
\includegraphics[scale=0.9]{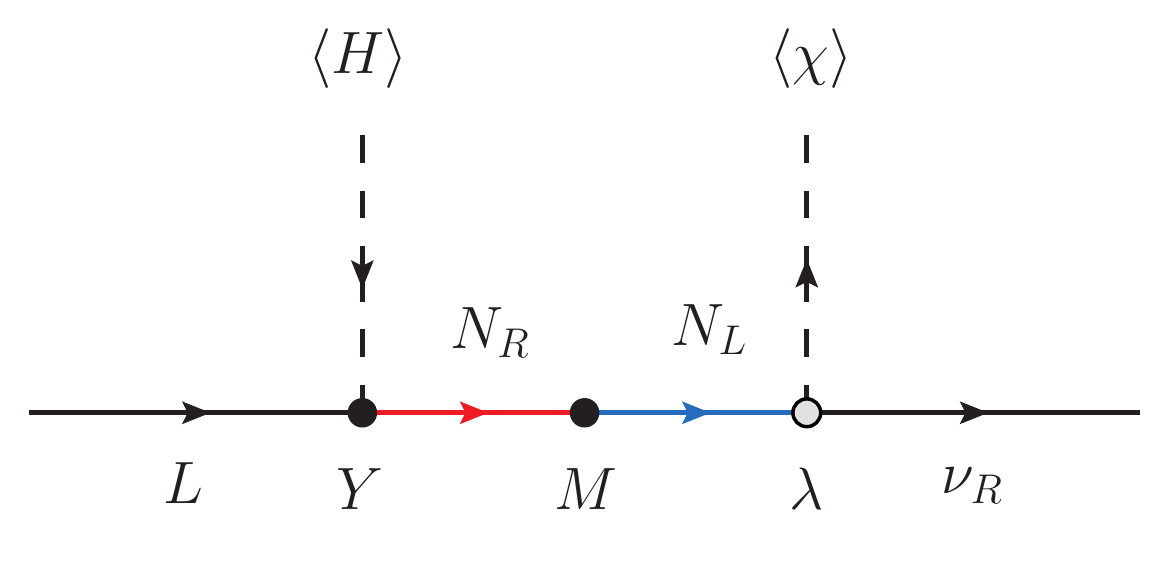}
\caption{Neutrino mass generation in the minimal Dirac inverse
  seesaw.
\label{fig:MinDirac}}
\end{figure}

Thanks to the residual $\mathbb{Z}_3$ symmetry, the neutrinos are Dirac
particles whose masses in the inverse seesaw limit $M \gg Y v \gg
\mu$ are given by
\begin{equation} \label{eq:MinDiracMass}
 m_\nu =   Y  \,v \, \frac{\mu}{M} \, ,
\end{equation}
as diagramatically shown in Fig.~\ref{fig:MinDirac}.

Since in the limit $\mu \to 0$ the $\rm U(1)_{B-L}$ symmetry is restored, the smallness of the $\mu$-term is protected. Its smallness will not be altered by higher order corrections, and therefore is perfectly natural in the 't Hooft sense. 
For example, if we take a small $\mu \sim 10$ eV, we can obtain $m_\nu \sim 0.1$ eV for $Y \sim 0.1$ and $M \sim 1$ TeV. 
We should also point out that the mass matrix in
Eq.~\eqref{eq:min-dirac-mass} is similar to the mass matrix obtained in the Dirac type-I
seesaw~\cite{Ma:2014qra,Ma:2015mjd,CentellesChulia:2016rms,CentellesChulia:2018gwr}. However,
in the type-I seesaw case, the off-diagonal terms in
Eq.~\eqref{eq:min-dirac-mass} are taken to be comparable to each other
i.e. $Y\, v \approx \lambda \, u \ll M$. Thus, for Dirac neutrinos, the
minimal inverse seesaw and the type-I seesaw are just two limits of
the same mass matrix.

Some final comments are in order. First, since the $\rm U(1)_{B-L}$
symmetry is anomaly free, it can be a gauge symmetry. This eliminates
the Goldstone boson associated to its breaking and leads to a much
richer phenomenology.
In contrast, the canonical Majorana inverse seesaw is not anomaly free and hence cannot be
gauged.
Second, the Dirac inverse seesaw model is also relatively simple in
terms of new fields added to the theory. In addition to the \sm \, 
particles, we have just added the right-handed neutrinos $\nu_R$, a
VL fermionic pair $N_L$ and $N_R$ and an extra singlet scalar
$\chi$, which is needed only if the spontaneous symmetry breaking is
desired.

\subsubsection{Generalized Dirac inverse seesaw}

\begin{figure}[t!]
\centering
\includegraphics[scale=0.7]{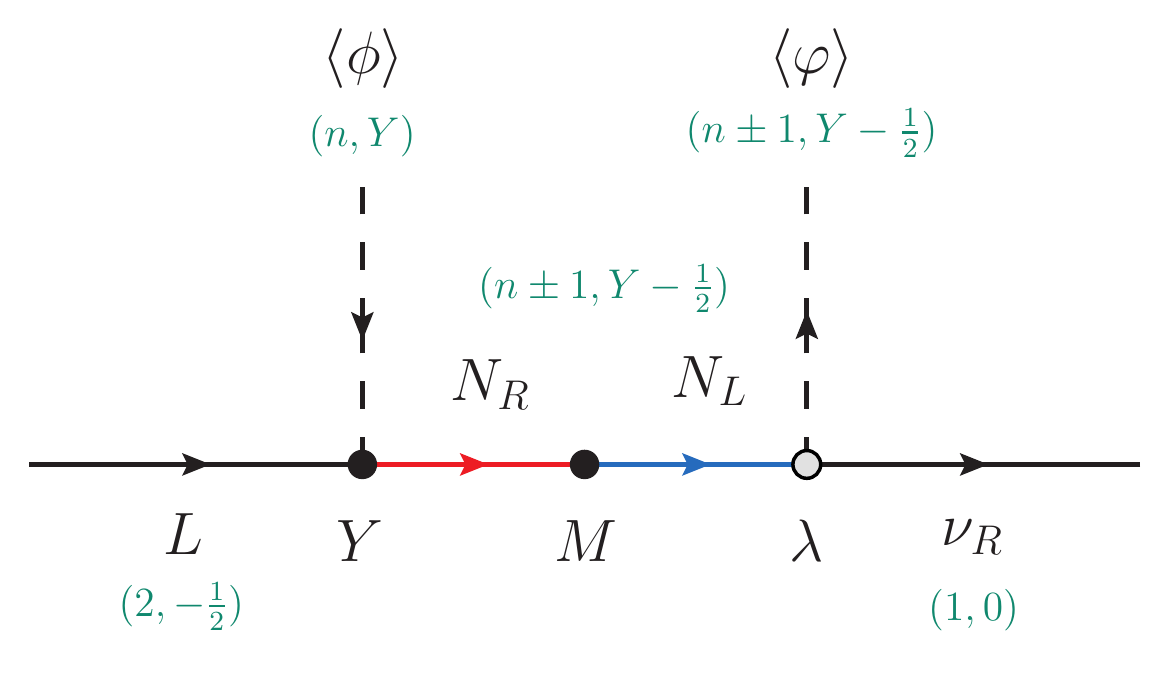}
\caption{Generalized Dirac inverse seesaw.
\label{fig:gendiracmin}}
\end{figure}

We will now generalize the $SU(2) \times U(1)_Y$ multiplicity of the Dirac inverse
seesaw. The Lagrangian of the model would be formally identical
to the one in Sec.~\ref{sec:simplestdirac}, but with higher $\rm
SU(2)_L$ multiplets and replacing $H \rightarrow \phi$ and $\chi \rightarrow \varphi$, where $\phi$ is now a generic scalar which doesn't break $U(1)_{B-L}$, which can be identified with the SM Higgs doublet $H$ in certain special scenarios. On the other hand, $\varphi$ is the scalar whose vev breaks $U(1)_{B-L}$ spontaneously, generating the $\mu$ term. The charges under $\rm U(1)_{B-L}$ will also be
identical and therefore the model would share the same appealing
features. The general $\rm SU(2)_L \times U(1)_Y$ charges can be seen
in Tab.~\ref{tab:generaldiracmin} and diagramatically in
Fig.~\ref{fig:gendiracmin}.

 \begin{table}[h!]
 \centering
\begin{tabular}{| c || c | c | c |}
  \hline 
&   \hspace{0.5cm} Fields          \hspace{0.5cm}  &     \hspace{0.5cm} $\rm SU(2)_L \otimes U(1)_Y$\hspace{0.5cm} &\hspace{0.5cm} $\rm U(1)_{B-L}$ \hspace{.05cm}$\to$\hspace{.05cm}  $\mathbb{Z}_3$ \hspace{.5cm}  \\
\hline \hline
\multirow{4}{*}{ \begin{turn}{90} Fermions \end{turn} }
&  \rule{0pt}{3ex} $L_i$        	  &   ($\mathbf{2}, {-1/2}$)       &    $-1  \,\, \to \,\,  \omega^2$  \\
&   $\nu_R$      	  &   ($\mathbf{1}, {0}$)          &    $(-4, -4, 5)  \,\, \to \,\,  \omega^2$ \\	
&   $N_L$             &   ($\mathbf{n \pm 1}, Y-1/2$)   &   $-1 \,\, \to \,\, \omega^2$     \\
&   $N_R$             &   ($\mathbf{n \pm  1}, Y-1/2$)  &   $-1 \,\, \to \,\, \omega^2$        \rule[-2ex]{0pt}{0pt}    \\	
\hline \hline 
\multirow{4}{*}{ \begin{turn}{90} \hspace{1.55cm}  Scalars \end{turn} }
&  \rule{0pt}{4ex} $H$  &  ($\mathbf{2}, {1/2}$)            &    $0 \,\, \to \,\, \omega^0 $      \\
&   $\varphi$        &  ($\mathbf{n \pm 1}, Y-1/2$)      &    $3 \,\, \to \,\, \omega^0 $        \\
&   $\phi$           &  ($\mathbf{n}, Y$)                &    $0 \,\, \to \,\, \omega^0 $  \rule[-2ex]{0pt}{0pt}   	       \\	
    \hline
  \end{tabular}
\caption{\label{tab:generaldiracmin} Particle content of the generalized Dirac inverse
  seesaw. As in the Majorana case, here also $\phi$ can be identified with the SM Higgs
  under certain conditions. Again, $\varphi$ breaks the symmetry and
  generates the $\mu$-term. The particle charges under the residual $\mathbb{Z}_3$ symmetry are given by cube roots of unity with $\omega = e^{2\pi I/3}; \, \omega^3 = 1$.}
\end{table}

Let us list a few viable examples in
Tab.~\ref{tab:diracminexamples}. The general strategy to build these
models is as follows. We first fix the $\rm SU(2)_L
\otimes U(1)_Y$ charge of the scalar $\phi$. Depending on the $\phi$
charges, the $N_L$ and $N_R$ fields can have only one option for their
$\rm SU(2)_L \otimes U(1)_Y$ transformations. Finally, given the
charges of both $\phi$ and the $N_L,N_R$ fields, the viable options
for the $\varphi$ gauge charges can be obtained. The final list of
viable models for the Dirac inverse seesaw with generalized multiplets
are listed in Tab.~\ref{tab:diracminexamples}.

\begin{table}[h!]
\hspace{-0.9cm}
\begin{tabular}{| c | c | c  | c | }
  \hline 
Name of the model    &    $N_L$ and $N_R$              &      $\phi$             &   $\varphi$ ($\mu$-term)       \\
\hline \hline
Standard Dirac Inverse seesaw            &     $(\mathbf{1}, 0)$     &   $(\mathbf{2}, 1/2)$  &   $(\mathbf{1}, 0)$ \\
\hline \hline
Type-III Dirac Inverse seesaw            &     $(\mathbf{3}, 0)$     &   $(\mathbf{2}, 1/2)$  &   $(\mathbf{3}, 0)$ \\
\hline
Type-III Dirac Inverse seesaw variant I  &     $(\mathbf{3}, -1)$    &   $(\mathbf{2}, -1/2)$ &   $(\mathbf{3}, -1)$ \\
\hline 
Type-III Dirac Inverse seesaw variant II &     $(\mathbf{3}, 1)$     &   $(\mathbf{4}, 3/2)$  &   $(\mathbf{3}, 1)$ \\
\hline \hline
Exotic or Type IV Dirac inverse seesaw   &     $(\mathbf{4}, Y-1/2)$ &   $(\mathbf{3}, Y = 0, \pm 1)$ &   $(\mathbf{4}, Y-1/2)$ \\
\hline \hline
Type-V Dirac Inverse seesaw              &     $(\mathbf{5}, 0)$     &   $(\mathbf{4}, 1/2)$  &    $(\mathbf{5}, 0)$  \\
\hline
Type-V Dirac Inverse seesaw variant I    &     $(\mathbf{5}, 1)$     &   $(\mathbf{4}, 3/2)$  &   $(\mathbf{5}, 1)$  \\
\hline
Type-V Dirac Inverse seesaw variant II   &     $(\mathbf{5}, -1)$    &   $(\mathbf{4}, -1/2)$ &   $(\mathbf{5}, -1)$ \\
\hline
Type-V Dirac Inverse seesaw variant III  &     $(\mathbf{5}, -2)$    &   $(\mathbf{4}, -3/2)$ &   $(\mathbf{5}, -2)$ \\
\hline 
\hline

 \end{tabular}
 \caption{ A few examples of the generalized Dirac inverse seesaw.
 \label{tab:diracminexamples}}  
\end{table}

In Tab.~\ref{tab:diracminexamples} we have restricted outselves to $n=4$ i.e. upto the case in which $\phi$ transforms as a quadruplet under $\rm SU(2)_L$. Nevertheless, the generalization to higher $n > 4$ is rather straightforward.
We have named the ``type'' of the model based on the $\rm SU(2)_L$ transformation of the $N_L, N_R$ fermions. The subclass numbering is based on the $\rm SU(2)_L$ transformation of $\phi$ as well as the $\rm SU(2)_L \otimes U(1)_Y$ transformation of the $\varphi$ field.

\subsubsection{Dirac ``Double'' Inverse Seesaw}
\label{subsec:dirac-double}

To build the double Dirac inverse seesaw, we start with the field and
symmetry inventory of the minimal Dirac inverse seesaw of
Section~\ref{sec:simplestdirac}. To it, we add new VL fermions,
$S_{L,R}$, and a new singlet scalar $\chi_2$. The only modification in
the $\rm B-L$ charges is that we will take $(\chi_1, \chi_2)$ to
transform as $(\chi_1, \chi_2) \sim (6, -9)$. Remember that in the
previous example we had $\chi \sim 3$. Moreover, we take the new
VL fermion to transform as $S_{L, R} \sim 5$, while the rest
of the fields share their transformation properties with the previous
model as shown in Tab.~\ref{tab:diracmodel1}.

With the above choice of $\rm U(1)_{B-L}$ charges it is easy to check
that the model is anomaly free and can be gauged if desired. Also, the
$\rm B-L$ charges of the scalars are chosen in such a way that their
VEVs break $\rm U(1)_{B-L} \to \mathbb{Z}_3$. This residual
$\mathbb{Z}_3$ symmetry remains unbroken thus ensuring the Dirac
nature of neutrinos.
Note that the fields $\nu_{R, 3}$ and $S_R$ share the same transformation
properties and are therefore indistinguishable, but we call them differently to follow the notational
conventions of the previous and following sections.

\begin{table}[h!]
\centering
\begin{tabular}{| c || c | c | c | }
  \hline 
& Fields   &   $\rm SU(2)_L \otimes U(1)_Y$   
&\hspace{.05cm}$\rm U(1)_{B-L}$\hspace{.05cm}$\to$\hspace{.05cm}$\mathbb{Z}_3$\hspace{.05cm}  \\
\hline \hline
\multirow{4}{*}{ \begin{turn}{90} \hspace{-1.25cm} Fermions \end{turn} }
&  \rule{0pt}{3ex} $L_i$        	  &   ($\mathbf{2}, {-1/2}$)       &    $-1 \,\, \to \,\, \omega^2$ \\
&   $\nu_R$        	  &   ($\mathbf{1}, {0}$)          &$(-4, -4, 5) \,\, \to \,\, \omega^2$   \\	
&   $N_L$        	  &   ($\mathbf{1}, {0}$)          &    $-1 \,\, \to \,\, \omega^2$ \\
&   $N_R$        	  &   ($\mathbf{1}, {0}$)          &    $-1 \,\, \to \,\, \omega^2$   \\	
&   $S_L$        	  &   ($\mathbf{1}, {0}$)          &    $5 \,\, \to \,\, \omega^2$ \\
&   $S_R$        	  &   ($\mathbf{1}, {0}$)          &    $5 \,\, \to \,\, \omega^2$  \rule[-2ex]{0pt}{0pt}  \\	
\hline \hline
\multirow{5}{*}{ \begin{turn}{90}\hspace{0.7cm} Scalars \end{turn} }
&   \rule{0pt}{3ex} $H$          	  &   ($\mathbf{2}, {1/2}$)       &    $0 \,\, \to \,\, \omega^0$ \\
&   $\chi_1$          &   ($\mathbf{1}, {0}$)         &    $6 \,\, \to \,\, \omega^0$   \\   
&   $\chi_2$          &   ($\mathbf{1}, {0}$)         &    $-9 \,\, \to \,\, \omega^0$\rule[-2ex]{0pt}{0pt} \\
    \hline
  \end{tabular}
\caption{Particle content of the Dirac analogue of the inverse
  seesaw. The $\rm U(1)_{B-L}$ charges of the fermions are fixed by an anomaly cancellation condition while the $\rm U(1)_{B-L}$ charges of the scalars are chosen such that the residual $\mathbb{Z}_3$ symmetry remains unbroken and the leading contribution to neutrino mass is induced by the double inverse seesaw.
\label{tab:diracmodel1}}
\end{table}

\begin{figure}[t!]
\centering
\includegraphics[scale=0.9]{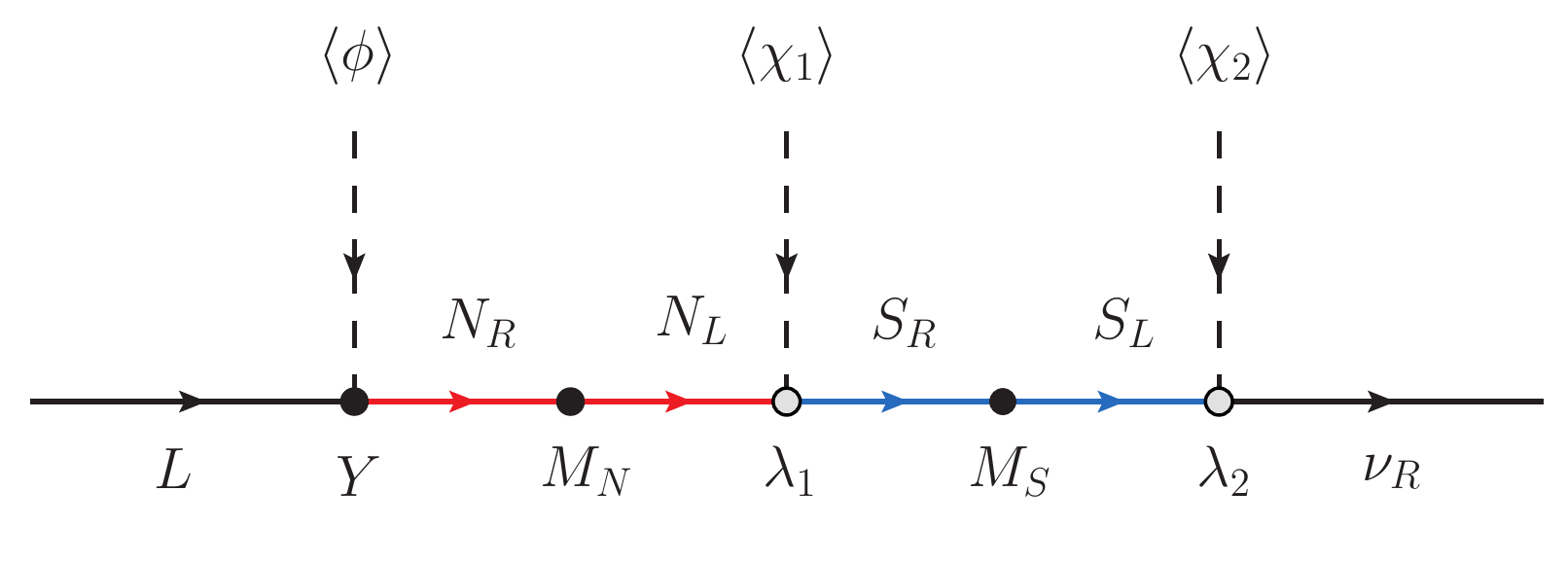}
\caption{Neutrino mass generation in the Dirac double inverse
  seesaw. The $\mu$-terms can be either explicit or spontaneously
  generated by the VEV of the scalars $\chi_1$ and $\chi_2$.
\label{fig:Dirac}}
\end{figure}

With all these ingredients, the Lagrangian of the model relevant to
neutrino mass generation is given by
\begin{align}
\label{eq:diracgenerallag}
\mathcal{L}_{\rm Dir} \, = & \, Y \, \bar{L} \, \widetilde H \, N_R \, + \, \lambda_2 \, \bar{S}_L \, \chi_2^* \, \nu_R \, + \, \lambda_1 \, \bar{N}_L \,  \chi_1^* \, S_R  + \lambda^\prime_1 \, \bar{S}_L \, \chi_1 \, N_R \, \\
& \, + \, M_N \, \bar{N}_L N_R \, + \, M_S \, \bar{S}_L S_R \, + \hc \, . 
\end{align}
Symmetry breaking is triggered by the scalar VEVs
\begin{equation}
\langle H \rangle = v \quad , \quad \langle \chi_1 \rangle = u_1 \quad , \quad \langle \chi_2 \rangle = u_2 \, ,
\end{equation}
which lead to the following $\mu$-terms,
\begin{equation}
\mu_1 = \lambda_1 \, u_1 \quad , \quad \mu^\prime_1 = \lambda^\prime_1 \, u_1 \, \quad , \quad \mu_2 = \lambda_2 \, u_2 \, .
\end{equation}
After symmetry breaking, Eq.~\eqref{eq:diracgenerallag} leads to the
mass Lagrangian in matrix form
\begin{equation}
 \mathcal{L}_{m} = \left ( \begin{matrix}
          \bar{\nu}_L & \bar{N}_L & \bar{S}_L
         \end{matrix}
 \right )
 \left ( \begin{matrix}
          0               & Y_N \, v     & 0 \\
          0               & M_N          & \mu_1 \\
          \mu_2      & \mu^\prime_1        & M_S          
          \end{matrix}
 \right )
 \left ( \begin{matrix}
          \nu_R \\
          N_R \\
          S_R
         \end{matrix}
 \right ) \, .
 \label{eq:double-dirac-lag}
\end{equation}
Again, the natural hierarchy among the parameters of the model is
\begin{equation}
\mu_1 , \mu^\prime_1, \mu_2 \ll Y \, v \ll M_N , M_S \, ,
\end{equation}
which leads to the light neutrino mass matrix
\begin{eqnarray}
 m_\nu &=&  \left ( \begin{matrix}
          Y \, v & 0
         \end{matrix}
 \right )
 \left ( \begin{matrix}

          M_N      & \mu_1\\
          \mu_1^\prime    & M_S
          \end{matrix}
 \right )^{-1}
 \left ( \begin{matrix}
          0 \\
          \mu_2
         \end{matrix}
 \right ) \, .
\end{eqnarray}
For one generation of light neutrinos, this is equivalent to
\begin{equation} \label{eq:2-DiracMass}
 m_\nu = Y_N \, \frac{v \, \mu_1 \, \mu_2}{\mu_1 \mu^\prime_1 - M_N M_S} \simeq - Y_N  \, v \, \frac{\mu_1 \mu_2}{M_N \, M_S} \, .
\end{equation}
This result is illustrated in the Feynman diagram shown in
Fig.~\ref{fig:Dirac}.

As it is clear from Eq.~\eqref{eq:2-DiracMass}, neutrino masses in
this case are suppressed by two $\mu$-terms and hence the name Dirac
\textit{double} inverse seesaw.
Note that the effect of $\mu^\prime$ is subleading in the neutrino mass
generation. 
Finally, we should remark that the spontaneous breaking of $\rm U(1)_{B-L}$ by the $u_1$ and $u_2$ VEVs leaves a residual $\mathbb{Z}_{3}$ symmetry.  As in the minimal Dirac model of Section~\ref{sec:simplestdirac}, here too all scalars transform trivially under this residual symmetry, while all fermions (except quarks which transform as $\omega$)  transform as
$\omega^2$. Again, this symmetry forbids all Majorana mass terms and
protects the Diracness of light neutrinos.

We finally point out that in order to induce a triple Dirac inverse
seesaw and beyond one would just need to sequentially add a new
VL fermion alongside a new scalar with a judicious choice of
symmetry breaking charges for the scalars. 
To obtain an $n$th order Dirac inverse seesaw as the leading contribution we need to add $n$ VL fermionic pairs and $n$ scalars. Of course appropriate symmetries, with particles carrying appropriate charges  under them, are required to ensure that neutrinos are Dirac particles with the leading contribution to their mass given by the $n$th order Dirac inverse seesaw as
\begin{eqnarray}
 m_\nu & \approx & Y \, v \, \prod_{i=1}^{i = n} \, \frac{\mu_i}{M_i} \, ,
\end{eqnarray}

A combination of both generalization methods, i.e. higher order multiplets of $SU(2)_L$ in double or more Dirac inverse seesaws is straightforward and will not be pursued here.

 \section{Radiative masses and the `scoto' connection}
 \label{sec:radanddark}
 
 At present, a plethora of cosmic observations all indicate that the bulk of matter in the Universe is in the form of dark matter, a hitherto unknown form of matter which interacts gravitationally, but has little or no electromagnetic interaction \cite{Aghanim:2018eyx}. 
These observations point to one of the most serious shortcomings of the Standard Model (SM) since in the SM there is no viable candidate for dark matter.
Thus, they inarguably point to the presence of new physics beyond the SM and it is a topic of active theoretical and experimental research.

To explain dark matter, the particle content of the SM needs to be extended. Furthermore, to account for dark matter stability new explicit \cite{Silveira:1985rk, Ma:2006km} 
or accidental symmetries \cite{Cirelli:2005uq} beyond those of the SM are also invoked.

There are particularly attractive scenarios that connect dark matter to neutrino physics in an intimate manner. The scotogenic model is one such model where the `dark sector' participates in the loop responsible for neutrino mass generation \cite{Ma:2006km}. This work has been followed up in multiple ocasions over the last two decades, analyzing the phenomenology of the scotogenic model and variants of it in multiple dimensions as well as connecting the neutrino mass generation and dark matter with other new physics scenarios, both for Majorana and Dirac neutrinos \cite{Ma:2012ez,Bouchand:2012dx,Farzan:2012sa,Brdar:2013iea,Hirsch:2013ola,Toma:2013zsa,Ma:2013yga,Fraser:2014yha,Vicente:2014wga,Ma:2014eka,Merle:2015gea,Ma:2015xla,Rocha-Moran:2016enp,Merle:2016scw,Ferreira:2016sbb,Ahriche:2016cio,Merle:2016scw,Yu:2016lof,Borah:2016zbd,Borah:2017dfn,Wang:2017mcy,Bonilla:2018ynb,Rojas:2018wym,Abada:2018zra,Hugle:2018qbw,Han:2018zcn,Reig:2018mdk,Hagedorn:2018spx,Leite:2019grf,Dasgupta:2019rmf,Ma:2019byo,Ma:2019yfo,Kumar:2019tat,Chen:2019nud,Ma:2019coj,CentellesChulia:2019gic,Okada:2019xqk,Baumholzer:2019twf,Babu:2019mfe,Pramanick:2019oxb,Rojas:2019llr,Avila:2019hhv,Nomura:2019lnr,Barreiros:2020gxu,CarcamoHernandez:2020ehn,Ma:2020sca,Escribano:2020iqq,Wong:2020obo,Leite:2020wjl,Jana:2020joi,Guo:2020qin,Borah:2020wut,Fujiwara:2020unw,Ahriche:2020pwq,Leite:2020bnb,Behera:2020lpd,Ma:2020lnm,Bernal:2021ezl,Kitabayashi:2021hox,Escribano:2021ymx,DeRomeri:2021yjo,deBoer:2021pon,Kang:2021jmi, Mandal:2021yph}.

 Recently, a relation between the Dirac nature of neutrinos and dark matter stability has also been proposed \cite{CentellesChulia:2016rms}. 
Furthermore, it has been shown that this relation is independent of the neutrino mass generation mechanism \cite{CentellesChulia:2018gwr}. 
It utilizes the SM lepton number $U(1)_L$ symmetry\footnote{One can equivalently use the anomaly free $U(1)_{B-L}$ symmetry.} or its appropriate $\mathcal{Z}_n$ subgroup, to forbid Majorana mass terms of neutrinos as well as to stabilize dark matter \cite{CentellesChulia:2016rms}.
In this approach, the Dirac nature of neutrinos and the stability of dark matter are intimately connected, having their origins in the same lepton number symmetry.

In this section we aim to combine and generalize these two approaches and develop a general formalism where the following conditions are satisfied:

\begin{enumerate}
\item[ I.] Neutrinos are Dirac in nature.
\item[ II.]   Naturally small neutrino masses are generated through finite loops, forbidding the tree-level neutrino Yukawa couplings.
\item[ III.] The dark sector participates in the loop. The lightest particle being stable is a good dark matter candidate.\\
\end{enumerate}
\vspace{-1cm}
Usually one needs at least three different symmetries besides those within the Standard Model to achieve this \cite{Bonilla:2016diq}.
However, we show that all of these requirements can be satisfied without adding any extra explicit or accidental symmetries. In our formalism we employ an anomaly free chiral realization of the  $U(1)_{B-L}$ spontaneously broken to a residual $\mathcal{Z}_n$ symmetry and show that just the $U(1)_{B-L}$ already present in the SM is sufficient.

Before going into the details of the formalism, let us briefly discuss the possibility of chiral solutions of a gauge $U(1)_{B-L}$ and the anomaly cancellation conditions.
It is well known that the accidental $U(1)_B$ and $U(1)_L$ symmetries of the SM are anomalous, but the $U(1)_{B-L}$ combination can be made anomaly free by adding three right-handed neutrinos $\nu_{R_i}$ with $(-1,-1,-1)$ vector charges under $U(1)_{B-L}$. 
However, chiral solutions to $U(1)_{B-L}$ anomaly cancellation conditions are also possible. The particularly attractive feature of chiral solutions is that by using them one can automatically satisfy conditions I and II, as shown in \cite{Ma:2014qra,Ma:2015mjd}, using the chiral solution $\nu_{R_i} \sim (-4,-4,5)$ under $U(1)_{B-L}$ symmetry.

\begin{figure}[th]
    \begin{subfigure}[b]{0.43\textwidth}
        \includegraphics[width=\textwidth]{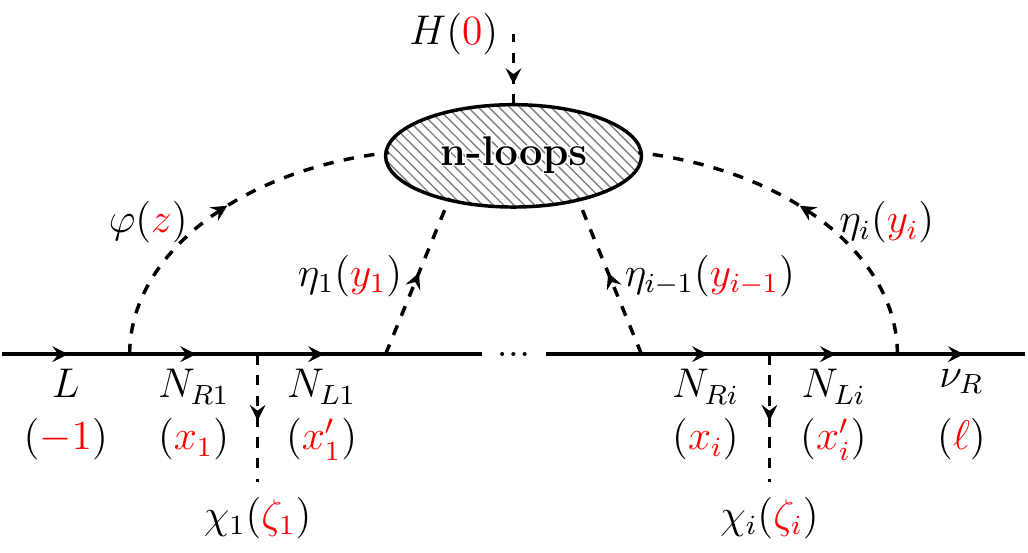}
        \caption{General $U(1)_{B-L}$ charge assignment.}
        \label{fig:genU1}
    \end{subfigure}
    ~ 
    \begin{subfigure}[t]{0.1\textwidth}
    \vspace*{-4.15cm}
    \hspace*{-0.45cm}
   \includegraphics[width=1.5\textwidth]{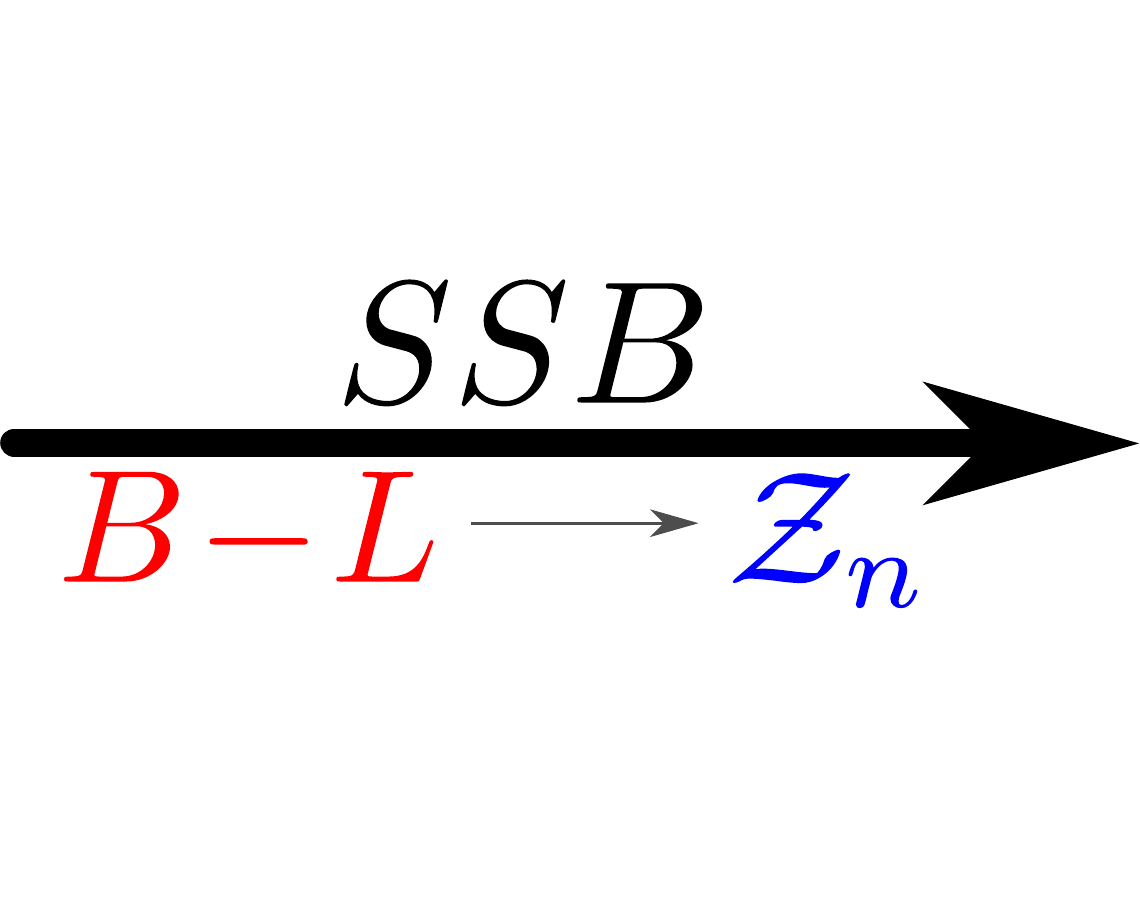}      \end{subfigure}
    ~ 
   \begin{subfigure}[b]{0.43\textwidth}
        \includegraphics[width=\textwidth]{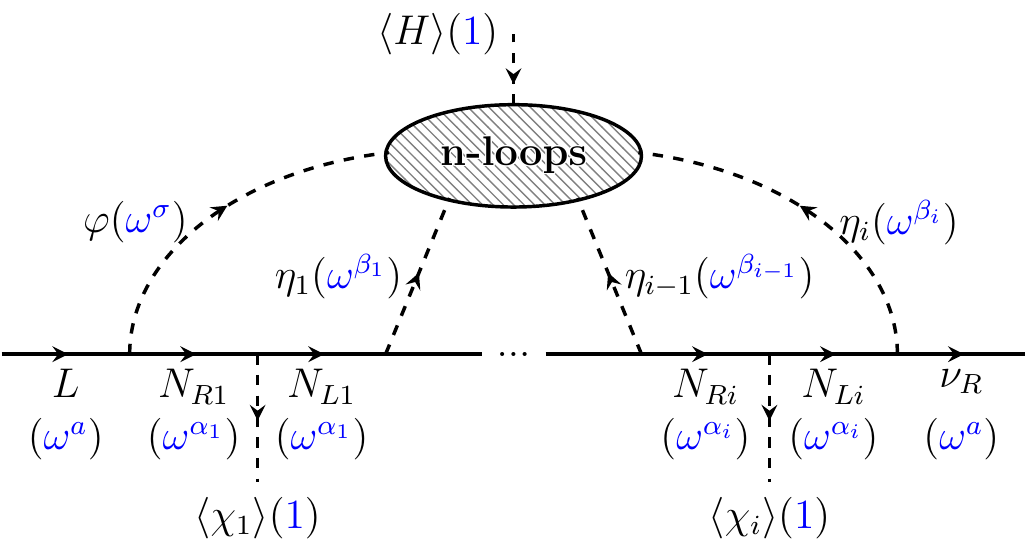}
        \caption{General residual $\mathcal{Z}_n$ charge assignment.}
        \label{fig:genZn}
    \end{subfigure} \\
    \caption{General charge assignment for any topology and its spontaneous symmetry breaking pattern.}\label{fig:gencase}
\end{figure}

Our general strategy is to use the chiral anomaly free solutions of $U(1)_{B-L}$ symmetry to generate loop masses for Dirac neutrinos and also have a stable dark matter particle mediating the aforementioned loop. Then, after symmetry breaking, once all of the scalars get a vacuum expectation value (vev), the $U(1)_{B-L}$ symmetry will be broken down to one of its $\mathcal{Z}_n$ subgroups, such that the dark matter stability and Dirac nature of neutrinos remain protected. This scheme is shown diagrammatically  in Fig.~\ref{fig:gencase}. 

In Fig.~\ref{fig:gencase} the SM singlet fermions $N_{Li}, N_{Ri}$, as well as the right-handed neutrinos $\nu_{R}$, have non-trivial chiral charges under $U(1)_{B-L}$ symmetry.\footnote{It is not necessary that all fermions $N_{Li}, N_{Ri}$ be chiral under $U(1)_{B-L}$ symmetry.} In order to generate the masses of these chiral fermions we have also added SM singlet scalars $\chi_i$ which also carry $U(1)_{B-L}$ charges. To complete the neutrino mass generation loop, additional scalars $\varphi, \eta_i$ are required. After spontaneous symmetry breaking (SSB) of the $U(1)_{B-L}$ symmetry, all of the scalars $\chi_i$ will acquire vevs that break the $U(1)_{B-L} \to \mathcal{Z}_n$ residual symmetry. The fermions $N_{Li}, N_{Ri}$ obtain masses through the vevs of the scalars $\chi_i$, while the neutrinos acquire a naturally small $n$-loop mass as shown in Fig. \ref{fig:gencase}. 

In order to satisfy all of the requirements listed above, several conditions must be applied. First of all, the model should be anomaly-free:
\begin{itemize}[label=\textcolor{orange}{\textbullet}]
\item The chiral charges of the fermions must be taken in such a way that the anomalies are canceled.
\end{itemize}
In order to obtain non-zero but naturally small Dirac neutrino masses we impose the following conditions:
\begin{itemize}[label=\textcolor{blue}{\textbullet}]
\item The tree-level Yukawa coupling $\bar{L} H^c \nu_R$ should be forbidden. This implies that apart from the SM lepton doublets $L_i$ no other fermion can have $U(1)_{B-L}$ charge of $\pm 1$. Furthermore, to ensure that the desired loop diagram gives the dominant contribution to the neutrino masses, all lower loop diagrams should also be forbidden by an appropriate choice of the charges of the fields. 
 \item The operator leading to neutrino mass generation, i.e. $\bar{L} H^c \chi_1 \dots \chi_i \nu_{R}$, should be invariant under the SM gauge symmetries as well as under $U(1)_{B-L}$. Following the charge convention of Fig.~\ref{fig:gencase}, i.e. $L \sim -1$, $\chi_i \sim \zeta_i$ and $\nu_R \sim \ell$, the charges of the vev carrying scalars $\chi_i$ should be such that $\sum_i \zeta_i = -1 - \ell$ .
 \item All of the fermions and scalars running in the neutrino mass loop must be massive. Since the fermions will be in general chiral, this mass can only be generated via the coupling with a vev carrying scalar. For example, in the diagram in Fig.~\ref{fig:gencase} we should have $-x_i + x'_i + \zeta_i = 0$.
\item To protect the Diracness of neutrinos, all of the Majorana mass terms for the neutrino fields at all loops must be forbidden in accordance with Eq.~\eqref{evenzndir}.
\end{itemize}

Additionally, for dark matter stability, we impose the following conditions:

\begin{itemize}
 \item After SSB, the $U(1)_{B-L}$ symmetry is broken down to a $\mathcal{Z}_n$ subgroup. Only even $\mathcal{Z}_n$ subgroups with $n > 2$ can protect dark matter stability. The odd $\mathcal{Z}_n$ subgroups invariably lead to dark matter decay.\footnote{For odd $\mathcal{Z}_n$ subgroups, there will always be an effective dark matter decay operator allowed by the residual odd $\mathcal{Z}_n$ symmetry. Even then it is possible that such an operator cannot be closed within a particular model, thus pinpointing the existence of an accidental symmetry that stabilizes dark matter. Another possibility is that the dark matter candidate decays at a sufficiently slow rate. Thus for residual odd $\mathcal{Z}_n$ symmetries, one can still have either a stable dark matter stabilized by an accidental symmetry or a phenomenologically viable decaying dark matter. In this work we will not explore such possibilities.} The symmetry breaking pattern can be extracted as follows. First all the $U(1)$ charges must be rescaled in such a way that all the charges are integers and the least common multiple (lcm) of all of the rescaled charges is $1$. Defining $n$ as the least common multiple of the charges of the scalars $\chi_i$, it is easy to see that the $U(1)$ will break to a residual $\mathcal{Z}_n$. This $n$ must be taken to be even as explained before, i.e. $n \equiv $ lcm$(\zeta_i) \in 2\mathbb{Z}$. 
 \item Dark sector particles should neither mix with nor decay to SM particles or to vev carrying scalars.
\item There are two viable dark matter scenarios depending on the transformation of the SM fermions under the residual symmetry. 
\begin{itemize}
\item When all SM fields transform as even powers of $\omega$, where $\omega^n = 1$, under the residual $\mathcal{Z}_n$, the lightest particle transforming as an odd power will be automatically stable, irrespective of its fermionic or scalar nature. We will show an explicit example of this simple yet powerful idea later.
\item In the case in which all SM fermions transform as odd powers of the residual subgroup, it can be shown that all of the odd scalars and the even fermions will be stable due to a combination of the residual $\mathcal{Z}_n$ and Lorentz symmetry. 
\end{itemize}
\end{itemize}

Given the long list of requirements, most of the possible solutions that lead to anomaly cancellation fail to satisfy some or most of them. 
Still we have found some simple one-loop and several two-loop solutions that can satisfy all of the conditions listed above.

In what follows we demonstrate the idea for a simple solution in which the $U(1)_{B-L}$ symmetry is broken down to a residual $\mathcal{Z}_6$ symmetry. 
However, in general, many other examples with different residual even $\mathcal{Z}_n$ symmetries can be found by applying the given framework.

\begin{center}{\bf Realistic example}\end{center}

\begin{figure}[th]
    \centering
    \begin{subfigure}[b]{0.4\textwidth}
        \includegraphics[width=\textwidth]{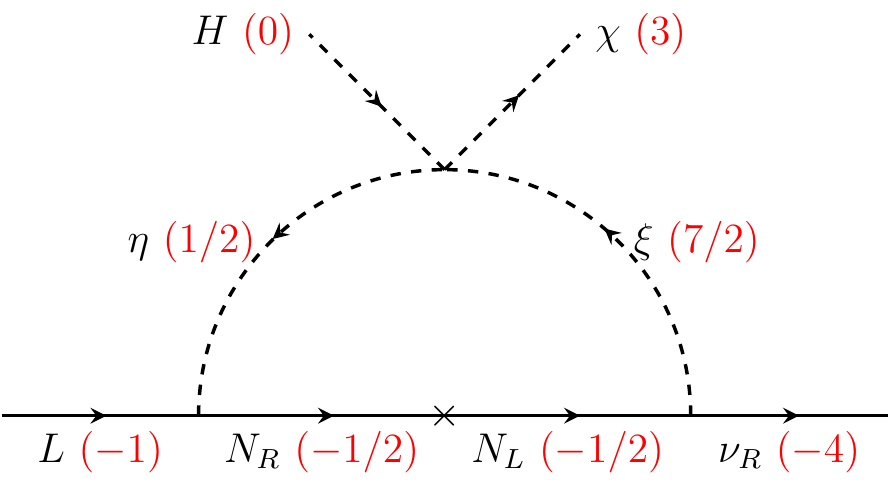}
        \caption{$U(1)_{B-L}$ charge assignment.}
        \label{fig:gull}
    \end{subfigure}
    ~ 
    \begin{subfigure}[t]{0.1\textwidth}
    \vspace*{-3.5cm}
    \hspace*{-0.4cm}
   \includegraphics[width=1.4\textwidth]{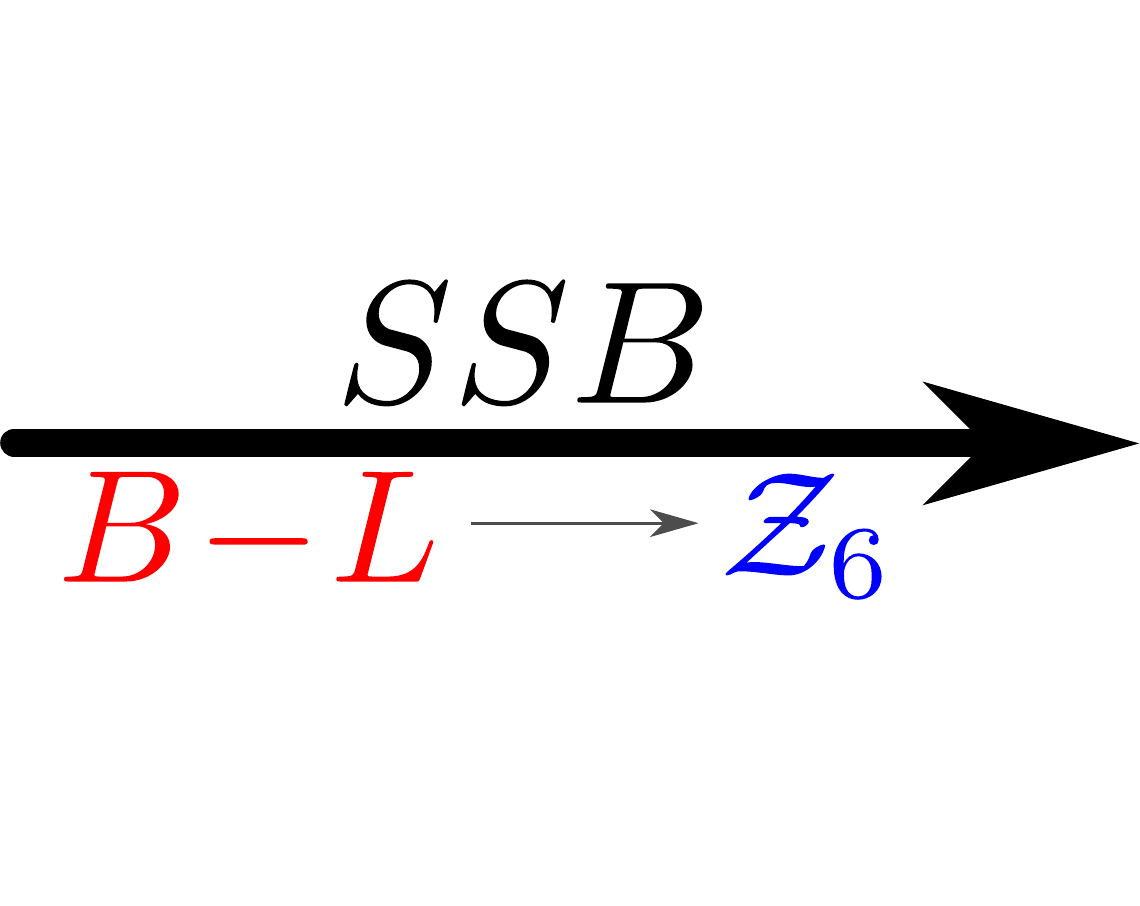}      \end{subfigure}
    ~ 
   \begin{subfigure}[b]{0.4\textwidth}
        \includegraphics[width=\textwidth]{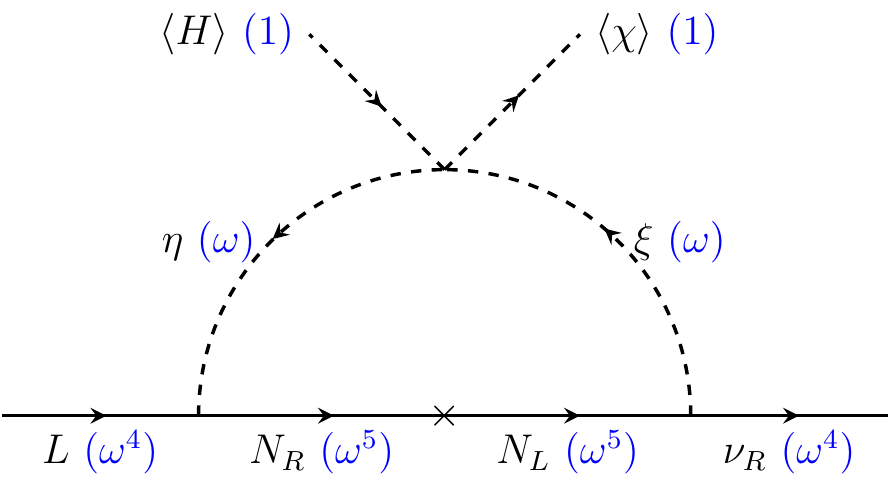}
        \caption{Residual $\mathcal{Z}_6$ charge assignment.}
        \label{fig:mouse}
    \end{subfigure} \\
    \caption{Charge assignment for the example model and its spontaneous symmetry breaking pattern.}\label{fig:z6}
\end{figure}

Let us consider an extension of the SM by adding an extra Higgs singlet $\chi$ with a $U(1)_{B-L}$ charge of $3$, along with an scalar doublet $\eta$, a singlet $\xi$ and two vector-like fermions $N_{L_l}$ and $N_{R_l}$, with $l=1,2$, all carrying non-trivial $U(1)_{B-L}$ charges as shown in Table \ref{modelZ6} and depicted in Fig.~\ref{fig:gull}.
\begin{table}
\begin{center}
\begin{tabular}{| c || c | c | c || c |}
  \hline 
&   Fields            &    $SU(2)_L \otimes U(1)_Y$            &     $U(1)_{B-L}$                       & 
  $\mathcal{Z}_{6}$                              \\
\hline \hline
\multirow{4}{*}{ \begin{turn}{90} Fermions \end{turn} } &
 $L_i$        	  &    ($\mathbf{2}, {-1/2}$)       &   {\color{red}${-1}$}    	  &	 {\color{blue}$\omega^4$}                     \\	
&   $\nu_{R_i}$       &   ($\mathbf{1}, {0}$)      & {\color{red} $({-4},{-4},\,{5})$ }   &  	 {\color{blue}($\omega^4, \omega^4, \omega^4)$}\\
&   $N_{L_l}$    	  &   ($\mathbf{1}, {0}$)      & {\color{red}${-1/2}$ }   &    {\color{blue} $\omega^5$}     \\
&  $N_{R_l}$     	  &  ($\mathbf{1}, {0}$) 	     & {\color{red} ${-1/2}$ } &  {\color{blue}$\omega^5$}     \\
\hline \hline
\multirow{4}{*}{ \begin{turn}{90} Scalars \end{turn} } &
 $H$  		 &  ($\mathbf{2}, {1/2}$)      &  {\color{red}${0}$ }    & {\color{blue} $1$}    \\
& $\chi$          	 &  ($\mathbf{1}, {0}$)        &  {\color{red}${3}$ }  &  {\color{blue} $1$}     \\		
& $\eta$          	 &  ($\mathbf{2}, {1/2}$)      &  {\color{red}${1/2}$}    &  {\color{blue}$\omega$}       \\
& $\xi$             &  ($\mathbf{1}, {0}$)        &  {\color{red}${7/2}$}      &	{\color{blue}$\omega$} \\	
    \hline
  \end{tabular}
\end{center}
\caption{Charge assignment for all of the fields. $\mathcal{Z}_6$ is the residual symmetry in this example, with $\omega^6=1$.}
 \label{modelZ6} 
\end{table}

The neutrino interactions are described by the following Lagrangian, 
\begin{equation}
{\cal L}_\nu=y_{il}\bar{L}_i\tilde{\eta}N_{R_l}+y'_{li}\bar{N}_{L_l}\nu_{R_i}\xi+M_{lm} \bar{N}_{R_l}N_{L_m}+h.c.,
\label{eq:YukInt}
\end{equation}
where $\tilde{\eta} = i \tau_2 \eta^*$, with the indices $i = 1,2,3$ and $l,m = 1,2$. The relevant part of the scalar potential for generating the Dirac neutrino mass is given by
\begin{equation}
\label{Veps}
{\cal V}\supset m_{\eta}^2 \eta^\dagger \eta + m_\xi^2 \xi^\dagger \xi + (\lambda_D H^\dagger \eta \chi \xi^* + h.c.),
\end{equation}
where $\lambda_D$ is a dimensionless quartic coupling. 
 
After spontaneous symmetry breaking of $U(1)_{B-L}$, the scalar $\chi$ gets a vev $\vev{\chi}=\text{u}$, giving mass to two neutrinos through the loop depicted in Fig.~\ref{fig:z6}. Note that only $\nu_{R_1}$ and $\nu_{R_2}$ can participate in this mass generation due to the chiral charges $(-4, -4, \, 5)$, i.e. $y'_{l3}=0$ in Eq.~\eqref{eq:YukInt}. The third right-handed neutrino $\nu_{R_3}$ remains massless and decouples from the rest of the model, although it is trivial to extend this simple model to generate its mass. 

The neutral component of the gauge doublet $\eta$ and the singlet $\xi$ are rotated into the mass eigenbasis with eigenvalues $m^2_i$ in the basis of ($\xi$,\,$\eta^0$). The neutrino mass matrix is then given in terms of the one-loop Passarino-Veltman function $B_0$ \cite{Passarino:1978jh} by,
 \begin{equation}
    (M_\nu)_{\alpha\beta} \sim \frac{1}{16\pi^2} \frac{\lambda_D \text{v} \text{u}}{m^2_\xi-m^2_\eta} y_{\alpha k} y'_{k\beta}  M_k \sum\limits_{i=1}^2 (-1)^i B_0(0,m_i^2,M_k^2),
    \label{numass}
\end{equation}

where $M_k$ ($k=1,2$) are the masses of the Dirac fermions $N_k$ and $\vev{H}=\text{v}$ the Standard Model vev.

As a benchmark point, we can take the internal fermion to be heavier than the scalars running in the loop, one of which will be the dark matter candidate. Then, Eq.~\eqref{numass} can be approximated by,
\begin{equation}
    m_\nu \sim \frac{1}{16 \pi^2} \frac{v u}{M} y y' \lambda_D.
\end{equation}
For comparison, we can take the Yukawa couplings to be of order $10^{-2}$ and the quartic coupling $\lambda_D \sim 10^{-4}$, like in the original scotogenic model \cite{Ma:2006km}. We can also take neutrino masses to be of order $0.1$ eV and $u \sim v$. With these choices, we can find the mass scale of the neutral fermions,
\begin{equation}
M \sim \frac{1}{16 \pi^2} \frac{v u}{m_\nu} y y' \lambda_D \sim 10^{4} \text{GeV}.
\end{equation}

Compared with the type-I seesaw scale $M \approx y^2 \frac{v^2}{m_\nu} \sim 10^{10}$ GeV we can see a five order of magnitude suppression coming from the loop and the possibility of a broader parameter space.

It is worth mentioning that since the $U(1)_{B-L}$ is anomaly free, it can be gauged. Then the physical Nambu-Goldstone boson associated to the dynamical generation of the Dirac neutrino mass, the Diracon \cite{Bonilla:2016zef, Alvarado:2021fbw}, is absent.
   
\begin{figure}[h!]
\centering
\includegraphics[width=0.4\textwidth]{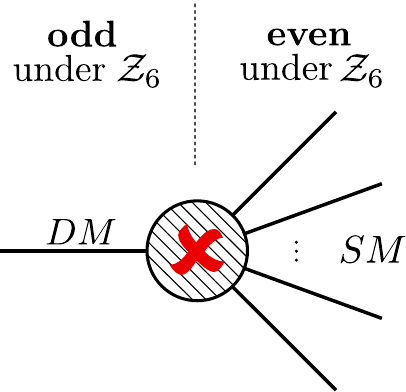}
\caption{The decay of dark matter (odd under $\mathcal{Z}_6$) to SM particles (all even under $\mathcal{Z}_6$) is forbidden by the residual $\mathcal{Z}_6$ symmetry. This argument can be generalized to any even $\mathcal{Z}_n$ symmetry.}
\label{U1mod}
\end{figure}

Regarding dark matter stability in this particular model, we can see that the lightest particle inside the loop is stable. This is true for both the fermionic and scalar dark matter candidates. As can be seen in Fig.~\ref{fig:mouse}, all of the internal loop particles are odd under the remnant $\mathcal{Z}_6$, while all of the SM particles are even. Therefore any combination of SM fields will be even under the residual subgroup, forbidding all effective operators leading to dark matter decay as shown graphically in Fig.~\ref{U1mod}.
   
To summarize, we have shown that by using the $U(1)_{B-L}$ symmetry already present in the Standard Model, it is possible to address the dark matter stability and relate it to the smallness of Dirac neutrino masses.
We have described a general framework in which these features are realized by exploiting the anomaly free chiral solutions of a global $U(1)_{B-L}$. This framework can be utilized in a wide variety of scenarios. 
We have presented a particular simple realization of this idea where neutrino masses are generated at the one-loop level and the $U(1)_{B-L}$ symmetry is broken spontaneously to a residual $\mathcal{Z}_6$ symmetry. 
The framework can also be used in models with higher-order loops as well as in cases where $U(1)_{B-L}$ symmetry is broken to other even $\mathcal{Z}_n$ subgroups.
Since the $U(1)_{B-L}$ is anomaly free, it can be gauged in a straightforward way, giving a richer phenomenology from the dark matter and collider point of view.

\chapter{Phenomenology of Dirac neutrino mass models}

As argued in Chapter \ref{chap:Dirac}, the `Dirac-Majorana confusion theorem' \cite{Schechter:1980gk, Kayser:1981nw, Kayser:1982br} states that phenomenological differences between Dirac and Majorana neutrinos will be proportional to $m_\nu$, posing an experimental challenge that is not expected to be solvable in the foreseeable future. Model independent analysis must therefore deal with the conclusions of this theorem \cite{Nieves:1985ir, Chhabra:1992be,Rodejohann:2017vup, Berryman:2018qxn, Kim:2021dyj}. While the detection of neutrinoless double beta decay would indoubtedly imply Majorana neutrinos, the reverse is not true and the non-observation of this process sheds no light on the mistery. However, as argued in \cite{Hirsch:2017col}, a weak black box theorem can be formulated for the case of Dirac neutrinos: if neutrinoless quadrupole beta decay \cite{NEMO-3:2017gmi} is detected before the double one, then neutrinos are most probably Dirac in nature. The contrary would imply extreme fine tuned cancellations.

We direct the interested reader to the references above and within them for more details on model independent proof of Dirac neutrinos. In this chapter we will briefly review the phenomenology of some appealing classes of testable Dirac neutrino mass models. In Sec.~\ref{sec:flavourpred} we will focus on predictive flavour models, which can be tested in upcoming oscillation experiments, while in Sec.~\ref{sec:LFV} we will study how low scale Dirac neutrino mass models feature lepton flavour violation.

\section{Flavour predictive Dirac neutrino models}
\label{sec:flavourpred}

One of the mysteries of particle physics is the
understanding of the pattern of fermion masses and mixings from first
principles. Indeed, the charged fermion mass pattern is not described
in the theory: the SM only allows us the freedom to fit the observed
charged fermion masses, while lacking the masses of neutrinos
altogether.
An approach towards addressing, at least partially, the charged
fermion mass problem, is the possibility of relating quarks and lepton
masses as a result of a flavour symmetry~\cite{Morisi:2011pt}, i.e.
\begin{equation}
  \label{eq:b-tau}
\frac{m_b}{{\sqrt{m_d m_s}}} = \frac{m_\tau}{{\sqrt{m_e m_\mu}}}~.
\end{equation}
Notice that this mass relation constitutes a consistent
flavour-dependent generalization of the conventional bottom-tau SU(5)
prediction, but does not require grand-unification. It provides a
partial solution to the charged fermion mass problem, which can be
shown to hold in some theories of flavour based on the
$A_4$~\cite{Morisi:2011pt,King:2013hj,Morisi:2013eca}.
$T_7$~\cite{Bonilla:2014xla} and $SO(10)$ \cite{Reig:2018ocz} symmetries.
Several alternative ideas for flavour predictions have come out, invoking non-Abelian flavour
symmetries~\cite{Ma:2001dn,Harrison:2002er,Giunti:2002ye,Harrison:2002kp,Babu:2002dz,Minakata:2004xt,Frampton:2004ud,Raidal:2004iw,deMedeirosVarzielas:2005qg,Altarelli:2005yx,Ma:2005pd,King:2006np,deMedeirosVarzielas:2006fc,Lam:2007qc,Grimus:2009pg,Kajiyama:2010sb,Morisi:2013qna,Goswami:2009yy,Ding:2011qt,Hartmann:2011pq,Feruglio:2012cw,Holthausen:2012dk,Morisi:2012fg,Hernandez:2012ra,King:2013eh,King:2014nza,Chen:2014tpa,Kile:2014kya,He:2015xha,CentellesChulia:2016fxr,Novichkov:2018nkm,NevesPenedo:2018zvi,Novichkov:2018ovf,Rahat:2018sgs,Barreiros:2018bju,Chen:2018zbq,Chen:2018eou,Chen:2018lsv,Novichkov:2019sqv,Ding:2019xna,Chen:2019fgb,Perez:2019aqq,CentellesChulia:2019ijn,Novichkov:2020ahb,Yao:2020qyy,CentellesChulia:2020bnf,Novichkov:2020eep,Okada:2020brs,Ding:2020vud,Okada:2020ukr,Ma:2021kfa,Novichkov:2021evw,Li:2021buv,King:2021fhl,Qu:2021jdy}, both in UV complete models or in a model-independent way and covering both possibilities for neutrino nature.
On the other hand, specially tantalizing is the idea that the stability of dark matter
can be directly traced to the Dirac nature of neutrinos.
One way to realize this idea is by means of a $Z_4$ Lepton Quarticity
symmetry~\cite{ CentellesChulia:2016rms, CentellesChulia:2016fxr}. Within such approach
the same $Z_4$ discrete lepton number symmetry ensures the stability
of dark matter and the absence of all the Majorana mass terms. Thus
owing to Lepton Quarticity, the Dirac nature of neutrinos and the
stability of dark matter are intimately related: the breakdown of this
symmetry will simultaneously imply loss of dark matter stability as
well as the \textit{Diracness} of neutrinos.
 
As an example of an elegant and predictive Dirac neutrino mass model we will flesh out the model in \cite{CentellesChulia:2017koy}, along the lines of the Lepton Quarticity models of dark matter \cite{CentellesChulia:2016rms,CentellesChulia:2016fxr}. Its testability potential in DUNE has been analyzed in \cite{Srivastava:2017sno}


 \section{Sample model: Bottom-Tau unification and neutrino oscillations}
\label{sec:model}


The particle content of
our model along with the $SU(2)_L \otimes Z_4 \otimes A_4$ charge
assignments of the particles are given in Table \ref{tab1}.
\begin{table}[ht]
\begin{center}
\begin{tabular}{|c c c c || c c c c|}
  \hline \hline
Fields          \hspace{0.5cm}    & $SU(2)_L$           \hspace{0.5cm}   & $A_4$           \hspace{0.5cm}   &  $Z_4$	    \hspace{0.5cm}    & Fields          \hspace{0.5cm}    & $SU(2)_L$           \hspace{0.5cm}   & $A_4$           \hspace{0.5cm}   &  $Z_4$	     \\
\hline \hline
\rule{0pt}{3ex}  $L_i$     \hspace{0.5cm}    & $\mathbf{2}$    \hspace{0.5cm}    & $\mathbf{3}$    \hspace{0.5cm}   &  $\mathbf{z}$   \hspace{0.5cm}   &		
$\nu_{e,R}$     \hspace{0.5cm}    & $\mathbf{1}$    \hspace{0.5cm}    & $\mathbf{1}$    \hspace{0.5cm}   &  $\mathbf{z}$      \\
$N_{i,L}$ \hspace{0.5cm}    & $\mathbf{1}$    \hspace{0.5cm}    & $\mathbf{3}$    \hspace{0.5cm}   &  $\mathbf{z}$   \hspace{0.5cm}   &        
$\nu_{\mu, R}$  \hspace{0.5cm}    & $\mathbf{1}$    \hspace{0.5cm}    & $\mathbf{1'}$   \hspace{0.5cm}   &  $\mathbf{z}$       \\
$N_{i,R}$       \hspace{0.5cm}    & $\mathbf{1}$    \hspace{0.5cm}    & $\mathbf{3}$    \hspace{0.5cm}   &  $\mathbf{z}$      \hspace{0.5cm}  &		
$ \nu_{\tau,R}$ \hspace{0.5cm}    & $\mathbf{1}$    \hspace{0.5cm}    & $\mathbf{1''}$  \hspace{0.5cm}   &  $\mathbf{z}$        \\
$ l_{i,R} $     \hspace{0.5cm}    & $\mathbf{1}$    \hspace{0.5cm}    & $\mathbf{3}$    \hspace{0.5cm}   &  $\mathbf{z}$      \hspace{0.5cm}  &		
$ d_{i,R} $     \hspace{0.5cm}    & $\mathbf{1}$    \hspace{0.5cm}    & $\mathbf{3}$    \hspace{0.5cm}   &  $\mathbf{z^3}$        \\
$ Q_{i,L} $  \hspace{0.5cm}    & $\mathbf{2}$    \hspace{0.5cm}    & $\mathbf{3}$    \hspace{0.5cm}   &  $\mathbf{z}^3$    \hspace{0.5cm}  &		
$ u_{i,R} $     \hspace{0.5cm}    & $\mathbf{1}$    \hspace{0.5cm}    & $\mathbf{3}$    \hspace{0.5cm}   &  $\mathbf{z^3}$    \rule[-2ex]{0pt}{0pt}     	\\
\hline
\hline
\rule{0pt}{3ex}  $\Phi_1^u$    \hspace{0.5cm}          & $\mathbf{2}$      \hspace{0.5cm}        & $\mathbf{1}$      \hspace{0.5cm}  	    &  $\mathbf{1}$        \hspace{0.5cm} &		
$\chi_i$        \hspace{0.5cm}        & $\mathbf{1}$      \hspace{0.5cm}   	    & $\mathbf{3}$    	\hspace{0.5cm} 	    &  $\mathbf{1}$            \\
$\Phi_2^u$    \hspace{0.5cm}        & $\mathbf{2}$      \hspace{0.5cm}        & $\mathbf{1'}$     \hspace{0.5cm}        &  $\mathbf{1}$        \hspace{0.5cm}     &		             
$\eta$          \hspace{0.5cm}        & $\mathbf{1}$	\hspace{0.5cm}        & $\mathbf{1}$      \hspace{0.5cm}        &  $\mathbf{z}^2$      	      \\
$\Phi_3^u$    \hspace{0.5cm}        & $\mathbf{2}$	\hspace{0.5cm}        & $\mathbf{1''}$    \hspace{0.5cm}        &  $\mathbf{1}$        \hspace{0.5cm}     &		
$\zeta$         \hspace{0.5cm}        & $\mathbf{1}$	\hspace{0.5cm}        & $\mathbf{1}$      \hspace{0.5cm}        &  $\mathbf{z}$       	 \\
$\Phi_i^d$      \hspace{0.5cm}        & $\mathbf{2}$	\hspace{0.5cm}        & $\mathbf{3}$      \hspace{0.5cm}        &  $\mathbf{1}$        \hspace{0.5cm}     &		           
                \hspace{0.5cm}        & 		 	\hspace{0.5cm}        & 			\hspace{0.5cm}        &     	       \rule[-2ex]{0pt}{0pt}     \\
    \hline
  \end{tabular}
\end{center}
\caption{Charge assignments for leptons, quarks,  scalars 
  ($\Phi_i^u$, $\Phi_i^d$ and $\chi_i$) as well as `dark matter sector' 
  ($\zeta$ and $\eta$). Here $\mathbf{z}$ is the fourth root
  of unity, i.e. $\mathbf{z}^4 = 1$.  }
 \label{tab1} 
\end{table}
Note that in Table~\ref{tab1} the $L_i = (\nu_i, l_i)^T$,
$i = e, \mu, \tau$ denote the lepton doublets, transforming as
indicated under the flavour symmetry.

Apart from the \sm fermions, the model also includes three
right--handed neutrinos $\nu_{i,R}$ which are singlets under the \SM
gauge group, singlets under $A_4$, but carry charge $\mathbf{z}$ under
$Z_4$. We also add three gauge singlet Dirac fermions
$N_{i,L}, N_{i,R}$; $i = 1, 2, 3$ transforming as triplets of $A_4$
and with charge $\mathbf{z}$ under $Z_4$, as shown in
Table~\ref{tab1}.
Notice that in the scalar sector we have two different sets of fields
$\Phi^u_i, \Phi^d_i$; $i = 1,2,3$, which are all doublets under the
SU(2)$_L$ gauge group, both sets transforming trivially under
$Z_4$. Under the $A_4$ flavour symmetry, $\Phi_i^d$ transforms as a
triplet, while $\Phi_i^u$ transform as singlets. In addition to the
above symmetries we also impose an additional $Z_2$
symmetry\footnote{This additional $Z_2$ symmetry is only required in a
  non-supersymmetric variant. Clearly the model can be easily
  supersymmetrized, in which case this additional $Z_2$ symmetry is no
  longer required.}.  Under this $Z_2$ symmetry, all the fields
transform as $1$ except for $\Phi^d_i$, $l_{i,R}$ and $d_{i, R}$,
which transform as $-1$. The role of this $Z_2$ symmetry is to prevent
the Higgs doublets $\Phi_i^d$ from coupling the up-type quarks and
neutrino sector, and the $\Phi^u_i$ Higgs doublets from the down-type
quarks and charged leptons.

In addition we need scalar singlets, for example the $\chi_i$,
$i = 1, 2, 3$. These are gauge singlets transforming as a triplet
under the $A_4$ and trivially under $Z_4$. We also add two other gauge
singlet scalars $\zeta$ and $\eta$ both of which transform trivially
under $A_4$ but carry $Z_4$ charges $\mathbf{z}$ and $\mathbf{z}^2$
respectively.
Notice that, since under the $Z_4$ symmetry the field $\eta$ carries a
charge $\mathbf{z}^2 = -1$, it follows that $\eta$ can be taken to be
real.

As discussed in \cite{CentellesChulia:2016rms, CentellesChulia:2016fxr} the lepton
quarticity symmetry $Z_4$ serves a double purpose. It not only ensures
that neutrinos are Dirac particles, but also guarantees the stability
of the scalar particle $\zeta$, making it a viable dark matter
WIMP. If the quarticity symmetry is broken either by an explicit soft
term or spontaneously, through non-zero vacuum expectation values
(vevs) to any of the scalars $\eta, \zeta$ which carry a non-trivial
$Z_4$ charge, then both the Dirac nature of neutrinos and stability of
dark matter is simultaneously lost.

We now turn our attention to the Yukawa sector of our model. In the
neutrino sector the Yukawa terms relevant for generating masses for
the neutrinos and the heavy neutral fermions $N_{L}, N_{R}$ are given
by
\begin{eqnarray}
\hspace{-3cm}\mathcal{L}_{\rm{Yuk}, \nu}  & = & y_1 \, \left [  \left( \begin{array}{c} \bar{L}_e  \\ \bar{L}_\mu \\ \bar{L}_\tau \end{array} \right)_{3} \, \otimes \,  \left( \begin{array}{c}  N_{1,R}   \\ N_{2, R} \\ N_{3, R} \end{array} \right)_{3}  \right]_1 \, \otimes \ \left ({\Phi}^u_1 \right)_1   
  \, + \, 
y_2 \,  \left [  \left( \begin{array}{c} \bar{L}_e  \\ \bar{L}_\mu \\ \bar{L}_\tau \end{array} \right)_{3} \, \otimes \,  \left( \begin{array}{c}  N_{1,R}   \\ N_{2, R} \\ N_{3, R} \end{array} \right)_{3}  \right]_{1''} \, \otimes \,  \left ({\Phi}^u_2 \right)_{1'}  
  \nonumber \\
& + &    y_3 \, \left [  \left( \begin{array}{c} \bar{L}_e  \\ \bar{L}_\mu \\ \bar{L}_\tau \end{array} \right)_{3} \, \otimes \,  \left( \begin{array}{c}  N_{1,R}   \\ N_{2, R} \\ N_{3, R} \end{array} \right)_{3}  \right]_{1'} \, \otimes \,   \left ({\Phi}^u_3 \right)_{1''}  
  \, + \,
y'_1 \,   \left [ \left( \begin{array}{c}  \bar{N}_{1,L}   \\ \bar{N}_{2, L} \\ \bar{N}_{3, L} \end{array} \right)_{3} \, \otimes \, \left( \begin{array}{c} \chi_1   \\ \chi_2 \\ \chi_3 \end{array} \right)_{3} \right]_1  \, \otimes \, \left(\nu_{e,R} \right)_1
  \nonumber \\
  & + &
y'_2 \,   \left [ \left( \begin{array}{c}  \bar{N}_{1,L}   \\ \bar{N}_{2, L} \\ \bar{N}_{3, L} \end{array} \right)_{3} \, \otimes \, \left( \begin{array}{c} \chi_1   \\ \chi_2 \\ \chi_3 \end{array} \right)_{3} \right]_{1''}  \, \otimes \, \left(\nu_{\mu,R} \right)_{1'}
   \, + \, 
y'_3 \,   \left [ \left( \begin{array}{c}  \bar{N}_{1,L}   \\ \bar{N}_{2, L} \\ \bar{N}_{3, L} \end{array} \right)_{3} \, \otimes \, \left( \begin{array}{c} \chi_1   \\ \chi_2 \\ \chi_3 \end{array} \right)_{3} \right]_{1'}  \, \otimes \, \left(\nu_{\tau,R} \right)_{1''}
    \nonumber \\
& + &     M \,   \left[ \left( \begin{array}{c}  \bar{N}_{1,L}   \\ \bar{N}_{2, L} \\ \bar{N}_{3, L} \end{array} \right)_{3} \,
\otimes \,\left( \begin{array}{c}  N_{1,R}   \\ N_{2, R} \\ N_{3, R} \end{array} \right)_{3} \right]_1 
      \, + \, 
c_1 \, \left[ \left( \begin{array}{c}  \bar{N}_{1,L}   \\ \bar{N}_{2, L} \\ \bar{N}_{3, L} \end{array} \right)_3  \, \otimes \, \left [  \left( \begin{array}{c}  \chi_{1}   \\ \chi_{2} \\ \chi_{3} \end{array} \right)_3 \, \otimes  \left( \begin{array}{c} N_{1,R}   \\ N_{2,R} \\ N_{3, R} \end{array} \right)_{3}   \right]_{3S} \right ]_1
    \nonumber \\
&+&    c_2 \, \left[ \left( \begin{array}{c}  \bar{N}_{1,L}   \\ \bar{N}_{2, L} \\ \bar{N}_{3, L} \end{array} \right)_3  \, \otimes \, \left [  \left( \begin{array}{c}  \chi_{1}   \\ \chi_{2} \\ \chi_{3} \end{array} \right)_3 \, \otimes  \left( \begin{array}{c} N_{1,R}   \\ N_{2,R} \\ N_{3, R} \end{array} \right)_{3}   \right]_{3A} \right ]_1 + h.c.
 \label{neutyuk}
\end{eqnarray}
where $3S$ and $3A$ denote the symmetric and antisymmetric $A_4$
triplet combinations obtained from the tensor product of two $A_4$
triplets. Notice also that $3S$ and $3A$ are not two different
irreducible representations of $A_4$, which only has one triplet, but
simply different contractions with the same transformation rule. Also,
$y_i, y'_i, c_1, c_2$; $i = 1, 2, 3$ are the Yukawa couplings which,
for simplicity, are taken to be real. The parameter $M$ is the gauge
and flavour-invariant mass term for the heavy neutral fermions. Here we like to
highlight the important role played by the $A_4$ flavour
symmetry. Owing to the $A_4$ charges of the left and right handed
neutrinos, a tree level Yukawa coupling between them of type
$y_{\nu} \, \bar L_L \nu_R \Phi^u_i$ is forbidden. Thus neutrino
masses can only appear through type-I Dirac seesaw mechanism as we now
discuss.
 
After symmetry breaking the scalars $\chi_i$ and $\Phi^u_i$ acquire
vevs $\vev{ \chi_i} = u_i$; $\vev{ \Phi^u_i} = v_i^u$; $i = 1, 2, 3$.
 The invariant mass term $M$ can be naturally much larger than the
 symmetry breaking scales, i.e.  $ M \gg v^u_i, u_i$. In this limit,
 for any numerical purpose the last two terms in Eq.~\ref{neutyuk} can
 be safely neglected. Under this approximation the $6\times 6$ mass
 matrix for the neutrinos and the heavy neutral fermions in the basis
 $( \bar{\nu}_{e,L}, \bar{\nu}_{\mu,L}, \bar{\nu}_{\tau,L},
 \bar{N}_{1,L}, \bar{N}_{2,L}, \bar{N}_{3,L} )$
 and
 $( \nu_{e,R}, \nu_{\mu,R}, \nu_{\tau,R}, N_{1,R}, N_{2,R}, N_{3,R}
 )^T$ is given by
  \begin{eqnarray}
   M_{\nu, N} \, \, = \, \left( 
\begin{array}{cccccc}
0           & 0                   &  0                  & a'_1	 	& 0            &   0      \\
0           & 0                   &  0                  & 0	        & a'_2	       &   0       \\
0           & 0                   &  0                  & 0	 	& 0	       &   a'_3    \\ 
y'_1 u_1    & y'_2 u_1            &  y'_3 u_1           & M             & 0	       &  0	    \\
y'_1 u_2    & \omega y'_2 u_2     & \omega^2 y'_3 u_2   & 0		& M            &  0	    \\
y'_1 u_3    & \omega^2 y'_2 u_3   & \omega y'_3 u_3     & 0		& 0	       &  M         \\
\end{array}
\right)
   \label{gneutmass}
  \end{eqnarray}
 where $\omega$ is the third root of unity, with $\omega^3 = 1$ and

   \begin{eqnarray}
   a'_1 & = & y_1 v_1^u  +  y_2 v_2^u  +  y_3 v_3^u \nonumber \\
   a'_2 &=&   y_1 v_1^u + \omega y_2 v_2^u + \omega^2y_3 v_3^u \nonumber \\
   a'_3 &=& y_1 v_1^u + \omega^2y_2 v_2^u +\omega y_3 v_3^u
  \end{eqnarray}

  As mentioned before, owing to the $A_4$ symmetry, a direct
    coupling between $\nu_L$ and $\nu_R$ is forbidden, leading to
    the vanishing of all entries in the upper left quadrant of
    Eq.~\ref{gneutmass}.  The mass matrix in Eq.~\ref{gneutmass} can
  be rewritten in a more compact form, as
  \begin{eqnarray}
   M_{\nu, N} \, \, = \, \left( 
\begin{array}{cc}
0     		     &\mathrm{diag}(a'_1, a'_2, a'_3)            \\
\mathrm{diag}(u_1, u_2, u_3) \sqrt{3}U_{\rm m}\mathrm{diag}(y'_1,y'_2,y'_3)    & M \mathrm{diag}(1,1,1)  \\
\end{array}
\right)
   \label{gneutmasscompact}
  \end{eqnarray}

  Where $U_{\rm m}$ is the usual magic matrix,
    \begin{eqnarray}
  U_{\rm m} = \dfrac{1}{\sqrt{3}} \, \left( 
\begin{array}{ccc}
1     & 1           &  1           \\
1     & \omega      &  \omega^2   \\
1     & \omega^2    &  \omega     \\  
\end{array}
\right)~.
   \label{magmat}
  \end{eqnarray}
  
\begin{figure}[!h]
 \centering
  \includegraphics[scale=0.4]{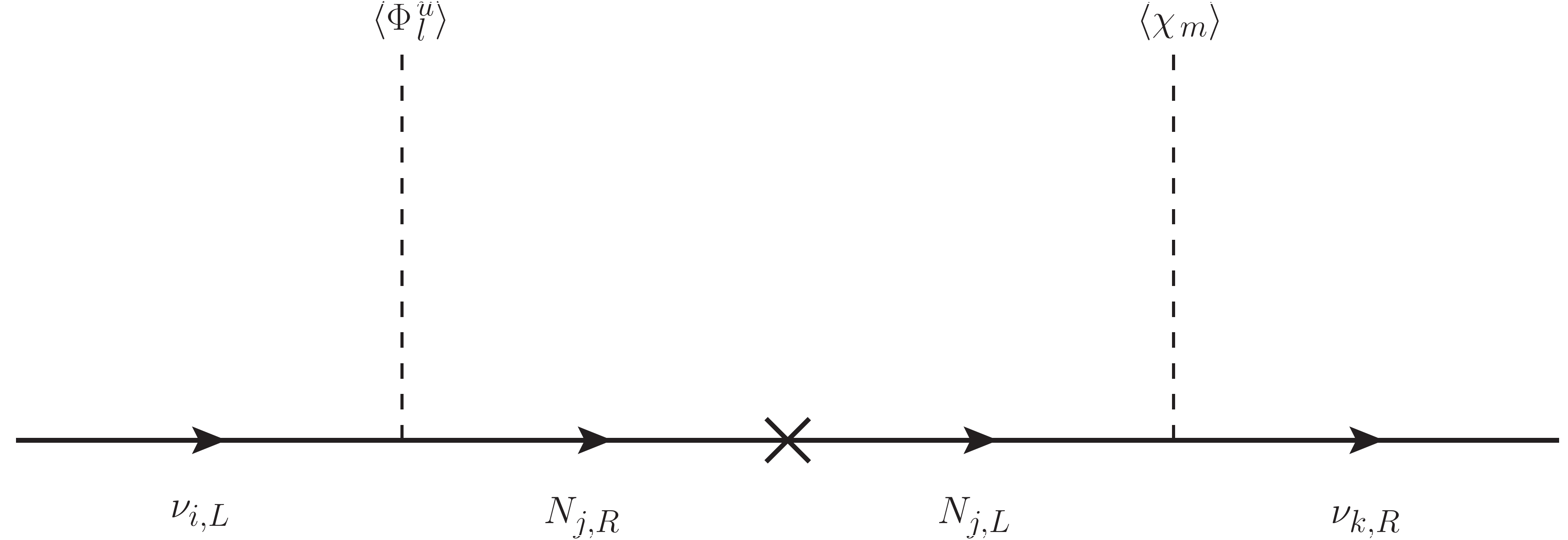}
  \caption{Feynman view of type-I Dirac seesaw mechanism in the model
    where the indices $i,j,k,m,l = 1,2,3$.} 
    \label{fig:feyn}
  \end{figure}

  Note that, in the limit $ M \gg v^u_i, u_i$ the mass matrix in
  Eq.~(\ref{gneutmass}) can be easily block diagonalized by the
  perturbative seesaw diagonalization method given in
  Ref.~\cite{Schechter:1981cv}. The resulting $3 \times 3$ mass matrix
  for light neutrinos can be viewed as the Dirac version of the well
  known type-I seesaw mechanism. The above mass generation mechanism
  can also be represented diagramatically as shown in
  Fig. \ref{fig:feyn}. 

  The $3\times 3$ matrix for the light neutrinos is
  \begin{eqnarray}
   M_{\nu} = \frac{1}{M}\mathrm{diag}(a_1, a_2, a_3)\sqrt{3}U_{\rm m}\mathrm{diag}(y'_1,y'_2,y'_3)~,
   \label{gnumass}
  \end{eqnarray}
  where $a_i = a'_i u_i$. We take the alignment $u_1 = u_2 = u_3 = u$
  for the vev of the $A_4$ triplet scalars $\chi_i$ similar
  to~\cite{Babu:2002dz, Ma:2001dn}. In this alignment limit of $A_4$
  triplet scalars we have 
   \begin{eqnarray}
   a_1 &=& (y_1 v_1^u + y_2v_2^u + y_3v_3^u) u \nonumber  \\
   a_2 &= &(y_1v_1^u + \omega y_2v_2^u + \omega^2y_3v_3^u) u \nonumber \\
   a_3 &=& (y_1v_1^u  + \omega ^2 y_2v_2^u + \omega y_3v_3^u ) u 
   \label{sima}
  \end{eqnarray}
  which simplifies the notation, although it does not change the form
  of the neutrino mass matrix in Eq.~(\ref{gnumass}). Notice that we
  have not imposed any alignment for the vevs of the $A_4$ singlet
  scalars $\Phi^u_i$~\footnote{Doing so for the different $A_4$
    singlet scalar vevs is not very natural. Indeed, unlike the case
    of $A_4$ triplet scalars, a priori the vevs of different $A_4$
    singlet scalars have no reason to obey any mutual alignment. }
  The light neutrino mass matrix of Eq.~\ref{gnumass} with the
  simplified $a_i$ of Eq.~\ref{sima} can be diagonalized by a
  bi-unitary transformation
\begin{eqnarray}
U_\nu^\dagger M_\nu V_\nu = D_\nu,
\label{biunit}
\end{eqnarray}
where $D_\nu$ is diagonal, real and positive. Owing to the $A_4$
flavour symmetry, the resulting rotation matrix acting on left handed
neutrinos $U_\nu$ in the standard parametrization (for both
hierarchies), leads to $\theta^\nu_{23} = \frac{\pi}{4}$ and
$\delta^\nu = \pm \frac{\pi}{2}$ while the other two angles can be
arbitrary. Thus, owing to the $A_4$ symmetry, $U_\nu$ in standard
parameterization leads to following mixing angles
\begin{eqnarray}
 \theta^\nu_{23} & = & 45^\circ,  \quad \quad \delta^\nu \, = \, \pm \, 90^\circ  \nonumber \\
 \theta^\nu_{12} & = & \rm{arbitrary} \quad \quad  \theta^\nu_{13} \, = \, \rm{arbitrary} 
 \label{unu}
\end{eqnarray}
Similar features of maximal $\theta_{23}$ and $\delta$ have been
obtained previously in the context of Majorana neutrinos
\cite{Babu:2002dz,Ma:2015pma}.  Although the angles $\theta^\nu_{12}$
and $\theta^\nu_{13}$ can independently take any value, they are strongly
correlated with each other. 

We have performed an extensive numerical scan for both type of
hierarchies in the whole parameter range taking all Yukawa couplings
in the perturbative range of $[-1, 1]$. We find that in the whole
allowed range for either type of ordering, one cannot simultaneously
fit both $\theta^\nu_{12}$ and $\theta^\nu_{13}$ in the current global
experimental range obtained from neutrino oscillation
experiments~\cite{deSalas:2020pgw}. This implies that in our model
$U_\nu$ alone cannot explain the current neutrino oscillation data.

However, the lepton mixing matrix $U_{LM}$ which is probed by neutrino
oscillation experiments is the product of the charged lepton rotation
matrix $U_l$ with the neutrino transformation matrix
$U_\nu$~\cite{Schechter:1980gr} i.e.
\begin{eqnarray}
 U_{LM} & = & U^\dagger_l \, U_\nu
 \label{ulm}
\end{eqnarray}
In our model the charged lepton mixing matrix $U_l$ is also
non-trivial and contributes to the full leptonic mixing matrix
$U_{LM}$. We now move to discuss the structure of mass matrices and
mixing matrices for charged leptons as well as the up and down type
quarks.  \\[-.2cm]

We now turn to the discussion with up type quark mass matrix.  The
invariant Yukawa Lagrangian relevant to generating up type quark mass
matrix is given by
   \begin{eqnarray}
  \mathcal{L}_{\rm{Yuk}, u}  & = & y^u_1 \, \left [  \left( \begin{array}{c} \bar{Q}_1  \\ \bar{Q}_2 \\ \bar{Q}_3 \end{array} \right)_{3} \, \otimes \,  \left( \begin{array}{c}  u_{R}   \\ c_{R} \\ t_{R} \end{array} \right)_{3}  \right]_1 \, \otimes \ \left ({\Phi}^u_1 \right)_1   
  \, + \,    y^u_2 \,  \left [  \left( \begin{array}{c} \bar{Q}_1  \\ \bar{Q}_2 \\ \bar{Q}_3 \end{array} \right)_{3} \, \otimes \,  \left( \begin{array}{c}  u_{R}   \\ c_{R} \\ t_{R} \end{array} \right)_{3}  \right]_{1''} \, \otimes \,  \left ({\Phi}^u_2 \right)_{1'} \nonumber \\
  & + & y_3^u \, \left [  \left( \begin{array}{c} \bar{Q}_1  \\ \bar{Q}_2 \\ \bar{Q}_3 \end{array} \right)_{3} \, \otimes \,  \left( \begin{array}{c}  u_{R}   \\ c_{R} \\ t_{R} \end{array} \right)_{3}  \right]_{1'} \, \otimes \,   \left ({\Phi}^u_3 \right)_{1''}   + \, \, h.c. 
  \label{uquarkyuk}
 \end{eqnarray}
 where $y^u_i$; $i = 1,2,3$ are the Yukawa couplings which for
 simplicity we take to be all real.  After spontaneous symmetry
 breaking Eq.~\ref{uquarkyuk} leads to a diagonal mass matrix given by
\begin{eqnarray}
 M_u & = & \left( 
\begin{array}{ccc}
y^u_1 v^u_1 + y^u_2 v^u_2  + y^u_3 v^u_3   & 0               &   0       \\
0            & y^u_1 v^u_1 + \omega \, y^u_2 v^u_2  + \omega^2 \, y^u_3 v^u_3         &   0     \\
0            & 0           &   y^u_1 v^u_1 + \omega^2 \, y^u_2 v^u_2  + \omega \, y^u_3 v^u_3           \\  
\end{array}
\right). \nonumber \\
\hspace{1cm}
 \label{upmassmat}
\end{eqnarray}

On the other hand, the Yukawa Lagrangian relevant to down type quarks
mass generation is given by
  \begin{eqnarray}
  \mathcal{L}_{\rm{Yuk}, d}  & = & y^d_1 \, \left[ \left( \begin{array}{c}  \bar{Q}_{1}   \\ \bar{Q}_{2} \\ \bar{Q}_{3} \end{array} \right)_3  \, \otimes \, \left [  \left( \begin{array}{c}  q_{d,R}   \\ q_{s, R} \\ q_{b, R} \end{array} \right)_3 \, \otimes  \left( \begin{array}{c} \Phi_1^d   \\ \Phi_2^d \\ \Phi_3^d \end{array} \right)_{3}   \right]_{3S} \right ]_1
    \nonumber \\
  &+&    y^d_2 \, \left[ \left( \begin{array}{c}  \bar{Q}_{1}   \\ \bar{Q}_{2} \\ \bar{Q}_{3} \end{array} \right)_3  \, \otimes \, \left [  \left( \begin{array}{c} q_{d,R}   \\ q_{s, R} \\ q_{b, R} \end{array} \right)_3 \, \otimes  \left( \begin{array}{c} \Phi_1^d   \\ \Phi_2^d \\ \Phi_3^d \end{array} \right)_{3}   \right]_{3A} \right ]_1 + h.c.
    \label{quarkyuk}
 \end{eqnarray}
 where $y^d_i$; $i = 1,2$ are the Yukawa couplings which for
 simplicity are taken to be real. The resulting mass matrix for down
 type quarks after spontaneous symmetry breaking is given by
  \begin{eqnarray}
   M_{d} & = & \left( 
\begin{array}{ccc}
0              & a_d \alpha      &   b_d       \\
b_d \alpha     & 0               &   a_d r     \\
a_d            & b_d r           &   0          \\  
\end{array}
\right)~.
   \label{gquarkmass}
  \end{eqnarray}
  \noindent
  where $\vev{ \Phi^d_i} = v^d_i$; $i = 1,2,3$ and
  $a_d = (y_1^d-y_2^d) v_2^d$, $b_l = (y_1^d +y_2^d) v_2^d$. Moreover,
  $\alpha$ and $r$ are ratios of the vevs of $\Phi^d_i$ and are given
  as $\alpha = \rfrac{v_3^d}{v_2^d}$ and $r = \rfrac{v_1^d}{v_2^d}$.

Finally, the invariant Yukawa terms for the charged leptons is given by
 \begin{eqnarray}
  \mathcal{L}_{\rm{Yuk}, l}  & = & y^l_1 \, \left[ \left( \begin{array}{c}  \bar{L}_{1}   \\ \bar{L}_{2} \\ \bar{L}_{3} \end{array} \right)_3  \, \otimes \, \left [  \left( \begin{array}{c}  l_{e,R}   \\ l_{\mu, R} \\ l_{\tau, R} \end{array} \right)_3 \, \otimes  \left( \begin{array}{c} \Phi_1^d   \\ \Phi_2^d \\ \Phi_3^d \end{array} \right)_{3}   \right]_{3S} \right ]_1
    \nonumber \\
  &+&    y^l_2 \, \left[ \left( \begin{array}{c}  \bar{L}_{1}   \\ \bar{L}_{2} \\ \bar{L}_{3} \end{array} \right)_3  \, \otimes \, \left [  \left( \begin{array}{c}  l_{e,R}   \\ l_{\mu, R} \\ l_{\tau, R} \end{array} \right)_3 \, \otimes  \left( \begin{array}{c} \Phi_1^d   \\ \Phi_2^d \\ \Phi_3^d \end{array} \right)_{3}   \right]_{3A} \right ]_1 + h.c.
    \label{lepyuk}
 \end{eqnarray}
 where $y^l_i$, $i = 1, 2$, are the Yukawa couplings which, for
 simplicity, we take to be real. After symmetry breaking the charged
 lepton mass matrix is given by
  \begin{eqnarray}
   M_{l} = \left( 
\begin{array}{ccc}
0               & a_l \alpha           &   b_l \\
b_l \alpha      & 0                    &   a_l r \\
a_l             & b_l r                &   0  \\  
\end{array}
\right)~.
   \label{glepmass}
  \end{eqnarray}
  where, just as in the down quark case, here also
  $a_l = (y_1^l-y_2^l) v_2^d$, $b_l = (y_1^l +y_2^l) v_2^d$.  
  The parameters $\alpha, r$ which are the ratios of the vevs of
  $\Phi^d_i$ i.e.  $\alpha = \rfrac{v_3^d}{v_2^d}$ and
  $r = \rfrac{v_1^d}{v_2^d}$ are the same as those defined after
  Eq.~\ref{gquarkmass}. This matrix is completely analogous to the
  down-type quark mass matrix. Note that while $\alpha$ and $r$ are the same
  both in the quark and in the lepton sector, as they are simply
  ratios between the vevs of $\Phi_i^d$, while $a_f$ and $b_f$,
  $f \in \{l, q\}$, are different.

  These mass matrices for charged leptons and down-type quarks
  correspond to those discussed
  in~\cite{Morisi:2011pt,King:2013hj,Morisi:2013eca,Bonilla:2014xla}
  and lead to the generalized bottom-tau relation of \ref{eq:b-tau}.
  In section \ref{sec:numerical-scan} we show that there is enough
  freedom to fit the charged lepton and down type quark masses within
  their 1-$\sigma$ range. Apart from fitting all the masses as well as
  leading to the generalized bottom-tau relations, the charged lepton
  mass matrix \ref{glepmass} also leads to non-trivial charged lepton
  rotation matrix $U_l$.  As we show in section
  \ref{sec:numerical-scan} this non-trivial contribution from $U_l$
  results in a lepton mixing matrix $U_{LM}$ consistent with the
  current global fits to neutrino oscillation data
  \cite{deSalas:2020pgw}. The lepton mixing
  matrix obtained from our model also implies normal ordering for
  neutrino masses and leads to an interesting correlation between the
  atmospheric mixing angle and CP violating phase, the two most
  ill--determined parameters in leptonic mixing matrix.
  
 \subsubsection{Flavour predictions: numerical results}
\label{sec:numerical-scan} 

In this section we discuss the phenomenological implications of our
model.  The important predictions emerging in our model are: 
\begin{itemize}
\item The flavour-dependent bottom-tau unification mass relation of Eq.~(\ref{eq:b-tau}).
\item A correlation between the two poorly determined oscillation parameters: the atmospheric angle $\theta_{23}$
and $\delta_{CP}$.
\item A normal ordering for the neutrinos.
\end{itemize}

In this section we discuss these numerical predictions in some detail, 
given the experimentally measured `down-type' fermion masses, solar and
reactor mixing angles as well as neutrino squared mass differences.

\subsubsection{Charged lepton and down-type quark masses}
\label{clmass}

We start our discussion by looking in more detail at the down type
quark and charged lepton mass matrices discussed previously in
Eqs.~\ref{gquarkmass} and \ref{glepmass}. This structure for the down
type quark and charged lepton mass matrices has been previously
discussed in several works
\cite{Morisi:2011pt,King:2013hj,Morisi:2013eca,Bonilla:2014xla}. In
this section for illustration purpose we first discuss the results
obtained in previous works by closely following the approach taken in
previous works like in \cite{Morisi:2009sc}. Subsequently, we will
generalize the analysis of previous works and discuss how the same
results can be obtained using a more general setup and more detailed
considerations.

We start from the charged lepton mass matrix obtained in
Eq.~\ref{glepmass}. The correct charged lepton masses are reproduced
if the vevs of the $A_4$ triplet fields $\Phi^d_i$ satisfy the
alignment limit $v^d(1, \epsilon_1, \epsilon_2)$, where
$v^d \gg \epsilon_1, \epsilon_2$. Then, in similar notation and spirit
as in Ref.~\cite{Morisi:2009sc}, we extract the three invariants of
the Hermitian matrix $S = M_l M_l^\dagger$: $Det(S)$, $Tr(S)$ and
$Tr(S)^2-Tr(S^2)$. We then compute their values in the diagonal
basis in terms of the charged lepton masses, $m_e$, $m_\mu$ and
$m_\tau$. The equations are
   \begin{eqnarray}
   \textnormal{Det} S &=& (m_e m_\mu m_\tau)^2 \nonumber \\
   \textnormal{Tr} S& =& m_e^2 + m_\mu^2 + m_\tau^2 \nonumber  \\
    (\textnormal{Tr} S)^2 - \textnormal{Tr} S^2 &=& 2m_e^2m_\mu^2 + 2m_e^2 m_\tau^2 + 2m_\mu^2 m_\tau^2
    \label{masseq}
  \end{eqnarray}
 
  The expressions in~Eq.\ref{masseq} can be readily solved in the
  vev alignment limit $v^d(1, \epsilon_1, \epsilon_2)$ discussed
  before.
  This amounts to the approximation 
  $$v_1^d \gg v_3^d~~~~~ {\rm and}~~~~~\dfrac{v_1^d}{v_2^d} \gg \dfrac{y_1^l+y_2^l}{y_1^l-y_2^l} \gg 1,$$
or equivalently  $$r \gg \alpha~~~~~{\rm and}~~~~~
  r \gg \dfrac{b_l}{a_l} \gg 1.$$
  The solutions for $r$, $a_l$ and $b_l$ are given as
\begin{eqnarray}
   r & \,\,=\,\, & \frac{m_\tau}{\sqrt{m_e m_\mu}} \sqrt{\alpha}  \label{rleq} \\
   a_l & \,\,=\,\, & \frac{m_\mu}{m_\tau} \sqrt{\frac{m_e m_\mu}{\alpha}}  \label{aleq} \\
    b_l & \,\,=\,\, & \sqrt{\frac{m_e m_\mu}{\alpha}}  \label{bleq}
\end{eqnarray}

Owing to the $A_4$ symmetry, the charged lepton mass matrix in
Eq.~\ref{glepmass} and down type quark mass matrix in
Eq.~\ref{gquarkmass} have the same structure. As a result the down
quark mass matrix can also be decomposed using equations analogous to
Eq.~\ref{masseq}. For down type mass matrix of \ref{gquarkmass} we
obtain
\begin{eqnarray}
r & = & \frac{m_b}{\sqrt{m_d m_s}} \sqrt{\alpha}  \label{rdeq} \\
a_d & = & \frac{m_s}{m_b} \sqrt{\frac{m_d m_s}{\alpha}}  \label{adeq} \\
b_d & = & \sqrt{\frac{m_d m_s}{\alpha}}  \label{bdeq}
\end{eqnarray}
 Note that the parameters $\alpha$ and $r$ are common for both the
 charged lepton sector as well as in the down-type quark sector, as
 they are simply ratios between vevs of the fields $\Phi^d_i$. Thus
 comparing Eqs.~\ref{rleq} and \ref{rdeq} we obtain the following mass
 relation
   \begin{eqnarray} \label{massrelation}
\frac{m_\tau}{\sqrt{m_e m_\mu}} & =  & \frac{m_b}{\sqrt{m_s m_d}}
  \end{eqnarray}

  The procedure sketched above can be performed in a more general way
  by solving the equations numerically. The relevant equations for the
  case of charged leptons are
\begin{eqnarray} \label{massequations}
(m_e m_\mu m_\tau)^2 & = & {a_l}^6r^2 \alpha^2 + 2 {a_l}^3 b_l^3 r^2 \alpha^2+b_l^6 r^2 \alpha^2 
  \nonumber \\ 
m_e^2 + m_\mu^2 + m_\tau^2 & = & ({a_l}^2 + b_l^2)(1+r^2+\alpha^2)  \nonumber \\
 2m_e^2m_\mu^2 + 2m_e^2 m_\tau^2 + 2m_\mu^2 m_\tau^2  & = &
({a_l}^2 + b_l^2)^2(1+r^2+\alpha^2)^2 -({a_l}^2 +b_l^2r^2)^2   \\
& &- (b_l^2+{a_l}^2 \alpha^2)^2
 - ({a_l}^2 r^2+ b_l^2 \alpha^2)^2 
\end{eqnarray}
Taking as input parameters the best fit values (at $M_Z$ scale) for
the charged lepton masses~\cite{Xing:2011aa} and imposing
$r > \frac{b}{a}$, there is a one-parameter family of solutions to
these equations.  These are related to the approximate solution
described before.
  We build the functions $r(\alpha)$, $a_l(\alpha)$ and $b_l(\alpha)$
  taking $\alpha$ as a free parameter. 
  In the correct range for the parameter $\alpha$, the unique solution
  is found to be near the limit $r \gg \dfrac{b}{a} \gg 1$ and
  therefore it again leads to the mass relation in
  Eq.~\ref{eq:b-tau}. Since the $(\alpha, r, a_l, b_l)$ are solutions
  of Eqs.~\ref{massequations}, the charged lepton masses are fitted
  exactly to their best-fit values.
  In order to underpin the relevant solution for down type quark
  masses, we also need to take into account not only the mass relation
  \ref{eq:b-tau} and the charged lepton masses, but also the
  constraints for the experimental measurements (along with
  renormalization group evolution to $M_Z$ scale) of all the down-type
  quark masses \cite{Xing:2011aa}. Here, we will impose the rather
  stringent 1-$\sigma$ bounds\footnote{Imposing 1-$\sigma$ is in
    fact rather stringent, and can easily be relaxed to a more
    conservative criterium e.g. 3-$\sigma$. We have deliberatively
    imposed the stringent 1-$\sigma$ bound in other to highlight the
    high precision obtained from our results. } on the down-type
quarks masses at $Z$ boson energy scale \cite{Xing:2011aa}.
    
Then, for each valid $( \alpha, r, a_l, b_l,)$ we take $a_q$ and $b_q$ as 
   \begin{eqnarray}
   a_q = \frac{m_s}{m_b} \sqrt{\frac{m_d m_s}{\alpha}} (1+\epsilon_1) \\
   b_q = \sqrt{\frac{m_d m_s}{\alpha}}(1+\epsilon_2)~,
  \end{eqnarray}
  where $\epsilon_1$ and $\epsilon_2$ are expected to be small.
  
\begin{figure}[!h]
 \centering
  \includegraphics[scale=0.75]{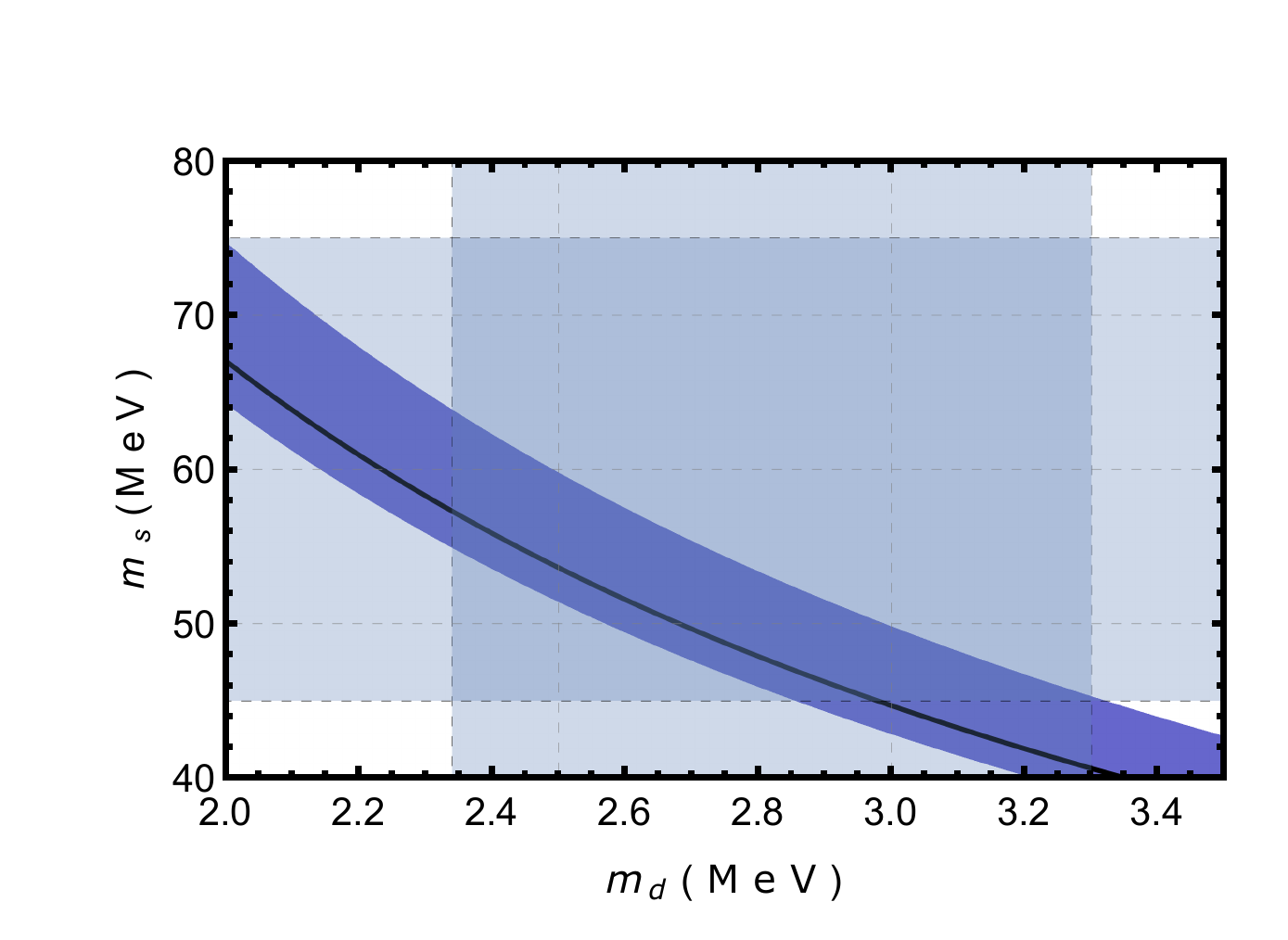}
  \caption{Prediction for the $s$ and $d$ quark masses (at $M_Z$
    scale) in our model. The dark blue area is the allowed region from
    our model for the $s$ and $d$ quark masses, while varying the mass
    of the $b$ quark in its 1-$\sigma$ range. The light blue area is
    the allowed 1-$\sigma$ range (at $M_Z$ scale) for the mass of the
    quarks $s$ and $d$ \cite{Xing:2011aa}.}
  \label{fig:msvsmd}
  \end{figure}

  Using this procedure gives sets of parameters
  $(\alpha, r, a_l, b_l, a_q, b_q)$ which give at the same time best
  fit values for charged lepton masses, down-type quarks inside the
  $1\sigma$ range and the mass relation in Eq.~\ref{eq:b-tau}.  In
  figure \ref{fig:msvsmd} we show the family-dependent bottom-tau mass
  prediction of our model for the s and d masses, along with their
  allowed 1-$\sigma$ ranges.

\subsubsection{The charged piece of the lepton mixing matrix $U_l$ }
\label{cul}

The charged lepton mass matrix Eq.~\ref{glepmass} not only leads to
correct lepton masses but also to non-trivial charged lepton rotation
matrix $U_l$ as we discuss now. Just like the neutrino mass matrix,
the charged lepton mass matrix can also be diagonalized by a
bi-unitary transformation as
\begin{eqnarray}
U_l^\dagger M_l V =\mathrm{diag}(m_e, m_\mu, m_\tau) 
\end{eqnarray}

The charged lepton mixing matrix $U_l$ in standard
parameterization can be written as
\begin{eqnarray}
 U_l = P U_{23} (\theta^l_{23}, 0) U_{13} (\theta^l_{23}, \delta^l) U_{12} (\theta^l_{12}, 0) P'
\end{eqnarray}
where $P$ and $P'$ are diagonal matrix of phases and $U_{ij}$ is the
usual complex rotation matrix appearing in the symmetric parametrization of fermion mixing given in~\cite{Schechter:1980gr}, e.g.
\begin{eqnarray}
  U_{12} (\theta_{12}, \delta) =  \, \left( 
\begin{array}{ccc}
\cos \theta_{12}      & e^{-i\delta} \sin \theta_{12}            &  0           \\
-e^{i\delta}\sin \theta_{12}     & \cos \theta_{12} &  0   \\
0     & 0    &  1     \\  
\end{array}
\right).
   \label{magmat}
  \end{eqnarray}
 with an analogous definitions for $U_{13}$ and $U_{23}$.  For the
  charged lepton mass matrix \ref{glepmass}, we find that
\begin{eqnarray}
\sin \theta^l_{12} & = & \sqrt{\frac{m_e}{m_\mu}} \frac{1}{\sqrt{\alpha}}+ \mathcal{O}(\frac{1}{\alpha^2}) \approx \mathcal{O} (\lambda_C) \nonumber \\
\sin \theta^l_{13} & = & \frac{m_u}{m_\tau^2} \sqrt{m_e m_\mu} \frac{1}{\sqrt{\alpha}} + \mathcal{O}(\frac{1}{\alpha^2}) \approx \mathcal{O} (10^{-5}) \nonumber \\
\sin \theta^l_{23} & = & \frac{m_e m_\mu^2}{m^3_\tau} \frac{1}{\alpha} + \mathcal{O}(\frac{1}{\alpha^2}) \approx \mathcal{O} (10^{-7})
 \label{leprot}
\end{eqnarray}
Where $\lambda_C \approx 0.22$ is the sine of the Cabbibo angle. In order to reproduce
adequate values for the CKM matrix elements we may introduce a
vector-like quark mixing with the up-type quarks, along the lines
followed recently in \cite{Bonilla:2016sgx}.

The diagonal phases in $P$ and $P'$ are all are exactly $0$ except for one
which is $\pi$. Performing the numerical computation reconfirms the
results obtained in Eq.~\ref{leprot} for the charged lepton mass
matrix i.e. $\theta^l_{12}$ is finite and its value depends on the
value of $\alpha$ in an inverse way, while $\theta^l_{13}$ and
$\theta^l_{23}$ are both negligible (in particular,
$\theta^l_{13} \sim 10^{-5}$ and $\theta^l_{23} \sim 10^{-7}$.  Then,
the charged lepton mixing matrix for our model is given as
\begin{eqnarray}
U_l \approx \left(\begin{matrix}
\cos \theta^l_{12} & \sin {\theta^l_{12}} & 0 \\
-\sin {\theta^l_{12}} & \cos \theta^l_{12} & 0 \\
0 & 0 & 1
\end{matrix} \right) \cdot 
\left(\begin{matrix}
-1 & 0 & 0 \\
0 & 1 & 0 \\
0 & 0 & 1
\end{matrix} \right) 
\end{eqnarray}
Thus in our model the lepton mixing matrix
$U_{LM} = U^\dagger_l U_\nu$ receives significant charged lepton
corrections which have interesting phenomenological consequences as we
discuss in next section.

\subsubsection{The lepton mixing matrix and neutrino mass ordering}
\label{lm}

As mentioned before in Section \ref{sec:model}, the light neutrino
mass matrix in Eq.~\ref{gnumass} leads to the neutrino mixing matrix
$U_\nu$ which in standard parameterization~\cite{Schechter:1980gr}
leads to
\begin{eqnarray}
U_\nu  =  P U_{23}\left(\rfrac{\pi}{4}, 0\right)& U_{13}\left( \theta_{13}^\nu, \rfrac{\pi}{2}\right)& U_{12} \left( \theta_{12}^\nu, 0\right) P'
\label{rnu}
\end{eqnarray}
As mentioned before, owing to the $A_4$ symmetry, we have that
$\theta_{23}^\nu = 45^{\circ}$ and $\delta^\nu_{CP} = 90^{\circ}$ for
both types of mass ordering: normal ordering (NO) or inverted
ordering (IO).  
Since neutrinos in our model are Dirac fermions, the phases in the
right in Eq.~\ref{rnu} i.e.  $P'$, are unphysical, while
$\theta_{13}^\nu$ and $\theta_{12}^\nu$ are strongly correlated
between each other. 
This result is completely general and follows from the $A_4$ symmetry,
independently of the mass ordering, NO or IO. However, the behavior
of the correlation between $\theta_{12}^\nu$ and $\theta_{13}^\nu$
does depend on the choice of NO or IO.

\begin{figure}[!h]
 \centering
  \includegraphics[scale=0.505]{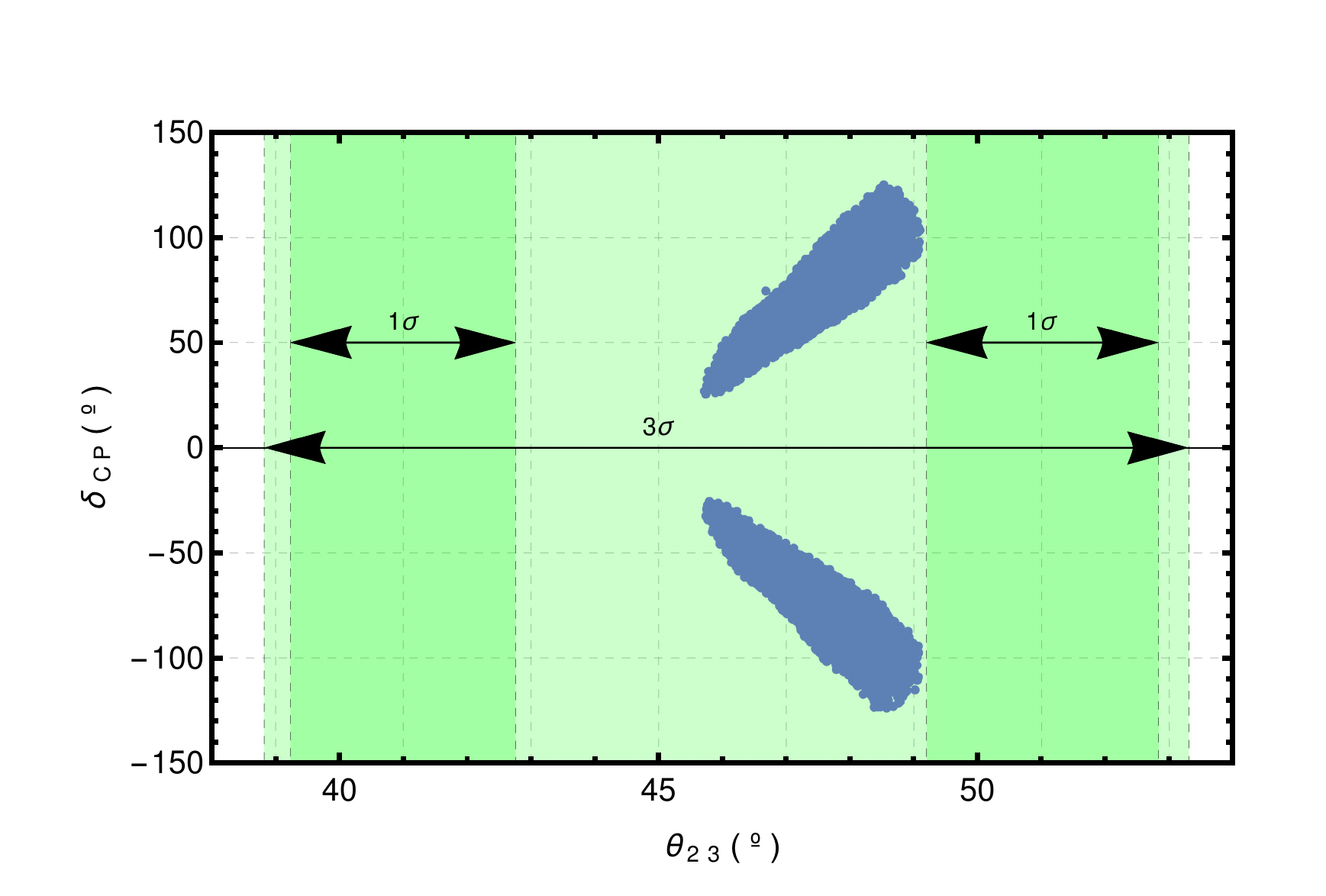}
  \includegraphics[scale=0.48]{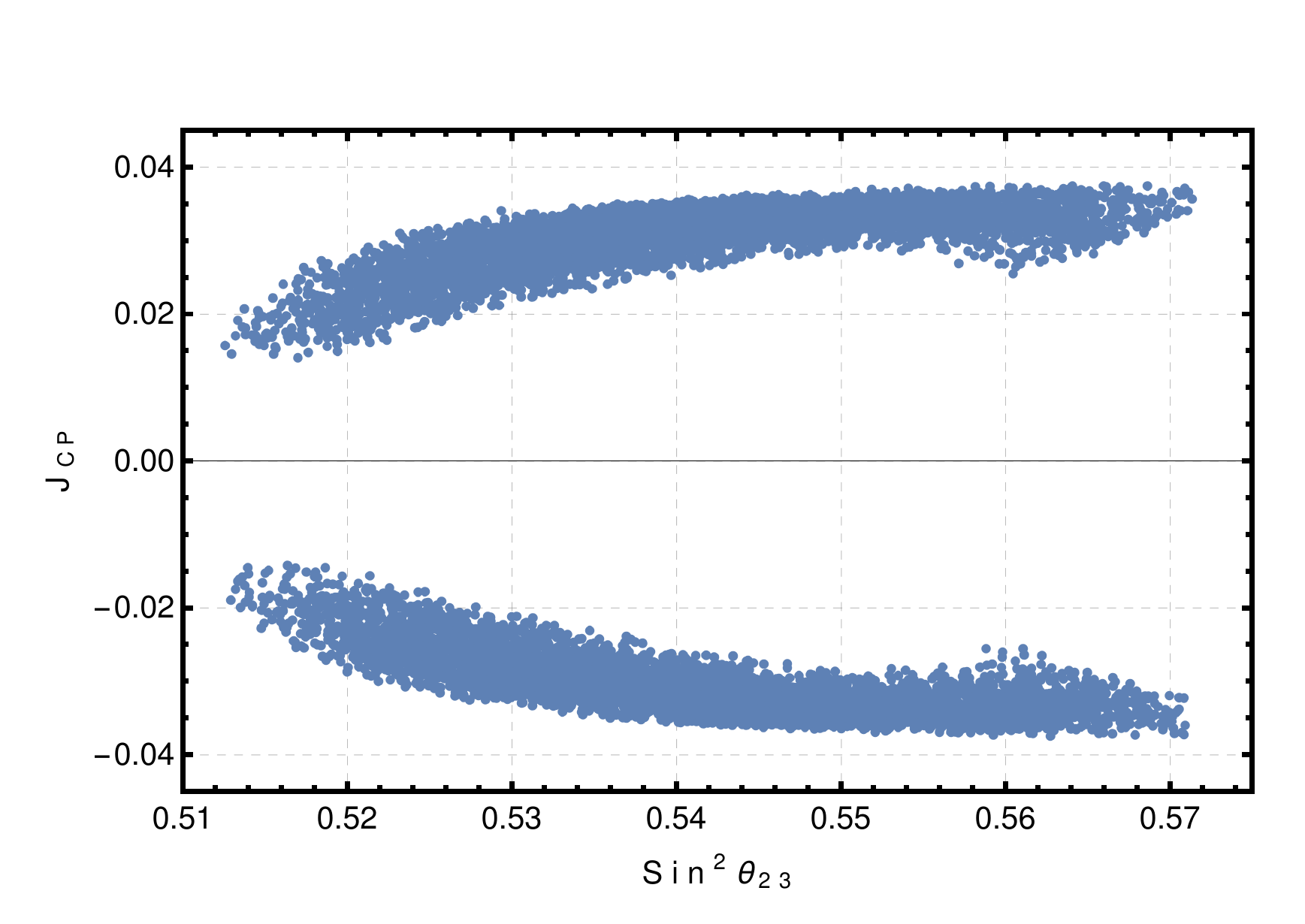}
  \caption{CP violation and $\theta_{23}$ predictions within the
    model. Left panel: $\delta_{CP}$ vs $\theta_{23}$. The green
    regions are the $1\sigma$ (dark) and $3\sigma$ (light) regions for
    $\theta_{23}$ from current oscillation fit. Right panel: Same
    correlation, now showing $J_{CP}$ vs $\sin^2 \theta_{23}$ and
    zooming in the region allowed by the model, fully consistent in
    the $2\sigma$ experimental range.}
    \label{fig:theta23vsdelta}
  \end{figure}

Taking into account the results in the previous sections, the lepton
mixing matrix is
\begin{eqnarray}
U_{LM} = U_l^\dagger U_\nu = U_{12}\left(\theta^l_{12}, 0\right)^\dagger  \cdot U_{\nu}
\end{eqnarray}
One can regard the matrix $U_l$ as a correction to the neutrino
mixing parameters obtained just by diagonalizing the neutrino mass
matrix.
For the NO case, the angle $\theta_{12}^l$ has to be big enough
($\sim >15 ^{\circ}$) so as to account for the correct mixing angles
of the lepton mixing matrix, but at the same time it has to remain
controlled ($< 20^{\circ}$) otherwise the down-type quark masses
cannot be fitted. This means that the parameter $\alpha$ has to be
between $0.04$ and $0.08$ in order to fit in the $3 \sigma$ allowed range of the neutrino oscillation parameters.
This lepton mixing matrix can fit the neutrino oscillation parameters
within $3 \sigma$ at the same time as the mass matrices fit the
down-type quarks and the neutrino squared mass differences in the
$1\sigma$ range and the charged lepton masses to their best fit value. Once the lepton
mixing matrix is written in the standard parametrization, two
interesting features arise. On the one hand, $\theta_{23}>45^{\circ}$
and, on the other, a strong correlation appears between the
atmospheric angle $\theta_{23}$ and $\delta_{CP}$, as shown in figure
\ref{fig:theta23vsdelta}.
  
  For IO, a different scenario arises. As in the case for NO, lepton
  corrections cannot be very big otherwise the down-type quark masses
  will not be fitted. However, in this case the structure of the correlation
  between $\theta_{12}^\nu$ and $\theta_{13}^\nu$ implies that for
  allowed charged lepton corrections, the reactor angle $\theta_{13}$
  is always outside the $3\sigma$ allowed range. Note that the model
  does not include any a priori theoretical bias in favour of normal
  ordering but it is a prediction of the model once one impose
  experimental constraints.


To summarize, we have proposed a $A_4 \otimes Z_4 \otimes Z_2$ flavour extension of
the \sm with naturally small Dirac neutrino masses. Our lepton
quarticity symmetry simultaneously forbids Majorana mass terms and
provides dark matter stability. The flavour symmetry plays a multiple
role, providing : 
\begin{itemize}
 \item A generalized family-dependent bottom-tau mass relation, Eq.~(\ref{eq:b-tau}) and Fig.~\ref{fig:msvsmd}.
 \item A natural realization of the type-I seesaw mechanism for Dirac neutrino masses, as the tree level Dirac Yukawa term between left and right handed neutrinos is forbidden.
 \item A very predictive flavour structure to the lepton mixing matrix. The latter directly correlates the CP phase $\delta_{CP}$ and the atmospheric angle $\theta_{23}$, as shown in Fig.~\ref{fig:theta23vsdelta}. CP must be significantly violated in neutrino oscillations, and the atmospheric angle $\theta_{23}$ lies in the second octant.
(v) Only the normal neutrino mass ordering is consistet at $3 \sigma$.

\end{itemize}

Our approach provides an adequate pattern of neutrino mass and mixing
as well as a viable stable dark matter.
This is achieved while providing testable predictions concerning the
currently most relevant oscillation parameters, the atmospheric angle
$\theta_{23}$ and the CP phase $\delta_{CP}$, as well as a successful
family generalization of bottom-tau unification, despite the absence
of an underlying Grand Unified Theory.
Our lepton quarticity approach can also potentially lead to other interesting
phenomena such as neutrinoless quadruple beta decay ($0 \nu 4 \beta$),
which has recently been searched for by the NEMO
collaboration~\cite{NEMO-3:2017gmi}. The intimate connection between
the Dirac nature of neutrinos and dark matter stability constitutes a
key feature of our model. Other phenomenological implications will be
taken up elsewhere.

 \section{Charged Lepton Flavour Violation}
 \label{sec:LFV}
 
Lepton Flavour Violation (LFV) processes such as $l_\alpha \rightarrow l_\beta \, \gamma$ or $l_\alpha \rightarrow 3 l_\beta$ and closely related quantities like the lepton anomalous magnetic moments have received a lot of attention for a few decades. In particular, a long standing discrepancy between the measured muon anomalous magnetic moment \cite{Muong-2:2006rrc, Muong-2:2021ojo} and the SM calculated prediction \cite{Aoyama:2012wj, Aoyama:2012wk, Laporta:2017okg, Aoyama:2017uqe, RBC:2018dos} hints towards some unknown new physics that can explain the SM deviation or a need to improve the already very precise SM calculations, in particular hadronization contributions. On the other hand, the SM with massless neutrinos predicts exactly a $0$ rate for LFV processes \footnote{Some extensions of the SM still generate sizeable LFV processes even in the massless limit \cite{Dittmar:1989yg, Gonzalez-Garcia:1990sbd}}, but models with massive neutrinos need to factor in stringent experimental upper bounds \cite{SINDRUM:1987nra,BaBar:2009hkt,MEG:2016leq}. Most importantly, models that try to explain the muon or electron anomalous magnetic moment will typically generate LFV. The bibliography on this topic is vast and we direct the reader to the existing reviews and references within for more complete information \cite{Jegerlehner:2009ry, Lindner:2016bgg,Cei:2020gbd,Aoyama:2020ynm, Keshavarzi:2021eqa}. In this section we show how a class of low scale Dirac neutrino mass models can generate observable LFV rates. The fact that LFV can proceed even in the limit of massless neutrinos was noted a long time ago~\cite{Bernabeu:1987gr,Branco:1989bn,Rius:1989gk} and hence it can be probed in experiment.
It was also proposed that in inverse seesaw schemes, the mediators of neutrino mass generation can be searched for at high energy colliders~\cite{Dittmar:1989yg,Gonzalez-Garcia:1990sbd}.
The latest restrictions come for the ATLAS and CMS experiments~\cite{CMS:2018iaf}. The phenomenology of charged LFV can be probed at high energies and also high intensity 
facilities~\cite{Gonzalez-Garcia:1991brm,Ilakovac:1994kj,Deppisch:2004fa,Deppisch:2005zm,DeRomeri:2016gum}. 

Here we present an illustration showing that, as expected, similar result holds for inverse Dirac seesaw models.
  
 As a paradigmatic example of a LFV process in a low scale Dirac neutrino model, let us estimate the $\mu^- \rightarrow e^- \gamma$ decay rate in the simplest Dirac inverse seesaw, described in Sec.~\ref{sec:simplestdirac} and \cite{CentellesChulia:2020dfh}. In this simple scenario the neutrino mass for the one generation limit is given by
 
 \begin{equation}
  m_\nu = Y_\phi Y_\chi \frac{v \, u}{M}
 \end{equation}
 
 In order to get $m_\nu \sim 0.1$ eV with Yukawas of order $Y_i \sim 1$ and $v \sim 100$ GeV we find the relation $u/M \sim 10^{-12}$. In the inverse seesaw limit we take $u \ll v$ and $M$ to be not far from the electroweak scale, for example $M \sim 1$ TeV and $u \sim 1$ eV. The $6\times 6$ lepton mixing matrix is given by
 
 \begin{equation}
  U_L = \left( \begin{matrix}U_l & U_\text{mix} \\ U_\text{mix}^\dagger & U_{\text{heavy}L} \end{matrix}\right)
  \label{eq:mixing}
 \end{equation}
 
 Where $U_l$ is the light neutrino mixing matrix, measured in oscillation experiments, while $U_\text{mix} \sim \mathcal{O}(v/M)$ is the mixing matrix between the light and heavy states. Since oscillation experiments yield no evidence for the non-unitarity of $U_l$ \cite{Miranda:2019ynh}, this mixing must be small.
 
 In the scalar sector of the model, there will be a tree level mixing between the physical scalars. The Lagrangian terms $-\mu_\phi^2 (\phi^\dagger \phi) + \lambda (\phi^\dagger \phi)^2 -\mu_\chi^2 (\chi^\dagger \chi) + \lambda_\chi (\chi^\dagger \chi)^2  + \lambda_\text{mix} (\phi^\dagger \phi)(\chi^\dagger \chi)^2 $ lead to a scalar mass matrix after spontaneous symmetry breaking of the form
 
 \begin{equation}
  m_h = \left(\begin{matrix}
\lambda v^2 & \lambda_{mix} u\, v    \\           
\lambda_{mix} u\, v               & \lambda_\chi u^2              \end{matrix}
 \right)  
 \end{equation}
 
 Which can be diagonalized by a standard $2\times 2$ rotation matrix of angles
 
 \begin{equation}
  \tan 2\alpha = \frac{\lambda_\text{mix} u v}{\lambda v^2 + \lambda_\chi u^2} \approx \frac{\lambda_\text{mix}}{\lambda} \frac{u}{v} \sim 10^{-11}
 \end{equation}

\begin{figure}
\begin{center}
	\includegraphics[scale=0.5]{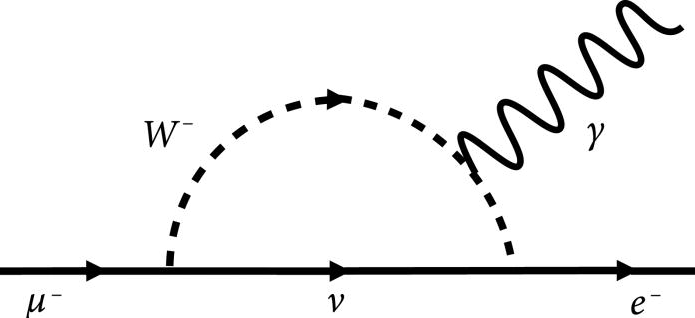}, 
	\caption{Leading contribution to $\mu \rightarrow e \, \gamma$ in the Dirac inverse seesaw model. Note that the neutrino mass Eigenstates, including the heavy neutral states, are running inside the loop. Therefore the amplitude of this process is dependent on the particular neutrino mass model under study, but independent on the neutrino nature.}
	\label{fig:Wmuegamma}
\end{center}
\end{figure}

 Owing to the small mixing, the heavy scalar $h$ will be be SM like and thus $m_h \sim 125 $GeV. The lightest scalar $H$, on the other hand, will have a mass of order $u = 1$ eV. The coupling of $H$ with the electrons will come at tree level via the term $\bar{L} \phi e_R$ and will be proportional to $\sin \alpha$, the mixing between $h$ and $H$, and will therefore be small enough to safely escape constraints like those coming from stellar cooling \cite{Escribano:2020wua}. 
 The mixing matrix of the pseudoscalar part is simply a matrix of $0$s. The pseudoscalar doublet will be 'eaten' by the $Z^0$ boson and the singlet will remain as 'pure' singlet without mixing. Therefore, the coupling between electrons and the massless pseudoscalar $D$, the Goldstone boson of the theory or the `Diracon' \cite{Bonilla:2016zef}, will come at the loop level and proportional to the mixing between the light and heavy neutrinos $\sim v/M$. Therefore, processes like $\mu^- \rightarrow e^- H$ are strongly suppressed. The leading LFV processes will be $\mu^- \rightarrow e^- D$ and $\mu^- \rightarrow e^- \gamma$ which both happen at the one loop level via exchange of a $W$ boson. We will concentrate on the latter, depicted in Fig.~\ref{fig:Wmuegamma}.

The amplitude $G_{\mu e}$ can be calculated as \cite{Ilakovac:1994kj, Deppisch:2004fa}

\begin{equation}
 G_{\mu e} = \sum_k \left(U_L \right)^*_{ek} (U_L)_{e k}) G_\gamma \left(\frac{M_{N_k}^2}{M_W^2}\right), \hspace{1cm} G_\gamma(x) = -\frac{2x^3+5x^2-x}{4(1-x)^3} - \frac{3x^3}{2(1-x)^4}
\end{equation}

As a proof of concept, let us for now perform a rough approximation to estimate the orders of magnitude of the magnitued involved. Since the light neutrino states are small, $m_\nu / M_W \rightarrow 0$ and $G_\gamma(0) \rightarrow 0$. Therefore we will neglect the light neutrino contribution. If the heavy states have a mass of the order TeV, then $M_N \gg M_W$ and $G_\gamma(x) \rightarrow 1/2$. On the other hand, the mixing between heavy and light states given by Eq.~\ref{eq:mixing} is of order $(U_L)_{e k} \sim Y v/M_N \sim  10^{-2}$ for $k \in \{4, 5, 6\}$. This yields an amplitude of order 

\begin{equation}
 G_{\mu e} \sim \frac{3}{2} \left(\frac{Y v}{M_N}\right)^2 \sim 10^{-3}
\end{equation}
In order to compare with experimental constraints we must compute the $\mu \rightarrow e \, \gamma$ branching ratio given by

\begin{equation}
\Gamma(\mu \rightarrow e \, \gamma) = \frac{ \alpha_W^3 s_W^2}{256 \pi^2} \left(\frac{m_\mu}{M_W}\right)^4 \frac{m_\mu}{\Gamma_\mu}|G_{\mu e}|^2 \sim 10^{-12}
\end{equation}

by taking the PDG values \cite{ParticleDataGroup:2020ssz} and the result above. The 2016 experimental limit given by MEG \cite{MEG:2016leq} is

\begin{equation}
 \Gamma(\mu \rightarrow e \, \gamma) < 4.2 10^{-13}
\end{equation}

Thus showing that this class of models has the potential to predict high LFV rates, testable in the close future. A more detailed phenomenological work following these lines is ongoing, so we will not expand further.


\chapter{Resumen de la tesis}
En esta secci\'on realizaremos un resumen en español de la tesis junto a  la metodolog\'ia empleada, los resultados m\'as importantes y la contribuci\'on personal del estudiante a cada uno de los art\'iculos en los que se basa la tesis.

\section{Introducci\'on y objetivos}
El Modelo Est\'andar de interacciones Electro-D\'ebiles ha sido un gran \'exito desde un punto de vista tanto te\'orico como experimental. Si bien este \'exito no se puede negar, es hora de avanzar y abordar las preguntas que el Modelo Est\'andar deja sin respuesta, como las masas de neutrinos, la naturaleza de la materia oscura, el problema de la jerarqu\'ia o la no observaci\'on de la violaci\'on de CP en el sector de las interacciones fuertes, entre otros.

De hecho, el Modelo Est\'andar predice neutrinos sin masa. Sin embargo, los experimentos de oscilaci\'on de neutrinos muestran claramente que al menos dos neutrinos son masivos y arrojan luz sobre su patr\'on de mezcla. Estos experimentos son m\'ultiples, la evidencia es s\'olida y las mediciones son muy precisas, alcanzando el nivel de error menor al 1\% en algunos observables. Sin embargo, las oscilaciones de neutrinos no pueden obtener informaci\'on acerca de cuestiones m\'as profundas respecto a la naturaleza de los neutrinos: ¿cu\'al es la escala absoluta de las masas de neutrinos? Las oscilaciones de neutrinos son sensible solo a las diferencias cuadr\'aticas de las masas de los autoestados masivos. ¿Su naturaleza es Dirac o Majorana? No hay diferencia en el fen\'omenos de las oscilaciones en uno u otro caso. Se necesitan observaciones complementarias para responder a estas preguntas, y ya se est\'a haciendo con experimentos en funcionamiento en el presente y tambi\'en experimentos proyectados para el futuro.

Tras una breve introducci\'on sobre el modelo est\'andar, las oscilaciones de neutrinos y los mecanismos de generaci\'on de masa de neutrinos de Majorana, el mecanismo 'seesaw', comenzamos la tesis dando una definici\'on general de un fermi\'on de Dirac: un fermi\'on cuya antipart\'icula tiene la misma masa pero cargas de signo contrario. Un ejemplo paradigm\'atico ser\'ia el electr\'on, cuya antipart\'icula de id\'entica masa y carga positiva, el positr\'on, fue descubierta experimentalmente en 1932. En realidad, todos los fermiones cargados conocidos hasta el momento (electrones, quarks y sus versiones pesadas de distintas generaciones) son part\'iculas de Dirac y debido a su carga el\'ectrica y/o de color no pueden ser de otra manera. Sin embargo, en esta descripcci\'on hemos dejado de lado a los neutrinos: el \'unico fermi\'on neutro cuya existencia conocemos. Puesto que el neutrino es neutro bajo el grupo gauge conservado, el grupo del electromagnetismo multiplicado por el grupo de color, surge una posibilidad adicional: que los neutrinos sean de tipo Majorana, es decir, que dicha part\'icula sea id\'entica a su antipart\'icula. Aunque solamente pueden distinguirse estas posibilidades experimentalmente, la comunidad suele asumir que son de Majorana debido a su neutralidad.

Desde el punto de vista te\'orico, est\'a claro que el Modelo Est\'andar debe ampliarse para incluir masas de neutrinos. La forma m\'as sencilla de hacerlo es agregar el compañero diestro del campo de neutrinos. En ese caso, si no se ampl\'ia el inventario de simetr\'ia del Modelo Est\'andar, se viola el n\'umero lept\'onico en dos unidades y los neutrinos resultan ser de tipo Majorana. Este es el llamado mecanismo del balanc\'in (‘seesaw’) tipo I, que explica de manera elegante las masas de neutrinos y su pequeñez con respecto a las dem\'as escalas de masas del Modelo Est\'andar. Se han desarrollado muchos mecanismos de masas de neutrinos diferentes, incluidas m\'ultiples variantes del mecanismo del balanc\'in (tipo II, tipo III, tipo lineal, tipo inverso...), as\'i como modelos radiativos. Sin embargo, la comunidad generalmente asume que los neutrinos son campos de Majorana, y la opci\'on de que sean de Dirac ha sido relativamente poco estudiada.

El objetivo de esta tesis ser\'a centrarse en la posibilidad, abierta experimentalmente, de que los neutrinos sean part\'iculas de Dirac. En la tesis damos los requisitos generales de simetr\'ia necesarios para tener neutrinos de Dirac.  En este sentido, es inmediato ver que el patr\'on de ruptura del n\'umero lept\'onico es un concepto central para determinar la naturaleza de los neutrinos. Sin embargo, es importante tener en cuenta que la naturaleza Dirac de los neutrinos se puede imponer con una simetr\'ia distinta del n\'umero lept\'onico, incluso utilizando grupos no abelianos. Esta simetr\'ia, sea una variante del n\'umero lept\'onico o alguna otra simetr\'ia a priori no relacionada, puede desempeñar al mismo tiempo diferentes roles: puede ser la simetr\'ia de Peccei-Quinn, estabilizar la materia oscura, explicar la jerarqu\'ia de masa de fermiones del Modelo Est\'andar, etc.

Despu\'es procedemos a revisar el zool\'ogico de mecanismos del balanc\'in (‘seesaw’) para neutrinos de Dirac. An\'alogamente a los mecanismos conocidos en la literatura para neutrinos de Majorana, estos son mecanismos elegantes que pueden explicar naturalmente la pequeñez de las masas de neutrinos. Algunos de ellos requieren de mediadores a escalas muy altas, lejos del alcance de los colisionares de part\'iculas actuales, pero otros, como los mecanismos del balac\'in inversos o pueden tener mediadores de escala TeV y fenomenolog\'ia observable.

Alternativamente, la pequeñez de las masas de neutrinos puede explicarse por su origen radiativo si se generan a trav\'es de bucles cu\'anticos. Es en este marco que la conexi\'on de la simetr\'ia de n\'umero lept\'onico (o m\'as precisamente, la combinaci\'on n\'umero bari\'onico menos numero lept\'onico) con la materia oscura se vuelve m\'as clara: esta simetr\'ia puede al mismo tiempo estabilizar la materia oscura, que corre en el bucle que da masa a los neutrinos, protege la naturaleza Dirac de los mismos y explica la pequeñez de su masas al prohibir el t\'ermino de masa a nivel de \'arbol, adem\'as de permitir ser una simetr\'ia local al estar libre de anomal\'ias. Esta es la contribuci\'on m\'as importante y elegante de la tesis.

Finalmente, para contrarrestar la noci\'on generalizada en la comunidad de que los modelos de masas de neutrinos no tienen fenomenolog\'ia testable en experimentos, mostramos dos modelos de contraejemplo. 
En uno de ellos utilizamos una simetr\'ia de sabor encima de un mecanismo de balanc\'in de Dirac de tipo I. Mientras que el mecanismo del balanc\'in de tipo I tiene mediadores de masas mucho m\'as elevadas que la escaal electrod\'ebil, y por tanto inaccesibles en experimentos de colisionadores, es importante hacer notar que esta caracter\'istica es compartida por el mecanismo tipo I de Majorana. Sin embargo, al imponer simetr\'ias de sabor extra obtenemos predicciones  muy fuertes en este sector: una relac\'on entre las masas de los quarks y los leptones, una correlaci\'on muy fuerte entre los par\'ametros de mezcla de los neutrinos y una predicci\'on para el ordenamiento de sus masas, que deber\'a ser ordenamiento normla, excluyendo el ordenamiento inverso. 
El segundo modelo de ejemplo es un mecanismo de balanc\'in inverso. Puesto que en este tipo de modelos la escala de los mediadores puede reducirse hasta la escala electrod\'ebil, es posible obtener predicciones testables en colisionadores. Adem\'as, al encontrarse la nueva f\'isica en una escala relativamente baja (orden TeV) habr\'a procesos visibles a bajas energ\'ias como por ejemplo procesos de violaci\'on del n\'umero lept\'onico de sabor. En esta secci\'on describimos un modelo de esta clase y estimamos las ratios de decaimiento de un mu\'on en un electr\'on y un fot\'on, proceso que no deber\'ia darse en el modelo est\'andar pero que diversos experimentos est\'an buscando.

\section{Metodolog\'ia}
El estudio te\'orico en profundidad de la posibilidad de que los neutrinos sean part\'iculas de Dirac requiere de un conocimiento adecuado de ciertas herramientas tanto te\'oricas como computacionales. En concreto, una lista incompleta de las herramientas de software necesarias son:

\begin{itemize}
 \item Wolfram Mathematica para ayudar con manipulaciones anal\'iticas de expresiones y ecuaciones matem\'aticas. Algunos de los paquetes que hemos utilizado est\'an integrados en Mathematica, como Susyno, Sym2Int, Sarah, SPheno y MicrOMEGAS. Usaremos los dos primeros para facilitar las partes de simetr\'ia de construcci\'on de modelos de nuestros proyectos y los tres \'ultimos para calcular los observables de la f\'isica de part\'iculas a partir de ellos: secciones transversales, masas f\'isicas, relaciones entre par\'ametros, abundancia de reliquias de materia oscura, etc.
 
 \item Python, ya que algunos de los paquetes usan este lenguaje. En particular, CosmoTransitions se basa en Python y se utiliza para calcular informaci\'on relacionada con las transiciones de fase de un modelo de f\'isica de part\'iculas determinado. Esto es clave para determinar si un modelo contar\'a con una transici\'on de fase de primer orden, qu\'e tan fuerte ser\'a y nos permitir\'a calcular si las ondas gravitacionales producidas ser\'an observables o no en experimentos futuros.

 \item GAP es un programa muy poderoso para realizar c\'alculos de teor\'ia de grupos. Esto es extremadamente \'util en el an\'alisis de simetr\'ia CP, as\'i como en la construcci\'on de modelos cuando la simetr\'ia es complicada.
\end{itemize}

Por otro lado, ser\'a necesario tener conocimientos avanzados en, entre otros, 

\begin{itemize}
 \item Mec\'anica cu\'antica. La mec\'anica cu\'antica es la teor\'ia fundamental que describe las interacciones de part\'iculas subat\'omicas a una escala microsc\'opica cuando las energ\'ias involucradas son bajas. 
 
 \item Relatividad especial. La relatividad especial es el marco te\'orico en el que se describen los fen\'omenos que implican velocidades altas, cercanas a la velocidad de la luz.

 \item Teor\'ia cu\'antica de campos. La teor\'ia cu\'antica de campos unifica la relatividad especial y la mec\'anica cu\'antica, es decir, que es el marco te\'orico adecuado para describir las interacciones entre part\'iculas a altas energ\'ias.

 \item Teor\'ia de grupos, que es la hierramienta matem\'atica necesaria para predecir diversos fen\'omenos f\'isicos gracias al teorema de Noether.

\item Cosmolog\'ia, que estudia la evoluci\'on del universo y tiene relaci\'on directa con la f\'isica de part\'iculas, en particular con la f\'isica de neutrinos y la materia oscura, adem\'as de con transiciones de fase en sectores de Higgs extendidos.
\end{itemize}

Adem\'as, ser\'a importante consultar la bibliograf\'ia especializada en cada tema de inter\'es.

\section{Resultados y contribuci\'on personal}

 En esta secci\'on resumiremos los resultados m\'as importantes en cada uno de los art\'iculos que forman parte de la tesis, adem\'as de la contribuci\'on personal del estudiante en el desarrollo de los mismos, que ha sido sustancial en todos los casos. Por supuesto, todos los proyectos son colaborativos y los dem\'as coautores tambi\'en aportaron una parte importante en el desarrollo de las ideas subyacentes, los c\'alculos, la interpretaci\'on y la escritura del manuscrito. Aunque es dif\'icil especificar que tareas concretas realiz\'o cada persona (todos los coautores est\'an involucrados en todos o casi todos los pasos), intentaremos resaltar los puntos donde la contribuci\'on del estudiante fue m\'as destacada.
 
\subsection{Dirac Neutrinos and Dark Matter Stability from Lepton Quarticity}

En este art\'iculo \cite{CentellesChulia:2016rms} encontramos un mecanismo elegante para obtener masas de neutrinos pequeñas con lo que se conoce como 'par\'ametros naturales', es decir, sin la necesidad de ajustar los par\'ametros del modelo de forma muy fina. Esto se puede conseguir de forma muy sencilla con un mecanismo al `seesaw' tipo I, con la particularidad de que los neutrinos son de Dirac en vez de Majorana. Esto requiere la adici\'on al modelo de una pareja de fermiones pesados, con una masa por encima de la escala electrod\'ebil. La inversa de dicha masa genera la escala correcta de las masas de los neutrinos. A su vez, el modelo tiene algunas condiciones de simetr\'ia, como se argument\'o a lo largo de la tesis. Por un lado, es necesario prohibir el operador de dimensi\'on $4$ $\bar{L} \phi^c \nu_R$, ya que de existir dar\'ia la contribuci\'on principal a las masas de los neutrinos y ser\'ia necesario tomar 'a mano' un acoplamiento pequeño para dicho operador (orden $10^{-12}$). Esto lo conseguimos simplemente con una simetr\'ia $Z_2$, que a su vez est\'a espont\'aneamente rota por el singlete escalar $\chi$, que permite el operador de dimensi\'on $5$ $\bar{L} \phi^c \chi \nu_R$. Por otro lado, si queremos que los neutrinos sean de Dirac debe existir una simetr\'ia que proteja su naturaleza de Dirac prohibiendo los t\'erminos de Majorana como $\bar{\nu}_R^c \nu_R$. Esto lo conseguimos con un subgrupo de la simetr\'ia de n\'umero lept\'onico $Z_4$. La belleza de dicha simetr\'ia es que autom\'aticamente permite la existencia de un mecanismo para la estabilidad de la materia oscura.

La contribuci\'on personal del estudiante a este art\'iculo fue hallar los grupos de simetr\'ia m\'inimos necesarios para conseguir el objetivo expuesto y las cargas de cada campo bajo dichos grupos de simetr\'ia. Tambi\'en realiz\'o los c\'alculos necesarios en el sector escalar para encontrar los bosones f\'isicos, sus masas y sus mezclas, as\'i como el acoplamiento con la materia oscura via 'portal de Higgs' y los c\'alculos sencillos, adem\'as de contribuir a la redacci\'on del manuscrito.

\subsection{Generalized Bottom-Tau unification, neutrino oscillations and dark matter: predictions from a lepton quarticity flavour approach}

En \cite{CentellesChulia:2017koy} tomamos una idea similar al anterior art\'iculo, pero esta vez cambiando la simetr\'ia $Z_2$ por un grupo no abeliano: $A_4$. Al extender dicha simetr\'ia tambi\'en para los quarks obtenemos una serie de predicciones muy interesantes, en particular una relaci\'on entre las masas de los leptones y los quarks, $\frac{m_b}{\sqrt{m_d m_s}} = \frac{m_\tau}{\sqrt{m_e m_\mu}}$, fuertes correlaciones entre algunos \'angulos de mezcla (en concreto la mejor predicci\'on se da entre $\theta_{23}$ y $\delta_{CP}$) y, debido a la estructura de sabor, el ordenamiento inverso de las masas de neutrinos est\'a excluido a m\'as de $3\sigma$, mientras que el ordenamiento normal sobrevive. Es, por tanto, un modelo de sabor altamente predictivo. Adem\'as este modelo conserva las propiedades del anterior, sobre el que se construye: los neutrinos son part\'iculas de Dirac, la pequeñez de sus masas se explica de forma natural con un mecanismo `seesaw' y existe una relaci\'on entre la naturaleza Dirac de los neutrinos y la estabilidad de la materia oscura.

La contribuci\'on personal del estudiante fue encontrar las cargas del grupo de simetr\'ia no abeliano $A_4$ que llevaran a las predicciones m\'as fuertes. Adem\'as, el estudiante se dio cuenta, a partir de la biblograf\'ia presente, que extendiendo la simetr\'ia $A_4$ a los quarks pod\'ia obtenerse la relaci\'on de masas. El estudiante realiz\'o todos los c\'alculos de teor\'ia de grupos, obtuvo los par\'ametros de mezcla a partir de las cargas de simetr\'ia y realiz\'o los gr\'aficos correspondientes, adem\'as de contribuir a la redacci\'on del manuscrito.

\subsection{Seesaw roadmap to neutrino mass and dark matter}

En \cite{CentellesChulia:2018gwr} obtenemos todas las realizaciones a nivel \'arbol de los operadores de masas de neutrinos de dimesni\'on $5$ tomando escalares que forman parte de distintos multipletes de $SU(2)_L$ hasta tripletes, siendo f\'acil su generalizaci\'on para multipletes mayores. Empezamos por el bien conocido operador de Weinberg $L L  \phi\phi$, que da lugar a los tres modelos de masas de neutrinos de Majorana m\'as sencillos y populares: los tipos I, II y III del mecanismo `seesaw'. Despu\'es procedemos a estudiar los diferentes operadors de masas de Dirac, y aunque hay muchos m\'as, el caso m\'as interesante es el del operador $\bar{L} \phi^c \chi \nu_R$, donde $\phi$ es el doblete de Higgs del modelo est\'andar y $\chi$ es un singlete. Este operador da lugar a los an\'alogos de Dirac directos de los mecanismos `seesaw' I, II y III. Adem\'as, es posible obtener variantes de dichos mecanismos simplemente modificando el multiplete de $SU(2)_L$ de los escalares involucrados en la generaci\'on de masas de neutrinos. 

El estudiante desarroll\'o de forma sistem\'atica todas las contracciones posibles de todos los operadores de dimensi\'on $5$ y realiz\'o los diagramas respectivos. Como en todos los dem\'as trabajos de investigaci\'on, colabor\'o con la discusi\'on e interpretaci\'on de resultados y la escritura del manuscrito.

\subsection{Seesaw Dirac neutrino mass through dimension-six operators}

Construyendo sobre los resultados del art\'iculo anterior, en \cite{CentellesChulia:2018bkz} pasamos a los operadores de dimensi\'on $6$ para masas de neutrinos de Dirac a nivel \'arbol. El n\'umero de diagramas y operadores se multiplica al aumentar la dimensionalidad, por lo que el an\'alisis es m\'as largo y complejo. Algunos de los diagramas pueden indentificarse con casos de dimensi\'on $5$ en los que una de las patas externas obtiene un valor esperado en el vac\'io inducido por otros dos campos. Esta inducci\'on es, de hecho, necesaria cuando una pata externa no es un singlete, ya que de lo contrario en el modelo existir\'a un bos\'on de goldstone sin masa que no forme parte mayoritaria de un singlete. Esto est\'a prohibido por las observaciones, ya que al no ser un singlete acoplar\'a con los bosones gauge y ser\'ia dif\'icil 'esconder' su efecto. Adem\'as, algunos casos aparentemente podr\'ian de forma inocente identificarse con una versi\'on espont\'anea de los modelos de Dirac `seesaw inverso' como ya existen para el caso Majorana. Sin embargo, evitamos una menci\'on expl\'icita a esto pues ese caso es m\'as sutil y lo relegamos a un trabajo futuro, \cite{CentellesChulia:2020dfh}.

De nuevo, la aportaci\'on principal del estudiante fue obtener sistem\'aticamente todas las contracciones posibles de todos los operadores de dimensi\'on $6$ y realizar los diagramas respectivos, participar de la redacci\'on del manuscrito, de la interpretaci\'on de los resultados  y las discusiones.

\subsection{Dark matter stability and Dirac neutrinos using only Standard Model symmetries}

\cite{Bonilla:2018ynb} es probablemente el art\'iculo m\'as completo de todos los presentados en esta tesis. En \'el se unifican y pulen todos los conceptos desarrollados en trabajos anteriores, para obtener un resultado extremadamente elegante y simple. Primero demostramos que los criterios de simetr\'ia generales necesarios para obtener (a) neutrinos de Dirac, (b) masas naturalmente pequeñas (c) estabilidad de la materia oscura y (d) la materia oscura forma parte de la generaci\'on de masas de los neutrinos via 'loop', pueden obtenerse todas ellas con un solo grupo de simetr\'ia. Adem\'as, aunque no es estrictamente necesario para obtener un modelo consistente y fenomenol\'ogicamente correcto, la soluci\'on cumple (e) que esta simetr\'ia no tiene anomal\'ias cu\'anticas, por lo que puede promocionar de simetr\'ia global a simetr\'ia local consiguiendo una fenomenolog\'ia m\'as rica y eliminar el bos\'on Goldstone de la teor\'ia.

Todo ello se consigue simplemente con la simetr\'ia $U(1)_{B-L}$ en la que los neutrinos $\nu_R$ transforman con unas cargas ex\'oticas $(-4, -4, 5)$, prohibiendo el t\'ermino $\bar{L} \phi^c \nu_R$ en el lagrangiano. El singlete escalar $\chi$ rompe la simetr\'ia desde $U(1)$ hasta $Z_6$, que es la que estabiliza al candidato de materia oscura. Aunque el mecanismo es completamente general mostramos un ejemplo particularmente sencillo donde la masa de los neutrinos se obtiene a trav\'es de un mecanismo 'scotog\'enico' a nivel 1 'loop'.

La contribuci\'on del estudiante en este caso es m\'as dif\'icil de identificar puesto que es un art\'iculo bastante m\'as conceptual que se desarroll\'o a trav\'es de m\'ultiples conversaciones y debates entre los coautores. El estudiante particip\'o activamente de dichas discusiones aportando ideas y demostraciones, adem\'as de contribuir a la redacci\'on del manuscrito.

\subsection{The Inverse Seesaw Family: Dirac And Majorana}

En \cite{CentellesChulia:2020dfh} desarrollamos una definici\'on general y rigurosa que permite diferenciar los modelos que podr\'ian entrar en la categor\'ia de `seesaw' inverso y cuales no. Aunque este concepto ha sido utilizado en muchas ocasiones en la bibliograf\'ia, su significado general no hab\'ia sido definido hasta ahora. Utilizamos dicha definici\'on para construir el cl\'asico y bien conocido modelo de `seesaw' inverso para el caso de Majorana y procedemos a generalizarlo en dos direcciones: por un lado construimos el `seesaw' inverso doble, triple etc, donde la supresi\'on de las masas de los neutrinos es m\'utiple, y por otro lado generalizamos los multipletes de $SU(2)_L$ que pueden utilizarse. Despu\'es realizamos el mismo ejercicio para el caso Dirac. Sorprendentemente, el modelo de `seesaw' inverso m\'as sencillo para neutrinos de Dirac es formalmente el mismo que el `seesaw' tipo I: la diferencia estriba en el espacio de par\'ametros y, en concreto, en la escala de ruptura del n\'umero lept\'onico, ya que para una escala muy por encima de la electrod\'ebil tendremos el seesaw de alta escala tipo I y para una escala de ruptura pequeña (orden 1 eV) tendremos un `seesaw' inverso.

La gran ventaja de todos estos modelos es, como en el modelo `seesaw' inverso cl\'asico, que al reducir la escala de los mediadores de la masa de los neutrinos hasta O(TeV) se puede obtener una fenomenolog\'ia m\'as rica que en el caso de un `seesaw' de escala alta. Adem\'as, en el caso de Dirac, puede sumarse a la fenomenolog\'ia de numero lept\'onico local, puesto que la simetr\'ia no es an\'omala al contrario que para el caso Majorana. A costa de complicar el modelo un poco m\'as pueden obtenerse m\'as predicciones siguiendo las generalizaciones propuestas: multipletes de orden mayor implicar\'an mayor probabilidad de detecci\'on en experimentos de colisionador y al hacer los mecanismos ‘doble’, ‘triple’ etc se consigue una supresi\'on a\'un mayor a la masa de los neutrinos, pudiendo de este modo reducir la escala de los mediadores.

La contribuci\'on del estudiante estuvo en la discusi\'on de la primera parte del art\'iculo, m\'as conceptual, y en hallar las posibles generalizaciones del mecanismo para Dirac y Majorana, as\'i como las f\'ormulas generales para la masa de los neutrinos y la redacci\'on del texto. Adem\'as, el estudiante realiz\'o c\'alculos fenomenol\'ogicos para los modelos m\'as simples aunque al final no se incluyeran en el manuscrito final.

\section{Conclusiones}

Los neutrinos son una puerta entreabierta hacia nueva f\'isica. Sabemos que son masivos, cuando el modelo est\'andar predice que no lo sean, pero a d\'ia de hoy no conocemos nada sobre el mecanismo que les otorga masa. Adem\'as, los neutrinos podr\'ian dar la explicaci\'on a diversos problemas aparentemente no relacionados en otros sectores f\'isicos. En la bibliograf\'ia hay miles de art\'iculos con propuestas que podr\'ian explicar la masa de estas part\'iculas, asumiendo la mayor\'ia de ellos que son part\'iculas de Majorana, es decir, que son su propia antipart\'icula. En esta tesis partimos desde el punto de vista contrario, es decir, que sean part\'iculas de Dirac, y exploramos las consecuencias de dicha suposici\'on.
Es claro que para tener neutrinos de Dirac es necesario extender el inventario de simetr\'ias del modelo est\'andar. Esto puede verse como una oportunidad, ya que dicha simetr\'ia extra puede servir otros prop\'ositos: por ejemplo, estabilizar un candidato de materia oscura. Estudiar el patr\'on de ruptura del n\'umero lept\'onico es por tanto clave para dilucidar esta materia. As\'i, la contribuci\'on m\'as relevante de la tesis es la presentaci\'on de un modelo radiativo (donde la masa de los neutrinos sucede no a nivel \'arbol sino a nivel ‘loop’) donde los neutrinos son de Dirac y el patr\'on de ruptura de la simetr\'ia ‘n\'umero bari\'onico menos n\'umero lept\'onico’ consigue al mismo tiempo, sin necesidad de simetr\'ias extra a) estabilidad de la materia oscura b) neutrinos de Dirac naturalmente ligeros c) inexistencia de simetr\'ias accidentales adicionales y d) un inventario de materia nueva minimal.

Adem\'as de discutir los posibles patrones de ruptura de simetr\'ia, en esta tesis describimos sistem\'aticamente una clase de mecanismos elegantes de generaci\'on de masas de neutrinos de Dirac: los mecanismos del balanc\'in (o seesaw), en analog\'ia a los ya existentes para Majorana. Tambi\'en discutimos en m\'as detalle algunos modelos fenomenol\'ogicamente m\'as interesantes en el sector de sabor.


\clearpage


\fancyhf{}
\fancyhead[LE,RO]{\thepage}

\backmatter

\cleardoublepage
\addcontentsline{toc}{chapter}{\refname}

\providecommand{\href}[2]{#2}\begingroup\raggedright\endgroup

\end{document}